\documentclass[ATLAS]{cernphprep}
\pdfoutput=1 % for arxiv submission

\usepackage{graphicx} % This is already loaded by the atlasnote class
\usepackage{atlasphysics} % Contains useful shortcuts. Uncomment to use
\usepackage{subfigure}
\usepackage{longtable}
\usepackage{multirow}
\usepackage{textcomp}
\usepackage{units}
\usepackage{cernunits,cernchemsym,heppennames2}

\usepackage{hyperref}
\usepackage{cite,fancyvrb}
\newcommand{\lumWe}{\ensuremath{315\;\mathrm{nb}^{-1}}}       % Wenu

\newcommand{\lumWmu}{\ensuremath{310\;\mathrm{nb}^{-1}}}      % Wmunu

\newcommand{\lumZe}{\ensuremath{316\;\mathrm{nb}^{-1}}}       % Zee

\newcommand{\lumZmu}{\ensuremath{331\;\mathrm{nb}^{-1}}}      % Zmumu

\newcommand{\lumWZ}{\ensuremath{320\;\mathrm{nb}^{-1}}}       % avg combined W+Z, e+mu

\newcommand{\lumWenolabel}{315}       % Wenu

\newcommand{\lumWmunolabel}{310}      % Wmunu

\newcommand{\lumZenolabel}{316}       % Zee

\newcommand{\lumZmunolabel}{331}      % Zmumu

\newcommand{\sigWeplus}{\ensuremath{ 6.27 \pm 0.26 \mathrm{(stat)} \pm 0.48 \mathrm{(syst)} \pm 0.69 \mathrm{(lumi)} }} % tot cross We+

\newcommand{\sigWeminus}{\ensuremath{ 4.23 \pm 0.22 \mathrm{(stat)} \pm 0.33 \mathrm{(syst)} \pm 0.47 \mathrm{(lumi)} }} % tot cross We-

\newcommand{\sigWe}{\ensuremath{    10.51 \pm 0.34 \mathrm{(stat)} \pm 0.81 \mathrm{(syst)} \pm 1.16 \mathrm{(lumi)} }} % tot cross W e+/-

\newcommand{\sigfidWeplus}{\ensuremath{ 2.92 \pm 0.12 \mathrm{(stat)} \pm 0.21 \mathrm{(syst)} \pm 0.32 \mathrm{(lumi)} }} % tot cross We+

\newcommand{\sigfidWeminus}{\ensuremath{ 1.93 \pm 0.10 \mathrm{(stat)} \pm 0.14 \mathrm{(syst)} \pm 0.21 \mathrm{(lumi)} }} % tot cross We-

\newcommand{\sigfidWe}{\ensuremath{     4.85 \pm 0.16 \mathrm{(stat)} \pm 0.34 \mathrm{(syst)} \pm 0.53 \mathrm{(lumi)} }} % tot cross W e+/-

\newcommand{\sigWmuplus}{\ensuremath{ 5.71 \pm 0.23 \mathrm{(stat)} \pm 0.30 \mathrm{(syst)} \pm 0.63 \mathrm{(lumi)} }} % tot cross Wmu+

\newcommand{\sigWmuminus}{\ensuremath{ 3.86 \pm 0.20 \mathrm{(stat)} \pm 0.20 \mathrm{(syst)} \pm 0.42 \mathrm{(lumi)} }} % tot cross Wmu-

\newcommand{\sigWmu}{\ensuremath{ 9.58 \pm 0.30 \mathrm{(stat)} \pm 0.50 \mathrm{(syst)} \pm 1.05 \mathrm{(lumi)} }} % tot cross Wmu+/-

\newcommand{\sigfidWmuplus}{\ensuremath{ 2.77 \pm 0.11 \mathrm{(stat)} \pm 0.12 \mathrm{(syst)} \pm 0.30 \mathrm{(lumi)} }} % tot cross Wmu+

\newcommand{\sigfidWmuminus}{\ensuremath{ 1.83 \pm 0.09 \mathrm{(stat)} \pm 0.08 \mathrm{(syst)} \pm 0.20 \mathrm{(lumi)} }} % tot cross Wmu-

\newcommand{\sigfidWmu}{\ensuremath{   4.60 \pm 0.15 \mathrm{(stat)} \pm 0.20 \mathrm{(syst)} \pm 0.51 \mathrm{(lumi)} }} % tot cross Wmu+/-

\newcommand{\sigWplus}{\ensuremath{5.93 \ \pm \ 0.17 \ \mathrm{(stat)} \ \pm 0.30 \ \mathrm{(syst)} \ \pm 0.65 \ \mathrm{(lumi)} }}  % tot cross We+ + Wmu+
\newcommand{\sigWminus}{\ensuremath{4.00 \ \pm \ 0.15 \ \mathrm{(stat)} \ \pm 0.20 \ \mathrm{(syst)} \ \pm 0.44 \ \mathrm{(lumi)} }}  % tot cross We+ + Wmu+
\newcommand{\sigW}{\ensuremath{9.96 \ \pm~0.23\mathrm{(stat)} \ \pm~0.50\mathrm{(syst)} \ \pm~1.10\mathrm{(lumi)} }}  % tot cross We+ + Wmu+

\newcommand{\sigZe}{\ensuremath{  0.75 \pm 0.09 \mathrm{(stat)} \pm 0.08 \mathrm{(syst)} \pm 0.08 \mathrm{(lumi)} }}  % tot cross Zee

\newcommand{\sigfidZe}{\ensuremath{ 0.33 \pm 0.04 \mathrm{(stat)} \pm 0.03 \mathrm{(syst)} \pm 0.04 \mathrm{(lumi)} }}  % tot cross Zee

\newcommand{\sigZmu}{\ensuremath{  0.87 \pm 0.08 \mathrm{(stat)} \pm 0.06 \mathrm{(syst)} \pm 0.10 \mathrm{(lumi)} }} % tot cross Zmumu

\newcommand{\sigfidZmu}{\ensuremath{ 0.43 \pm 0.04 \mathrm{(stat)} \pm 0.02 \mathrm{(syst)} \pm 0.05 \mathrm{(lumi)} }} % tot cross Zmumu

\newcommand{\sigZnb}{\ensuremath{ 0.82 \pm 0.06 \mathrm{(stat)} \pm 0.05 \mathrm{(syst)} \pm 0.09 \mathrm{(lumi)} }} % tot cross combined Zee plus Zmumu

\newcommand{\CWeplus}{0.656}  % CW+   (e)

\newcommand{\CWeminus}{0.662} % CW-   (e)

\newcommand{\CWe}{0.659}      % CW+/- (e)

\newcommand{\CWmuplus}{0.765} % CW+   (mu)

\newcommand{\CWmuminus}{0.748}% CW-   (mu)

\newcommand{\CWmu}{0.758}     % CW+/- (mu)

\newcommand{\CZe}{0.651}      % CZ    (e)

\newcommand{\CZmu}{0.773}     % CZ    (mu)

\newcommand{\ntrigWZe}{\ensuremath{6.5\times 10^6}}  % W/Z electron channel

\newcommand{\ntrigWZmu}{\ensuremath{5.1\times 10^6}} % W/Z positron channel

\newcommand{\JFQCDSFe}{2.4}

\newcommand{\JFQCDSFmu}{1.6}

\newcommand{\ntotWeplus}{637}     % total number of W^+ enu

\newcommand{\ntotWeminus}{432}    % total number of W^- enu

\newcommand{\ntotWe}{1069}        % total number of W->enu

\newcommand{\ntotWmuplus}{710}    % total number of W^+ munu

\newcommand{\ntotWmuminus}{471}   % total number of W^- munu

\newcommand{\ntotWmu}{1181}       % total number of W->munu

\newcommand{\ntotW}{2250}         % total number of W (e+mu)

\newcommand{\ntotZenowindow}{78}  % total number of Zee (no mass window)

\newcommand{\ntotZe}{70}          % total number of Zee  (in mass window)

\newcommand{\ntotZmunowindow}{117}% total number of Zmumu (no mass window)  

\newcommand{\ntotZmu}{109}        % total number of Zmumu (in mass window)

\newcommand{\ntotZ}{179}          % total number of Z (e+mu) (in mass window)

\newcommand{\ntotTightWe}{4003}   % number of events after tight

\newcommand{\ntotMETWe}{1116}     % number of events after MET cut

\newcommand{\ntotpTWmu}{7052}     % number of events after after pT>20

\newcommand{\ntotIsolWmu}{2920}   % number of events after isolation

\newcommand{\ntotMETWmu}{1220}    % number of events after after MET cut

\newcommand{\nWtaunuWe}{25.9}         % Wtanunu bkg We+-

\newcommand{\nZeeWe}{1.9}             % Zee bkg  We+-

\newcommand{\nZtautauWe}{1.6}         %Ztautau bkg We+-

\newcommand{\nttbarWe}{4.1}           % ttbar bkg We+- 

\newcommand{\nEWttbarbkgWeplus}{18.8 $\pm$ 0.2 $\pm$ 1.7}     % sum of EW and ttbar bkg: We+

\newcommand{\nEWttbarbkgWeminus}{14.7  $\pm$ 0.2 $\pm$ 1.3}    % sum of EW and ttbar bkg: We-

\newcommand{\nEWttbarbkgWe}{33.5  $\pm$ 0.2 $\pm$ 3.0}         % sum of EW and ttbar bkg: We+-

\newcommand{\nQCDbkgWeplus}{14.0 $\pm$ 2.1 $\pm$ 7.1}                     % Jochen's # QCD evts We+

\newcommand{\nQCDbkgWeminus}{14.0 $\pm$ 2.1 $\pm$ 7.1}                    % Jochen's # QCD evts We-

\newcommand{\nQCDbkgWe}{28.0 $\pm$ 3.0 $\pm$ 10.0}                        % Jochen's # QCD evts We+-

\newcommand{\nQCDbkgWestat}{28.0 $\pm$ 3.0(stat)}     % Jochen's # QCD evts We+- (with a stat label)

\newcommand{\nQCDWe}{\ensuremath{30.8 \pm 6.1\mathrm{(stat)}}}          % # QCD according the JF17

\newcommand{\nQCDbkgTightWe}{\ensuremath{48.0 \pm 17.0\mathrm{(stat)}}} % Jochen's # events after tight (iso method)

\newcommand{\ntotsignalWeplus}{604.2 $\pm$ 25.2 $\pm$ 7.6}    % background subtracted signal We-

\newcommand{\ntotsignalWeminus}{403.2 $\pm$ 20.8 $\pm$ 7.5}    % background subtracted signal We-

\newcommand{\ntotsignalWe}{1007.5 $\pm$ 32.7 $\pm$ 10.8}       % background subtracted signal We+-

\newcommand{\nWtaunuWmu}{33.6}    % Wtanunu bkg    

\newcommand{\nZmumuWmu}{38.4}     % Zmumu bkg

\newcommand{\nttbarWmu}{4.2}      % ttbar bkg

\newcommand{\nZtautauWmu}{1.4}    % Ztautau bkg

\newcommand{\nEWttbarbkgWmuplus}{ $42.5 \pm 0.2 \pm 2.9 $}  % sum of EW and ttbar bkg: Wmu+

\newcommand{\nEWttbarbkgWmuminus}{$35.1 \pm 0.2 \pm 2.4 $}  % sum of EW and ttbar bkg: Wmu-

\newcommand{\nEWttbarbkgWmu}{     $77.6 \pm 0.3 \pm 5.4 $}  % sum of EW and ttbar bkg: Wmu+-

\newcommand{\nQCDWmulabel}{\ensuremath{21.1 \pm 4.5\mathrm{(stat)} \pm 8.7 \mathrm{(syst)}}}        % QCD bkg in W+- with label

\newcommand{\nQCDWmuAlternate}{\ensuremath{9.7 \pm 0.4\mathrm{(stat)}}}    % QCD bkg (alternate method)

\newcommand{\nQCDWmuAlternateTwo}{\ensuremath{13.5 \pm 0.9\mathrm{(stat)} \pm 12.7 \mathrm{(syst)}}}    % QCD bkg (alternate method)

\newcommand{\nnonEWWmuplus}{  $12.0 \pm 3.0 \pm 4.6 $}          % QCD bkg in W+

\newcommand{\nnonEWWmuminus}{ $10.9 \pm 2.4 \pm 4.1 $}          % QCD bkg in W-

\newcommand{\nnonEWWmu}{      $22.8 \pm 4.6 \pm 8.7 $}          % QCD bkg in W+-

\newcommand{\nCosmic}{\ensuremath{1.7 \pm 0.8}}    % Cosmic bkg

\newcommand{\ntotsignalWmuplus}{ $655.6 \pm 26.6 \pm 6.2 $}   % background subtracted signal Wmu+

\newcommand{\ntotsignalWmuminus}{$425.0 \pm 21.7 \pm 5.4 $}   % background subtracted signal Wmu-

\newcommand{\ntotsignalWmu}{    $1080.6 \pm 34.4 \pm 11.2 $}   % background subtracted signal Wmu+-

\newcommand{\nWenuZe}{0.11}                                                 % Wenu bkg

\newcommand{\nZtautauZe}{0.06}                                              % Ztautau bkg

\newcommand{\nttbarZe}{0.10}                                                % ttbar bkg

\newcommand{\nEWttbarbkgZe}{$0.27  \pm 0.00 \pm 0.03$}                      % sum EW&ttbar bkg: Zee

\newcommand{\rejectZe}{\ensuremath{0.137\pm 0.001 \mathrm{(stat)}}}         % loose->medium rejection used in QCD estimation 

\newcommand{\nQCDbkgZe}{$0.91 \pm 0.11 \pm 0.41$}                           % tot # of QCD bkg data 

\newcommand{\nQCDbkgZelabel}{\ensuremath{0.91 \pm 0.11\mathrm{(stat)}}}     % tot # of QCD bkg data (with stat label)

\newcommand{\nQCDbkgZesyslabel}{\ensuremath{0.91 \pm 0.11\mathrm{(stat)}\pm 0.41\mathrm{(sys)}}}  % tot # of QCD bkg data (with stat and sys label)

\newcommand{\nQCDbkgZeMClabel}{\ensuremath{0.87 \pm 0.04\mathrm{(stat)}} }  % tot # of QCD bkg MC  (with stat label)

\newcommand{\ntotsignalZe}{$68.8 \pm 8.4 \pm 0.4 $}                         % bkg-subtracted signal We+

\newcommand{\nWmunuZmu}{0.01}                                        % Wmunu bkg 

\newcommand{\nZtautauZmu}{0.09}                                      % Ztautau bkg

\newcommand{\nttbarZmu}{0.11}                                        % ttbar bkg

\newcommand{\nEWttbarbkgZmu}{$0.21  \pm 0.01 \pm 0.01 $}             % sum EW and ttbar bkg: Zmumu

\newcommand{\nQCDbkgZmu}{$0.04 \pm 0.01 \pm 0.04$}                   % tot number QCD bkg = bbbar 

\newcommand{\ntotsignalZmu}{$108.8 \pm 10.4 \pm 0.0 $}               % bkg-subtracted signal Zmumu

\newcommand{\Zg}{\mbox{$Z /\gamma^*$}}

\newcommand{\stotWp}{\mbox{$\sigma_{W^+}^{\rm{tot}}$}}
\newcommand{\stotWm}{\mbox{$\sigma_{W^-}^{\rm{tot}}$}}
\newcommand{\stotW}{\mbox{$\sigma_{W}^{\rm{tot}}$}}
\newcommand{\stotWpm}{\mbox{$\sigma_{W^{(\pm)}}^{\rm{tot}}$}}
\newcommand{\stotZ}{\mbox{$\sigma_{Z}^{\rm{tot}}$}}
\newcommand{\stotZg}{\mbox{$\sigma_{Z/\gamma^*}^{\rm{tot}}$}}
\newcommand{\stotWZ}{\mbox{$\sigma_{W(Z)}^{\rm{tot}}$}}
\newcommand{\sfidWp}{\mbox{$\sigma_{W^+}^{\rm{fid}}$}}
\newcommand{\sfidWm}{\mbox{$\sigma_{W^-}^{\rm{fid}}$}}

\newcommand{\sfidWpm}{\mbox{$\sigma_{W^{(\pm)}}^{\rm{fid}}$}}

\newcommand{\sfidZg}{\mbox{$\sigma_{Z/\gamma^*}^{\rm{fid}}$}}
\newcommand{\sfidWZ}{\mbox{$\sigma_{W(Z)}^{\rm{fid}}$}}
\newcommand{\cw}{\mbox{$C_W$}}
\newcommand{\cz}{\mbox{$C_Z$}}

\begin{document}

\begin{titlepage}
\PHnumber{-2010-037}
\PHdate{7 Oct. 2010}
\DEFCOL{}

\title{Measurement of the $\ensuremath{W}\rightarrow \ell \nu$ and $\Zg \to \ell \ell$ production cross sections  in proton-proton collisions at $\sqrt{s}=7$~TeV with the ATLAS detector}

\vspace*{1.0cm}

\Collaboration{The ATLAS Collaboration%
 \thanks{See Appendix for the list of collaboration members}}
\ShortAuthor{The ATLAS Collaboration}
\vspace*{2.0cm}

\begin{abstract}
First measurements of the $\ensuremath{W}\rightarrow \ell \nu$ and $\Zg \to \ell \ell$ ($\ell=e,\mu$) production cross sections in  proton-proton collisions at $\sqrt{s}=7$~TeV are presented using data recorded by the ATLAS experiment at the LHC. The results are based on \ntotW~$\ensuremath{W}\rightarrow \ell \nu$ and \ntotZ~$\Zg \to~\ell\ell$ candidate events selected from a data set corresponding to an integrated luminosity of approximately
 \lumWZ. The measured total $W$ and $Z/\gamma^*$-boson production cross sections times the respective leptonic branching ratios for the combined electron and muon channels are $\stotW$ $\cdot$ BR($W \to \ell \nu$) = 9.96~$\pm$~0.23(stat) $\pm$~0.50(syst) $\pm$~1.10(lumi)~nb and
$\stotZg$ $\cdot$ BR($\Zg \to \ell \ell$)~=~$\sigZnb$~nb (within the invariant mass window $66<m_{\ell\ell}<116$~GeV). The $W/Z$ cross-section ratio is measured to be  11.7 $\pm$ 0.9(stat) $\pm$ 0.4(syst). In addition, measurements of the $W^+$ and $W^-$ production cross sections
and of the lepton charge asymmetry are reported. Theoretical predictions based on NNLO QCD calculations are found to agree with the measurements. 
\end{abstract}
\end{titlepage}
%}

%%%%%%%%%%%%%%%%%%%%%%%%%%%%%%%%%%%%
%            Content               %
%%%%%%%%%%%%%%%%%%%%%%%%%%%%%%%%%%%%

\clearpage

\section{Introduction}
\label{sec:intro}

Measurements of the inclusive production cross sections of the \Wboson~and \Zboson~bosons at hadron colliders constitute an important test of the Standard Model. The theoretical calculations involve parton distribution functions (PDF) and different couplings of the partons to the weak bosons. They are affected by significant higher-order QCD corrections. Calculations of the inclusive $W$ and $Z$ production cross sections have been carried out at next-to-leading order (NLO)~\cite{KubarAndre:1978uy,Altarelli:1979ub,Kubar:1980zv} and next-to-next-to leading order (NNLO) in perturbation theory~\cite{Rijken:1994sh,Hamberg:1990np,vanNeerven:1991gh,Harlander:2002wh,Anastasiou:2003ds}.

The production of \Wboson~and \Zboson~bosons at hadron colliders was measured previously by the UA1~\cite{UA1W2} and UA2~\cite{UA2W} experiments at $\sqrt{s}$ = 0.63 TeV at the CERN Sp$\bar{\rm{p}}$S and by the CDF~\cite{Abe:1995bm,Abe:1998gq,CDFW} and D0~\cite{Abbott:1999tt,D0W} experiments at $\sqrt{s}$ = 1.8 TeV and $\sqrt{s}$ = 1.96~TeV at the Fermilab Tevatron proton-antiproton colliders. In contrast to proton-antiproton collisions, the cross sections for \Wplus~and \Wminus~production are expected to be different in proton-proton collisions due to different valence quark distributions of the $u$ and $d$ quarks. Most recently, the RHIC collider experiments~\cite{Adare:2010xa,Aggarwal:2010vc} have reported the first observation of \Wboson~production in proton-proton collisions at $\sqrt{s} = 0.5$~TeV.

\Wboson~and \Zboson~bosons are expected to be produced abundantly at the Large Hadron Collider (LHC)~\cite{LHC:2008}. The projected large dataset and the high LHC energy will allow for detailed measurements of their production properties in a previously unexplored kinematic domain. These conditions, together with the proton-proton nature of the collisions, will provide new constraints on the parton  distribution functions and will allow for precise tests of perturbative QCD. Besides the measurements of the $W$ and $Z$ boson production cross sections, the measurement
of their ratio $R$ and of the asymmetry between the $W^+$ and $W^-$ cross sections constitute important
tests of the Standard Model. The ratio $R$ can be measured with a higher
relative precision because both experimental and theoretical uncertainties partially cancel.
With larger data sets this ratio can be used to provide constraints on the
$W$-boson width $\Gamma_W$ \cite{CDFW}.

This paper describes the first measurement of the $W^+, W^-$ and $\Zg$ boson production cross sections
in proton-proton collisions at $\sqrt{s}$ = 7 TeV by the ATLAS~\cite{DetectorPaper:2008}
experiment at the LHC. The measurements are based on data corresponding to an integrated luminosity of
approximately \lumWZ. The inclusive $\Zg$-production-cross section is measured within the mass range 66 $< m_{\ell \ell} <$ 116 GeV.
In addition to the individual cross-section measurements, first measurements of the ratio $R$ of the $W$ to
$Z$ cross sections
and  of the $W \to \ell \nu$ charge asymmetry are presented. Throughout this paper the label ``$Z$'' refers to $Z/\gamma^*$.

The paper is organized as follows: after a short description of the ATLAS detector, the data set and the Monte-Carlo samples in Sections 2 and 3, the identification of electrons, muons and the measurement of the transverse missing energy are discussed in Section 4. In Section 5, the selection of $W \to \ell \nu$ and $Z \to \ell \ell$ candidates is presented. Section 6 is devoted to a detailed discussion of backgrounds in these samples. The measurement of the $W \to \ell \nu$ and $Z \to \ell \ell$ cross sections and of their ratio is presented in Section 7 together with a comparison to theoretical predictions.  The measurement of the $W \to \ell \nu$ charge asymmetry is discussed in Section 8.

\section{The ATLAS detector}

   The ATLAS detector~\cite{DetectorPaper:2008} at the LHC comprises a thin superconducting solenoid surrounding the inner-detector and three large superconducting toroids arranged with an eight-fold azimuthal coil symmetry placed around the calorimeters, forming the basis of the muon spectrometer.

   The Inner-Detector (ID) system is immersed in a $2$\,T axial magnetic field and provides tracking information for charged particles in a pseudorapidity range matched by the precision measurements of the electromagnetic calorimeter; the silicon tracking detectors, pixel and silicon microstrip (SCT), cover the pseudorapidity
range $|\eta|< 2.5$.\footnote{The nominal interaction point is defined as the origin of the coordinate system, while the anti-clockwise beam direction defines the $z$-axis and the $x-y$ plane is transverse to the beam direction. The positive $x$-axis is defined as pointing from the interaction point to the centre of the LHC ring and the positive $y$-axis is defined as pointing upwards.  The azimuthal angle $\phi$ is measured around the beam axis and the polar angle $\theta$ is the angle from the beam axis. The pseudorapidity is defined as $\eta = -\ln \tan(\theta/2)$. The distance~$\Delta R$ in the $\eta-\phi$ space is defined as $\Delta R = \sqrt{(\Delta\eta)^2+(\Delta\phi)^2}$.}
The highest granularity is achieved around the vertex region using the pixel detectors. The Transition Radiation Tracker (TRT), which surrounds the silicon detectors, enables track-following up to $|\eta| = 2.0$. Electron identification information is provided by the detection of transition radiation in the  TRT straw tubes.

   The calorimeter system covers the pseudorapidity range $|\eta| < 4.9$. It is based on two different detector technologies, with liquid argon (LAr) and  scintillator-tiles as active media. The electromagnetic (EM) calorimeter, consisting of lead absorbers and liquid argon as the active material, is divided into one barrel ($|\eta| < 1.475$) and two end-cap components ($1.375 < |\eta| < 3.2$). It uses an accordion geometry to ensure fast and uniform response. It has a fine segmentation in both the lateral and longitudinal directions of the particle showers. At high energy, most of the EM shower energy is collected in the second layer which has a lateral cell granularity of
$\Delta \eta \times \Delta \phi$ = 0.025 $\times$ 0.025. The first layer is segmented into eight strips per cell in the $\eta$ direction which extend over four cells in $\phi$. A third layer measures the tails of very high energy EM showers and helps in rejecting hadron showers.
In the region  $|\eta| < 1.8$, a presampler detector consisting of a thin layer of LAr
is used to correct for the energy lost by electrons, positrons, and photons upstream of the calorimeter.
The hadronic tile calorimeter is placed directly outside the EM calorimeter envelope. This steel/scintillating-tile detector consists of a barrel covering the region $|\eta| < 1.0$, and two extended barrels in the range $0.8 < |\eta| < 1.7$.  The copper Hadronic End-cap Calorimeter (HEC), which uses LAr as active material, consists of two independent wheels per end-cap ($1.5<|\eta|<3.2$), located directly behind the end-cap electromagnetic calorimeter. The Forward Calorimeter (FCal), which also uses LAr as the active material, consists of three modules in each end-cap: the first, made of copper, is optimised for electromagnetic measurements, while the other two, made of tungsten, measure primarily the energy of hadronic interactions~\cite{aad0912}.

   Muon detection is based on the magnetic deflection of muon tracks in the large superconducting air-core toroid magnets, instrumented with separate trigger and high-precision tracking chambers. A system of three toroids, a barrel and two end-caps, generates the magnetic field for the muon spectrometer in the pseudorapidity range $|\eta|<2.7$. Over most of the $\eta$-range, a precision measurement of the track coordinates in the principal bending direction of the magnetic field is provided by Monitored Drift Tubes (MDTs). At large pseudorapidities, Cathode Strip Chambers (CSCs) with higher granularity are used in the innermost plane (station) over $2.0 < |\eta| < 2.7$, to withstand the demanding rate and background conditions expected with the LHC operation at the nominal luminosity. The muon trigger system,  which covers the pseudorapidity range $|\eta| < 2.4$, consists of Resistive Plate Chambers (RPCs) in the barrel ($|\eta|<1.05$) and Thin Gap Chambers (TGCs) in the end-cap regions ($1.05 < |\eta| < 2.4$), with a small overlap in the $|\eta| = $1.05 region.

The first-level (L1) trigger system uses a subset of the total detector information to make a decision on whether or not to record each event, reducing the data rate to a design value of approximately 75~kHz. Details about the L1 calorimeter and muon trigger systems used in the $W$ and $Z$~analyses are provided in Section~\ref{sec:mcsamples}. The subsequent two levels, collectively known as the high-level trigger, are the Level-2 (L2) trigger and the event filter. They provide the reduction to a final data-taking rate designed to be approximately 200~Hz.

\section{Data and Monte-Carlo samples}
\label{sec:mcsamples}

The data were collected over a four-month period, from March to July 2010.  Application of basic beam, detector, and data-quality requirements resulted in total integrated luminosities of \lumWe~for 
the  $W~\to~e\nu$, \lumWmu~for the $\ensuremath{W}\rightarrow \mu \nu$, \lumZe~for the $\ensuremath{Z}\rightarrow ee$, and \lumZmu~for the $\ensuremath{Z}\rightarrow \mu\mu$ channels. The uncertainty on the absolute luminosity determination is $\pm$11\%~\cite{lumiconf}.

Events in this analysis are selected using only the hardware-based L1 trigger, i.e. without use of the high-level trigger. The L1 calorimeter trigger selects photon and electron candidates within $|\eta| < 2.5$  using calorimeter information in trigger towers of dimension $\Delta\eta \times \Delta \phi= 0.1 \times 0.1$.  The calorimeter trigger used in this analysis accepts electron and photon candidates if the transverse energy from a cluster of trigger towers is above  approximately 10~GeV.  The L1 muon trigger searches for patterns of hits  within $|\eta| < 2.4$ consistent with high-\pT\ muons originating from the interaction region.
The algorithm requires a coincidence of hits in the different trigger stations along a road which follows the path of a muon from the interaction point through the detector. The width of the road is related to the $\pT$~threshold to be applied. The muon trigger used in this analysis  corresponds to a threshold of approximately 6~GeV.
As a result of these trigger decisions, a total of \ntrigWZe~and \ntrigWZmu~events are triggered in the electron and muon channels, respectively.

In order to compare the data with theoretical expectations and to estimate the backgrounds from various physics
processes, Monte-Carlo simulations were performed.
For the $W$ and $Z$ signal processes, dedicated $W \to \ell \nu $ and $Z \to \ell \ell$ signal samples were generated.
For the backgrounds the following processes were considered:

\begin{itemize}
\item  $W \to \tau \nu$:
this process is expected to contribute, in particular via leptonic tau decays, $ \tau \to \ell \nu \nu$,
to both electron and muon final states in the $W$ analysis.
\item $Z \to \ell \ell $: $ Z \to \mu \mu$ decays with one muon outside of the muon-spectrometer acceptance generate apparent missing transverse energy and  constitute an important background in the $W \to \mu \nu$ analysis.
Due to the larger $\eta$ coverage of the calorimeter system, this effect is less severe for the
corresponding $Z \to ee$ decays in the $W \to e \nu$ analysis.
\item  $Z\rightarrow \tau \tau$:
these decays contribute a smaller background to both the $W$ and $Z$ analyses via single or double leptonic tau decays.
\item \ttbar\ production:
the production of top pairs constitutes an additional background to both the $W$ and $Z$ analyses. The relative size, compared to
the backgrounds from $W$ and $Z$ decays, depends on the channel considered.
\item Jet production via QCD processes:
the production of jets via QCD processes (referred to as ``QCD background'' in the following) is another
important background contribution.
It has significant components from semi-leptonic decays of heavy quarks, hadrons misidentified as leptons and, in the case of the electron channel, electrons from conversions.
For the $Z \to \mu \mu$ analysis, dedicated  $b\bar{b}$ and $c\bar{c}$ samples were generated in addition, to
increase the statistics for these background components.
\end{itemize}

An overview of all signal and background processes considered and of the generators used for the simulation is given in
Table~\ref{MC}.
All signal and background samples were
generated at $\sqrt{s}=7$~TeV, then processed with the GEANT4~\cite{geant4}~simulation of the ATLAS detector~\cite{:2010wq}, reconstructed and passed through the same analysis chain as the data.
For the comparison to data, all cross sections, except the dijet cross section, are normalised to the results
of higher order QCD calculations (see Table~\ref{MC}). More details on the calculations for the $W$ and $Z$ processes
and on the assigned uncertainties are presented in Section~\ref{WXsection_comp}.
For the $t \bar{t}$ production cross section, an uncertainty of $\pm$6\% is assumed.

For the QCD background, no reliable prediction can be obtained from a leading order Monte-Carlo simulation. For the comparisons
of differential distributions to data, as presented in Section~\ref{sec:selection}, this background is normalised to data.
However, for the final cross-section measurement, except for the $Z \to \mu \mu$ analysis,
data-driven methods are used to determine the residual contributions of
the QCD background to the final $W$ and $Z$ samples, as discussed in Section~\ref{sec:wbkg}.

During the period these data were recorded, the average pile-up varied from zero to about two extra interactions per event, with most of the data being recorded with roughly one extra interaction per event. To account for this, the $\ensuremath{W}\rightarrow \ell \nu$, $\ensuremath{Z}\rightarrow \ell\ell$, and QCD-dijet Monte-Carlo samples were generated with on average two extra primary interactions and then weighted to the primary vertex multiplicity distribution observed in the data.

All data distributions in this paper are shown with statistical uncertainties only, based on Poisson statistics~\cite{Nakamura:2010zzi},
unless otherwise stated.

\begin{table}[t]
\begin{footnotesize}
\begin{center}
\begin{tabular}{ l|l|r r c}
\hline
\hline
\raisebox{-0.4ex}{Physics process} & Generator& {\raisebox{-0.4ex}{$\sigma \cdot$ BR [nb]}} &  \\
\hline
\hline
$W \to \ell \nu$ ~~($\ell = e,\mu$)           &PYTHIA~\cite{pythia}   & 10.46$\pm$0.52  &NNLO & \cite{Hamberg:1990np,Anastasiou:2003ds}  \\
 \hspace{0.5cm} $W^+\rightarrow \ell^+ \nu$        & & 6.16$\pm$0.31  &NNLO & \cite{Hamberg:1990np,Anastasiou:2003ds}  \\
 \hspace{0.5cm} $W^-\rightarrow \ell^- \overline{\nu}$ &         & 4.30$\pm$0.21  &NNLO & \cite{Hamberg:1990np,Anastasiou:2003ds}  \\
$\Zg \to \ell \ell$ ~~($m_{\ell\ell}>60$~GeV)         &PYTHIA    & 0.99$\pm$0.05 &NNLO&\cite{Hamberg:1990np,Anastasiou:2003ds} \\
\hline
$W \rightarrow \tau \nu$               &PYTHIA   & 10.46$\pm$0.52  &NNLO & \cite{Hamberg:1990np,Anastasiou:2003ds}  \\
$W \rightarrow \tau\nu \rightarrow  \ell \nu \nu \nu$         &PYTHIA & 3.68$\pm$0.18 &NNLO&\cite{Hamberg:1990np,Anastasiou:2003ds} \\
$\Zg \to \tau \tau$ ~~($m_{\ell\ell}>60$~GeV)         &PYTHIA    & 0.99$\pm$0.05 &NNLO&\cite{Hamberg:1990np,Anastasiou:2003ds} \\
$t\bar{t}$         &MC@NLO\cite{mcatnlo,mcatnlo2}, & 0.16$\pm$0.01  &NLO+NNLL& \cite{Bonciani:1998vc,Moch:2008qy,Beneke:2009ye}\\
                                                             &  POWHEG~\cite{powheg}         &     &         &      \\
Dijet ($e$ channel, $\hat{p}_{\mathrm{T}}>15$~GeV)      &PYTHIA  & 1.2$\times 10^6$  &LO&\cite{pythia}\\
Dijet ($\mu$ channel, $\hat{p}_{\mathrm{T}}> 8$~GeV)      &PYTHIA  & 10.6$ \times 10^6$&LO&\cite{pythia} \\
$b\overline{b}$ ($\mu$ channel, $\hat{p}_{\mathrm{T}}>18$~GeV, $p_{\rm{T}}(\mu) >$ 15 GeV) & PYTHIA &$73.9$ &LO&\cite{pythia} \\
$c\overline{c}$ ($\mu$ channel, $\hat{p}_{\mathrm{T}}>18$~GeV, $p_{\rm{T}}(\mu) >$ 15 GeV) & PYTHIA & $28.4$&LO&\cite{pythia}  \\
\hline
\hline

\end{tabular}
\caption{Signal and background Monte-Carlo samples as well as the generators used in the simulation.
For each sample the production cross section, multiplied by the relevant branching ratios (BR),
to which the samples were normalised is given. For the electroweak ($W$ and $Z$ boson production) and for the $\ttbar$ production,  contributions
from higher order QCD corrections are included. The inclusive QCD jet and heavy quark cross sections are given at leading order (LO).
These samples were generated with requirements on the transverse momentum of the partons involved in the hard-scattering process, $\hat{p}_{\mathrm{T}}$.
All Monte-Carlo samples result in negligible statistical uncertainties, unless otherwise stated.}
\label{MC}
\end{center}
\end{footnotesize}
\end{table}

\section{Reconstruction of electrons, muons and missing transverse energy}

\subsection{Track reconstruction in the inner detector}

The reconstruction of both electrons and muons uses reconstructed charged tracks in the inner detector.
A detailed description of the track reconstruction has already been presented in Ref.~\cite{minbiaspaper}.
The inner tracking system measures charged particle tracks at all $\phi$ over the pseudorapidity region
$\abseta <$ 2.5 using the pixel, SCT and TRT detectors. Tracks are reconstructed using a pattern recognition
algorithm that starts with the silicon information and adds hits in the TRT.
This ``inside-out'' tracking procedure selects track candidates with transverse momenta above
100~MeV~\cite{atlas_idtracking}.
One further pattern recognition step is then run, which only looks at hits not previously used.
It starts from the TRT and works inwards adding silicon hits as it progresses.
In this second step, tracks from secondary interactions, such as photon conversions and long-lived hadron decays,
with transverse momenta above 300~MeV are recovered.

\subsection{Electrons}
\label{sec:electrons}

The ATLAS standard electron reconstruction and identification algorithm~\cite{eg900} is designed to provide various levels of background rejection for high identification efficiencies for calorimeter transverse energy \ET~$>$~20~GeV, over the full acceptance of the inner-detector system.
Electron reconstruction begins with a seed cluster of \ET~$>2.5$~GeV in the second layer of the electromagnetic calorimeter.
A matching track, extrapolated to the second EM calorimeter layer, is searched for in a broad window of $\Delta \eta\times\Delta \phi=0.05\times0.1$ amongst all reconstructed tracks with \pT~$>$~0.5~GeV. The closest-matched track to the
cluster barycentre is kept as that belonging to the electron candidate. The final electron candidates have cluster sizes of $\Delta \eta\times\Delta \phi=0.075\times 0.175$ in the barrel calorimeter and $0.125\times 0.125$ in the end-cap. The transverse energy of these electron candidates is obtained from the corresponding calorimeter clusters.

  The electron identification selections are based on criteria using calorimeter and tracker information and have been optimised in 10 bins in $\eta$ and 11 bins in \ET. Three reference sets of requirements (``loose'', ''medium'', and ``tight'') have been chosen, providing progressively stronger jet rejection at the expense of some identification efficiency loss.  Each set adds additional constraints to the previous requirements:
\begin{itemize}
  \item {\bf ``Loose'':} this basic selection uses EM shower shape information from the second layer of the EM calorimeter (lateral shower containment and shower width) and energy leakage into the hadronic calorimeters as discriminant variables. This set of requirements provides high and uniform identification efficiency but a low background rejection.
  \item {\bf ``Medium'':} this selection provides additional rejection against hadrons by evaluating the energy deposit patterns in the first layer of the EM calorimeter (the shower width and the ratio of the energy difference associated with the largest and second largest energy deposit over the sum of these energies), track quality variables (number of hits in the pixel and silicon trackers, transverse distance of closest approach to the primary vertex (transverse impact parameter)) and a cluster-track matching variable ($\Delta\eta$ between the cluster and the track extrapolated to the first layer of the EM calorimeter).
  \item {\bf ``Tight'':} this selection further rejects charged hadrons and secondary electrons from conversions by fully exploiting the electron identification potential of the ATLAS detector.  It makes requirements on the ratio of cluster energy to track momentum, on the number of hits in the TRT, and on the ratio of high-threshold hits\footnote{The TRT readout discriminates at two thresholds. The lower one is set to register minimum-ionizing particles and the higher one is intended for the detection of
transition radiation.} to the total number of hits in the TRT.
Electrons from conversions are rejected by requiring at least one  hit in the first layer of the pixel detector. A conversion-flagging algorithm is also used to further reduce this contribution. The
impact-parameter requirement applied in the medium selection is further tightened at this level.
\end{itemize}

\Zee\ and \Wen\ signal Monte-Carlo samples were used to estimate the medium and tight electron identification efficiencies within the relevant kinematic and geometrical acceptance ($\ET~>~20$~GeV within the range $|\eta|<2.47$ and  excluding the transition region between the barrel and end-cap calorimeters, $1.37<|\eta|<1.52$). The efficiencies are estimated to be 94.3\% and 74.9\% respectively, relative to the basic reconstruction efficiency of~97\% which requires a very loose matching between the candidate electron track and the electromagnetic cluster. Using QCD dijet background Monte-Carlo samples, the corresponding rejections against
background from hadrons or conversion electrons in generated jets with $\ET > 20$~GeV within the relevant kinematic and geometrical acceptance are found to be
5700 and 77000, respectively.

Given the limited available statistics of~\Zee\ decays, the electron performance cannot yet be evaluated in detail with collision data. The overall uncertainty on the electron energy scale is estimated to be~$\pm 3$\%, based on extrapolations from test-beam measurements. The uncertainty on the electron energy resolution is also based on extrapolations from test-beam measurements and has a negligible impact on the measurements reported here.

The material in front of the electromagnetic calorimeter affects the reconstruction and identification efficiencies as well as the correct identification of the charge of the reconstructed electron. This has been studied in detail with dedicated simulations including additional material in the inner detector and in front of the electromagnetic calorimeter. The amount of additional material which might be present is currently best constrained by track efficiency measurements in minimum bias data~\cite{minbiaspaper} and studies of photon conversions. The probability for wrongly identifying the charge of the electron depends strongly on the amount of material it traverses in the inner detector and therefore on~\eta. It is expected to be~$(1.9 \pm 0.3)\%$ for the medium electron identification cuts (this affects the selection of $Z$-boson candidates as discussed in~Section~\ref{sec:zbkg})
and~$(0.6 \pm 0.3)\%$ for the tight identification cuts (this affects the measurement of the $W$-boson asymmetry as discussed in~Section~\ref{Wasymmetry}).

The most precise current estimate of the electron identification efficiencies is obtained from a sample of \Wen\ candidates which were selected using tight cuts on the missing transverse energy and the topology of the event and requiring only that an electron candidate be reconstructed through the very loose match between a track and an electromagnetic cluster mentioned above.
The residual background from QCD dijets was estimated using a calorimeter isolation technique similar to that described in~Section~\ref{Z selection}.
The results obtained for the medium efficiency were~$0.900 \pm 0.014 \rm{(stat)} \pm 0.040 \rm{(syst)}$ compared to~0.943 from the Monte Carlo. For the tight efficiency, the corresponding results
were~$0.742 \pm 0.013 \rm{(stat)} \pm 0.030 \rm{(syst)}$ compared to~0.749 from the Monte Carlo. These measurements
confirm that, within the current uncertainties, the electron identification efficiencies are well modelled by
the simulation and are used to evaluate the systematic uncertainties
discussed in~Section~\ref{WXsection_effic}.

\subsection{Muons}
\label{sec:muons}

The ATLAS muon identification and reconstruction algorithms take advantage of multiple sub-detector technologies which provide complementary approaches and cover pseudorapidities up to 2.7~\cite{Performance08}.

The \emph{stand-alone muon} reconstruction is based entirely on muon-spectrometer information, independently of whether or not the muon-spectrometer track is also reconstructed in the inner detector.
The muon reconstruction is initiated locally in a muon chamber by a search for straight line track segments in the bending plane. Hits in the precision chambers are used and the segment candidates are required to point to the centre of ATLAS. When available, the hit coordinate $\phi$ in the non-bending plane
measured by the trigger detectors is associated to the segment. Two or more track segments in different muon stations are combined to form a muon track candidate using three-dimensional tracking in the magnetic field.
The track parameters (\pt , $\eta$, $\phi$, transverse and longitudinal distances of closest approach to the primary vertex) obtained from the muon spectrometer track fit are extrapolated to the interaction point taking into account both multiple scattering and energy loss in the calorimeters. For the latter, the reconstruction utilises either a parameterisation or actual measurements of
calorimeter energy losses, together with a parameterisation of energy loss in the inert material.
The average muon energy loss in the calorimeters is 3~GeV.
The stand-alone muon reconstruction algorithms use the least-squares formalism to fit tracks in the muon spectrometer, with most material effects directly integrated into the $\chi^2$ function.

The \emph{combined muon} reconstruction associates a  stand-alone muon-spectrometer track to an inner-detector track. The association is performed using a $\chi^2$-test, defined from the difference between the respective track parameters weighted by their combined covariance matrices. The parameters are evaluated at the point of closest approach to the beam axis. The combined track parameters are derived either from a statistical combination of the two tracks or from a refit of the full track. To validate the results presented in this paper, these two independent reconstruction chains were exercised and good agreement was observed. The results presented here are based on the statistical combination of muon-spectrometer and inner-detector measurement.

Detailed studies of the muon performance in collision data were performed. The muon momentum scale and resolution were extracted by fitting the invariant mass distribution of the \Zboson~candidates described in Section~\ref{sec:wkine} to  a Breit-Wigner function convolved with a Gaussian function. The fitted mean value indicates that the muon-momentum scale is within
$\pm$1\% around the nominal value. From the fitted width the muon-momentum resolution, for muons from $Z$ decays,
is extracted to be $(4 \pm 2)$\% in the barrel and $(7 \pm 3)\%$ in the end-cap regions. These results are consistent
with those obtained from the single muon studies reported in Ref.~\cite{muconf}.

Two complementary approaches were used to measure the muon reconstruction efficiency in data. The first technique determines the efficiency of isolated combined muons relative to inner-detector tracks matched to muon hits in the muon spectrometer, resulting in an  efficiency measured in data of $0.994 \pm 0.006$(stat) $\pm$ 0.024(syst), compared to 0.986 from the Monte-Carlo simulation. In the second approach, events are selected requiring an isolated combined muon passing the same selection as for the \Zboson~analysis (see Section~\ref{sec:wkine}). The second muon of the \Zboson~candidate is then selected as an inner-detector track with opposite charge. The invariant mass of the muon-track pair is required to be within 10~GeV of the nominal \Zboson~mass. The combined muon efficiency, measured relative to this sample of tracks, is
0.933 $\pm$ 0.022(stat) $\pm$ 0.013(syst), while the value from Monte-Carlo simulation is 0.924.
The difference in the efficiency values obtained from the two methods is due to the different sensitivity to the geometrical acceptance of the muon spectrometer, as the first method explicitly requires muon hits.
Both techniques confirm that the reconstruction efficiency is well modelled by the simulation and the latter is used to assign the systematic uncertainty on that value.

\subsection{Missing transverse energy}
\label{sec:met}

   The transverse missing energy (\MET~) reconstruction used in the electron channel is  based on calorimeter information. It relies on a cell-based algorithm which sums the electromagnetic-scale energy deposits of calorimeter cells inside three-dimensional topological clusters~\cite{caloclus}. The EM scale corresponds to the energy deposited in the calorimeter calculated under the assumption that all processes are purely electromagnetic in nature. These clusters are then corrected to take into account the different response to hadrons than to electrons or photons, dead material losses and out of cluster energy losses~\cite{REF_LH}.
These topological clusters are built around energy $E>4\sigma_{\rm noise}$ seeds, where $\sigma_{\rm noise}$ is the Gaussian width of the cell energy distribution in randomly triggered events,  by iteratively gathering neighbouring cells with $E>2\sigma_{\rm noise}$ and, in a final step, by adding all direct neighbours of these accumulated secondary cells.

For the electron channel, the components of the missing transverse energy
are calculated by summing over all topological cluster cell energy components $E_{x,y}^i$:
\begin{equation}
E^{\rm{miss}}_{x,y} |_e \ = \   -  \sum_{i} E_{x,y}^i \ .
\label{eq:topo}
\end{equation}

   The \MET~  used in the muon channel is calculated by adding the reconstructed momenta of isolated and non-isolated
muons measured in the pseudorapidity range $|\eta| < 2.7$:
\begin{equation}
E^{\rm{miss}}_{x,y} |_\mu \ = \   -  \sum_{i} E_{x,y}^i
\ - \ \sum^{\rm isolated}_k p_{x,y}^k  \  -  \  \sum^{\rm non-isolated}_j p_{x,y}^j ,
\end{equation}
where non-isolated muons are those within a distance $\Delta R \leq 0.3$ of a jet in the event. The \pT\ of an isolated muon is determined from the combined measurement of the inner detector and muon spectrometer, as explained in Section~\ref{sec:muons}. The energy lost by an isolated muon in the calorimeters is removed from the calorimeter term. For a non-isolated muon, the energy lost in the calorimeter cannot be separated from the nearby jet energy.  The muon-spectrometer measurement of the muon momentum after energy loss in the calorimeter is therefore used, unless there is a significant mismatch between the spectrometer and combined measurements. In this case the combined measurement minus the parameterised energy loss in the calorimeter is used.  For values of the pseudorapidity outside the fiducial volume of the inner detector ($2.5 < |\eta| < 2.7$), there is no matched track requirement and the muon spectrometer stand-alone measurement is used instead.

Systematic uncertainties on the measurement of \met\ result mainly from uncertainties on the
energy scale of topological clusters.
From a comparison of the momentum and energy measurement of charged particles~\cite{JES},
the topological cluster energy scale is known to $\pm20\%$ for $\pT \sim $ 500 MeV and $\pm 5\%$
at high \pT. Other contributions result from uncertainties  due to the imperfect modelling of the overall $\met$ response
(low energy hadrons) and resolution, modelling of the underlying event and pile-up effects.

\section{Selection of  $\ensuremath{W}\rightarrow \ell \nu$ and  $\ensuremath{Z}\rightarrow \ell \ell$ candidates}
\label{sec:selection}

\subsection{Event selection}
\label{sec:evt_selW}

Collision candidates are selected by requiring a primary vertex with at least three tracks, consistent with the beam-spot position. To reduce contamination from cosmic-ray or beam-halo events,  the muon analysis requires the primary vertex position along the beam axis to be within 15~cm of the nominal position (this primary vertex distribution has a measured longitudinal RMS of 6.2~cm).

An analysis of a high-statistics sample of minimum-bias events has shown that events can occasionally contain very localised high-energy calorimeter deposits not originating from the proton-proton collision, but e.g. from sporadic discharges in the hadronic end-cap calorimeter or, more rarely, coherent noise in the electromagnetic calorimeter. Cosmic-ray muons undergoing a hard bremsstrahlung are also a potential source of localised energy deposits uncorrelated with the primary proton-proton collisions. The occurrence of these events is very rare but can potentially  impact significantly the \MET~ measurement by creating high-energy tails~\cite{CLEANING}. To remove such events, dedicated cleaning requirements  have been developed using a minimum-bias event sample. Using Monte-Carlo simulation, it was verified that these criteria remove less than 0.1\% of minimum-bias events, 0.004\% of  $W\rightarrow \ell \nu$, and 0.01\% of dijet events.

For the electron channel only, the event is rejected if the candidate electromagnetic cluster is located in any problematic
region of the EM calorimeter. Due to hardware problems~\cite{aad0912}, the signal cannot be read out from $\sim$2\%
 of the EM calorimeter cells.

\subsection{Selection of high transverse-energy leptons}
\label{sec:results_presel}

 Electron candidates selected with the identification level ``tight'' for the \Wboson~analysis and ``medium'' for the \Zboson~analysis  (according to the algorithm described in Section~\ref{sec:electrons})  are required to have a cluster \eT~$>$~20~GeV within the range $|\eta|<2.47$, excluding the transition region between the barrel and end-cap calorimeters ($1.37<|\eta|<1.52$).
Muon candidates selected according to the algorithm described in Section~\ref{sec:muons} are required to have a combined muon with \pT~$>$~20~GeV and a muon-spectrometer track with \pT~$>$~10~GeV within the range $|\eta|<2.4$.
To increase the robustness against track reconstruction mismatches, the difference between the inner-detector and muon-spectrometer
\pT\ corrected for the mean energy loss in upstream material, is required to be less than 15~GeV. The difference between the $z$ position of the muon track extrapolated to the beam line and the $z$ coordinate of the primary vertex is required to be less than 1~cm.

% Figure 1
\begin{figure}[t]
\begin{center}
\subfigure[]{\label{presel_et_ele}\includegraphics[width=0.47\textwidth]{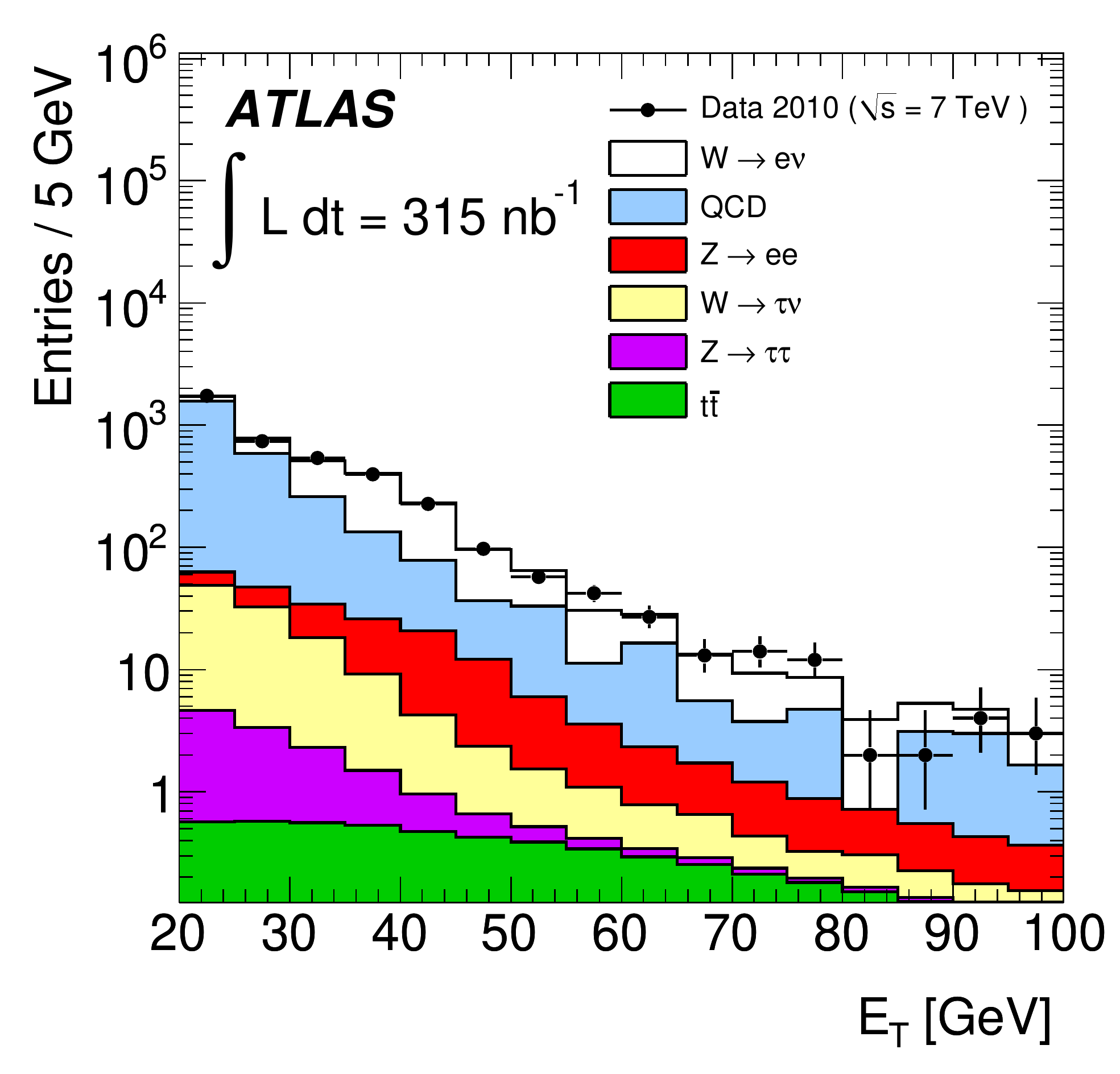}}
\subfigure[]{\label{presel_pt_muon}\includegraphics[width=0.47\textwidth]{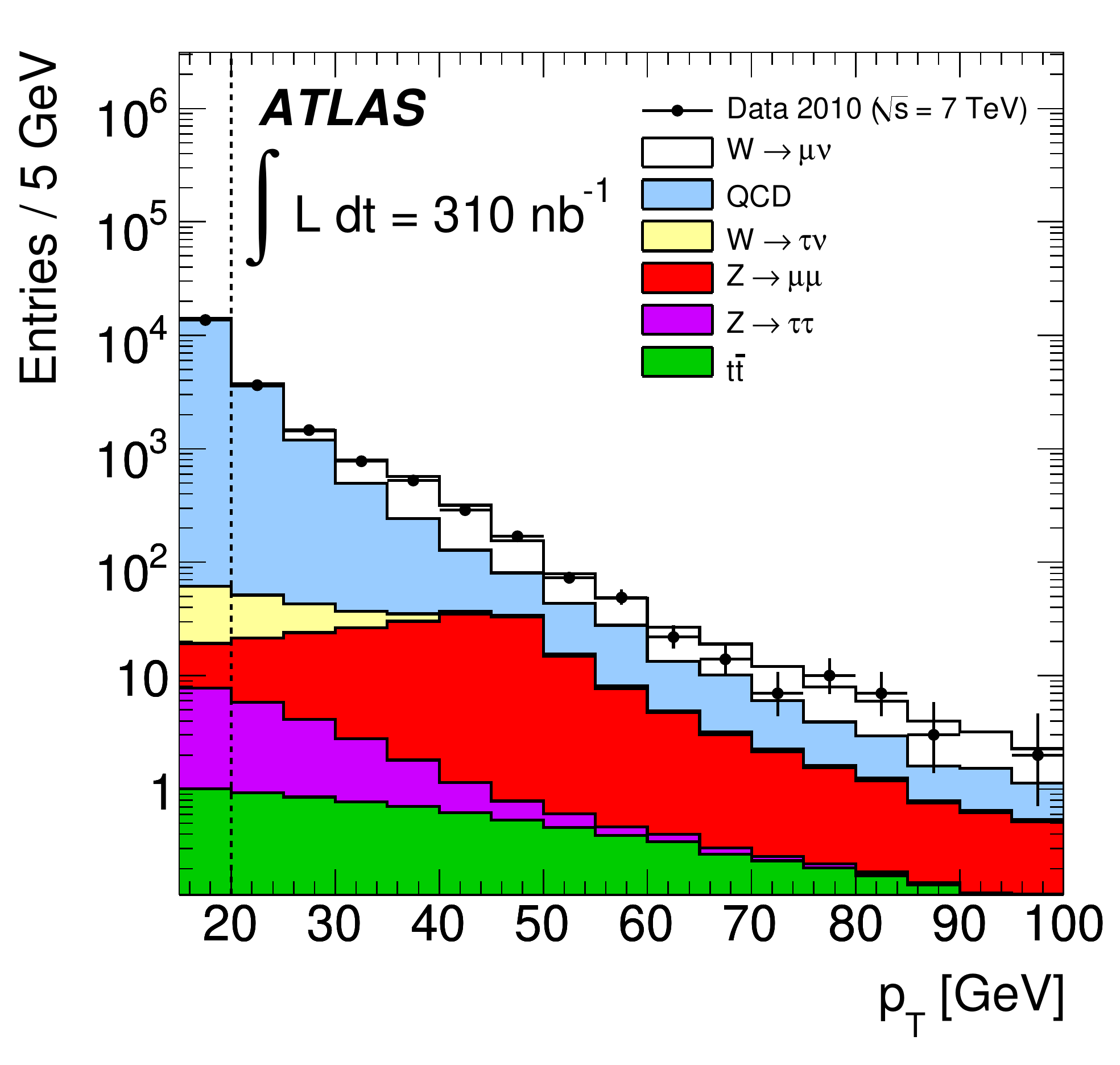}}
\caption{\it\small Calorimeter cluster $E_T$ of ``tight'' electron candidates~(a) and combined \pT~ of muon candidates~(b) for data and Monte-Carlo simulation, broken down into the signal and various background components. The vertical line in (b) indicates the analysis cut. The transverse momentum region between 15 and 20 GeV of the muon sample is used in the estimation of the QCD background (see Section~\ref{sec:mubkg}). \label{fig:kine_ele}}
\end{center}
\end{figure}

Figure~\ref{fig:kine_ele} shows the \eT~and  \pT~spectra of these ``tight'' electron and combined muon candidates and compares these to the signal and background Monte-Carlo samples described in Section~\ref{sec:mcsamples}. Comparisons of the dijet Monte-Carlo distributions to equivalent data distributions have shown that the  dijet Monte Carlo for this high-\pT~lepton selection over-estimates the amount of background by a factor of approximately \JFQCDSFe~for the electron channel and a factor of  \JFQCDSFmu~for the muon channel. The difference between these values is
likely explained by the different composition of the QCD background in the two analyses. For the electron case, this normalisation factor is obtained by comparing data and Monte-Carlo samples of high transverse-momentum electron candidates which are dominated by QCD background. For the muon case, this normalisation factor is obtained from a non-isolated muon data sample and is then applied to the isolated muon sample used in this analysis.

Unless otherwise stated, all Monte-Carlo distributions shown in this paper  have been normalised to the integrated luminosity of the data as described in Section~\ref{sec:mcsamples}, using the cross sections as given in Table~\ref{MC} and taking into account these scale factors for the QCD background.  At this stage of the selection, the event samples are dominated by QCD background. These distributions show agreement in shape between data and Monte-Carlo simulation.

\subsection{Lepton isolation}
\label{sec:results_isol}

The use of an isolation parameter to enhance the signal-to-background ratio was investigated. Separate isolation requirements must be considered for the electron and muon channels, since the electron can undergo bremsstrahlung, while a muon is primarily defined by its track.

A calorimeter-based isolation parameter defined as the total calorimeter transverse energy in a cone of $\Delta R<0.3$ surrounding the candidate electron cluster, divided by the cluster \eT, is considered for the electron channel. This variable is exploited for background estimations in this channel, but is not used in the event selection.

In the muon analysis, a track-based isolation defined as the sum of the transverse momenta of tracks with
$\pT >$ 1~GeV in the inner detector
within a cone of $\Delta R <0.4$ around the muon track, divided by the $\pT$ of the muon, is considered.
An isolation requirement of  $\sum \pT^{\mathrm{ID}}/\pT<0.2$ is imposed in the muon selection given that, after all other selections are made to identify \Wboson\ candidates, this requirement rejects over 84\% of the expected QCD background while keeping (98.4$\pm$1.0)\% of the signal events.

\subsection{Kinematic selection}
\label{sec:wkine}

Additional requirements beyond those  in Sections~\ref{sec:results_presel} and \ref{sec:results_isol} are imposed to better discriminate $\ensuremath{W}\rightarrow \ell \nu$ and  $\ensuremath{Z}\rightarrow \ell \ell$ events from background events. A summary of all requirements is as follows:

\begin{itemize}
\item An electron with \eT~$>20$~GeV or a combined muon  with $\pT >$ 20 GeV; \\
For the $\ensuremath{W}\rightarrow e\nu$~analysis, events containing an additional ``medium'' electron are vetoed.
If more than one combined muon candidate is reconstructed, the one with the highest $p_{\rm{T}}$  is chosen.

\item Isolation for the muon channel:    $\sum \pT^{\mathrm{ID}}/\pT<0.2$;\\
For the electron channel, no isolation criterion is used.
\item For the \Wboson~analysis only:
\begin{itemize}
   \item Missing transverse energy $\MET >$~25 GeV;
   \item Transverse mass of the lepton-\MET system, $m_{\rm T} >$ 40 GeV; \\
   The transverse mass is defined as
   $m_{\rm{T}} = \sqrt{ 2  \  \pT^{\ell} \  \MET \ (1-\cos \Delta \phi ) }$, where $\Delta \phi$ is the azimuthal
separation between the directions of the lepton and the missing transverse energy.
\end{itemize}
\item For the \Zboson~analysis only:
\begin{itemize}
   \item A pair of oppositely-charged leptons (each lepton with $\pt >$ 20 GeV) of the same flavour;
   \item Invariant mass window of lepton pair: $66<m_{\ell\ell}<116$~GeV;
   \item Veto on events with three or more ``medium'' electrons (for the $Z \to ee$ analysis).
\end{itemize}
\end{itemize}

% Figure 2
\begin{figure}[!]
\begin{center}
\subfigure[]{\label{etmiss_ele}\includegraphics[width=0.47\textwidth]{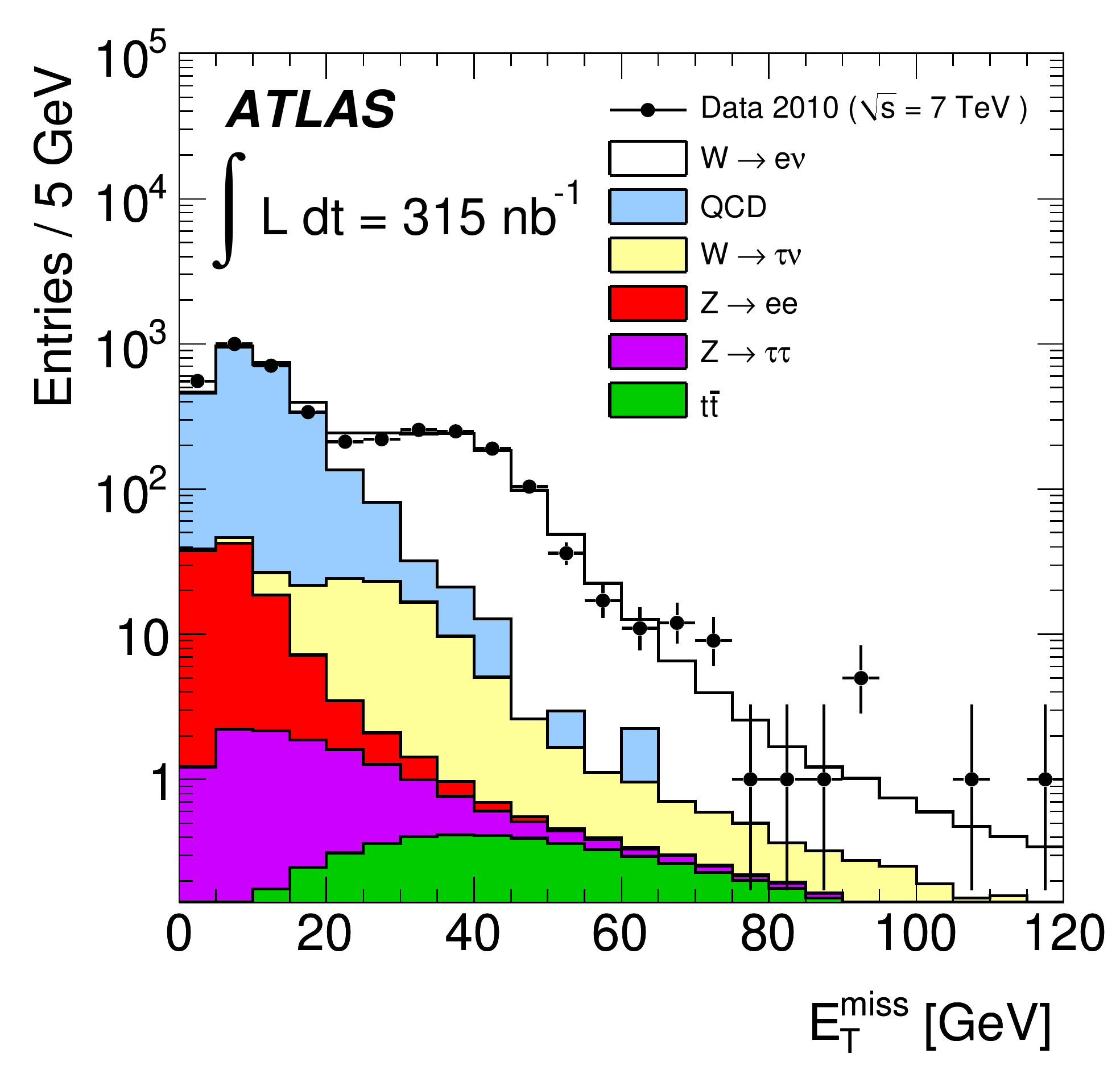}}
\subfigure[]{\label{etmiss_muon}\includegraphics[width=0.47\textwidth]{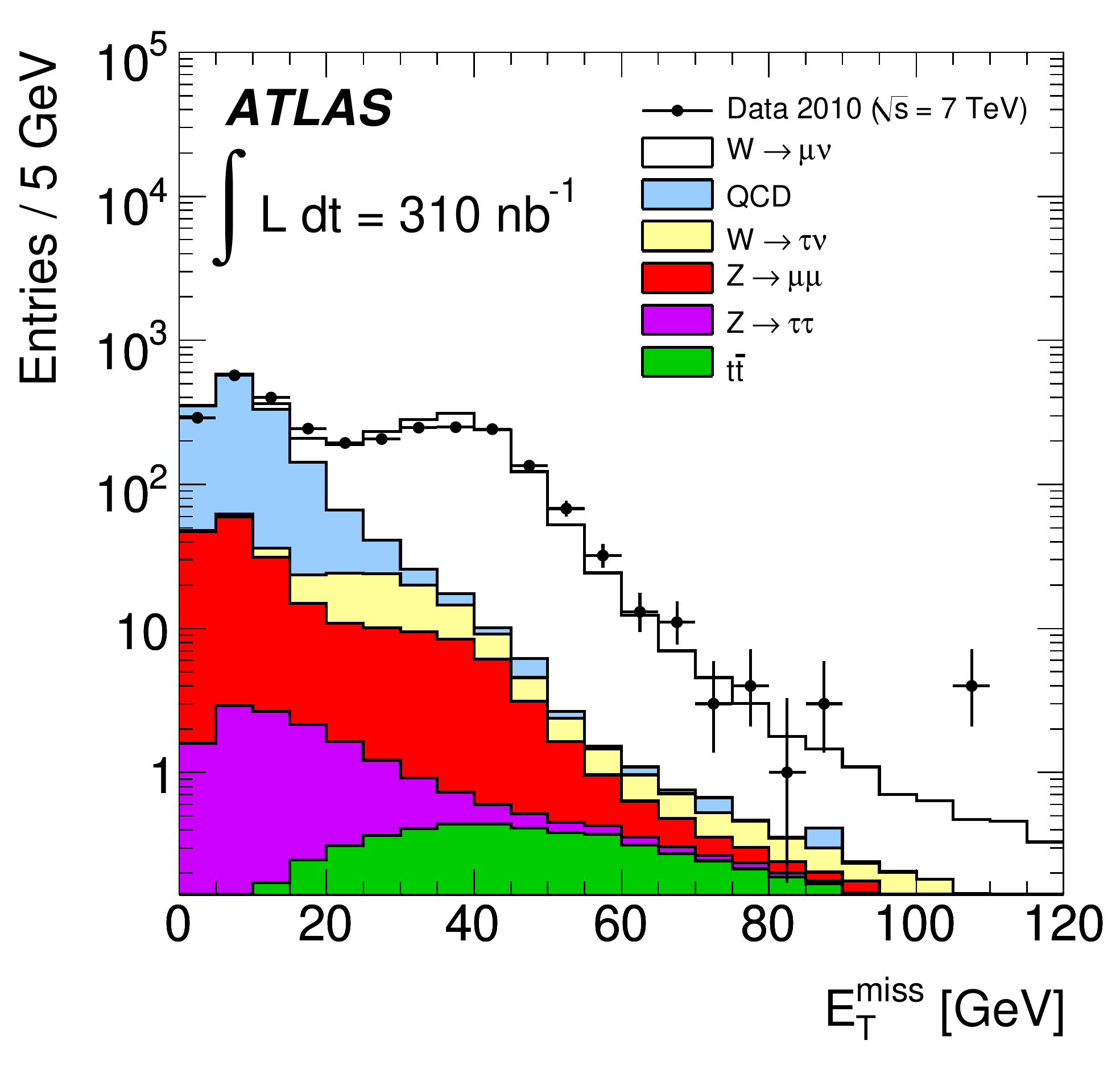}}
\caption{\it\small Distributions of the missing transverse energy, \MET, of electron~(a) and muon~(b) candidates  for data and Monte-Carlo simulation, broken down into the signal and various background components.  \label{fig:etmiss}}
\end{center}
\end{figure}

%Figure 3
\begin{figure}[t]
\begin{center}
\subfigure[]{\label{mt_wo_ele}\includegraphics[width=0.47\textwidth]{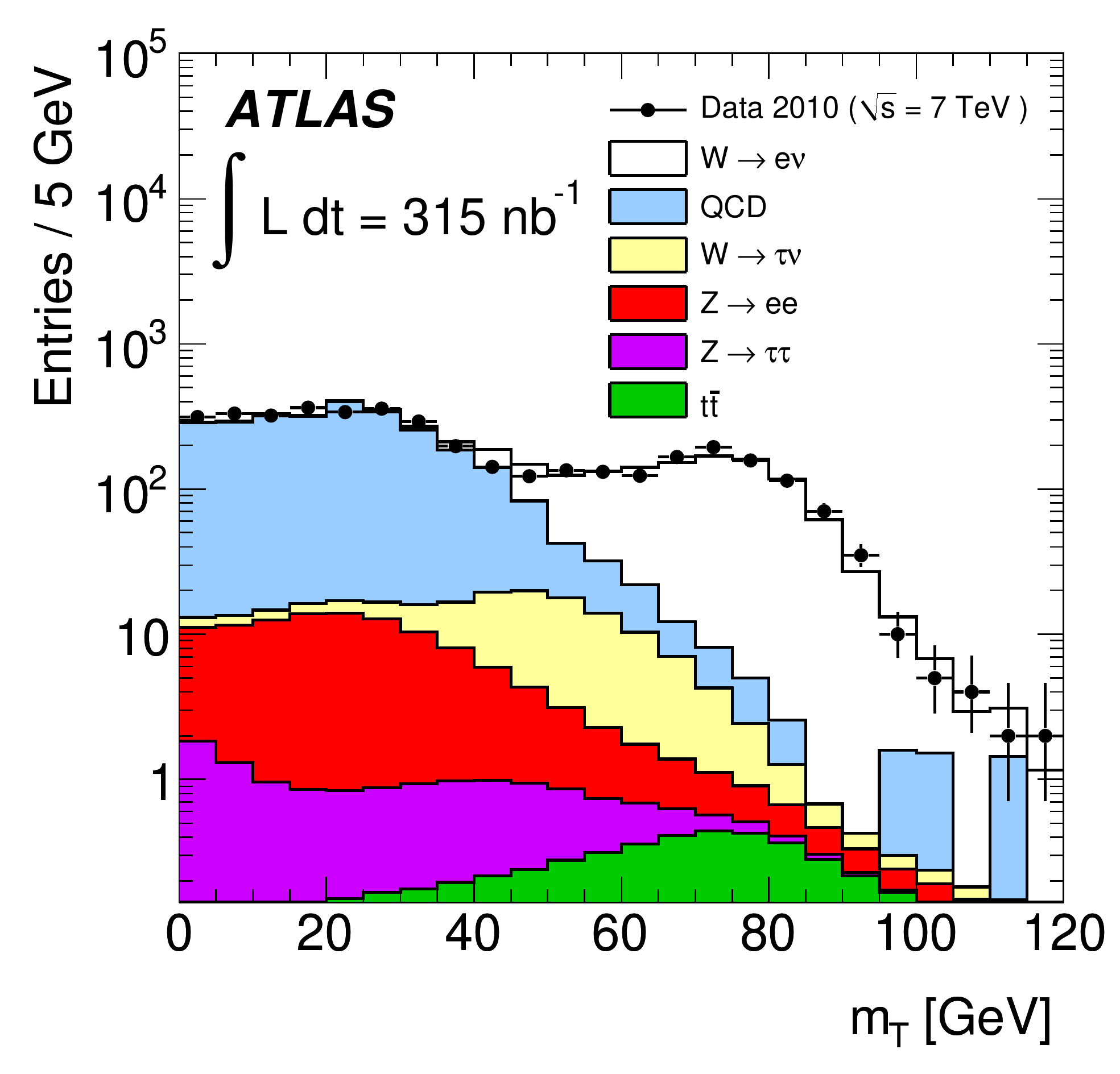}}
%\subfigure[]{\label{mt_w_ele}\includegraphics[width=0.47\textwidth]{./figures/electron/mt_w_ele}}
\subfigure[]{\label{mt_wo_muon}\includegraphics[width=0.47\textwidth]{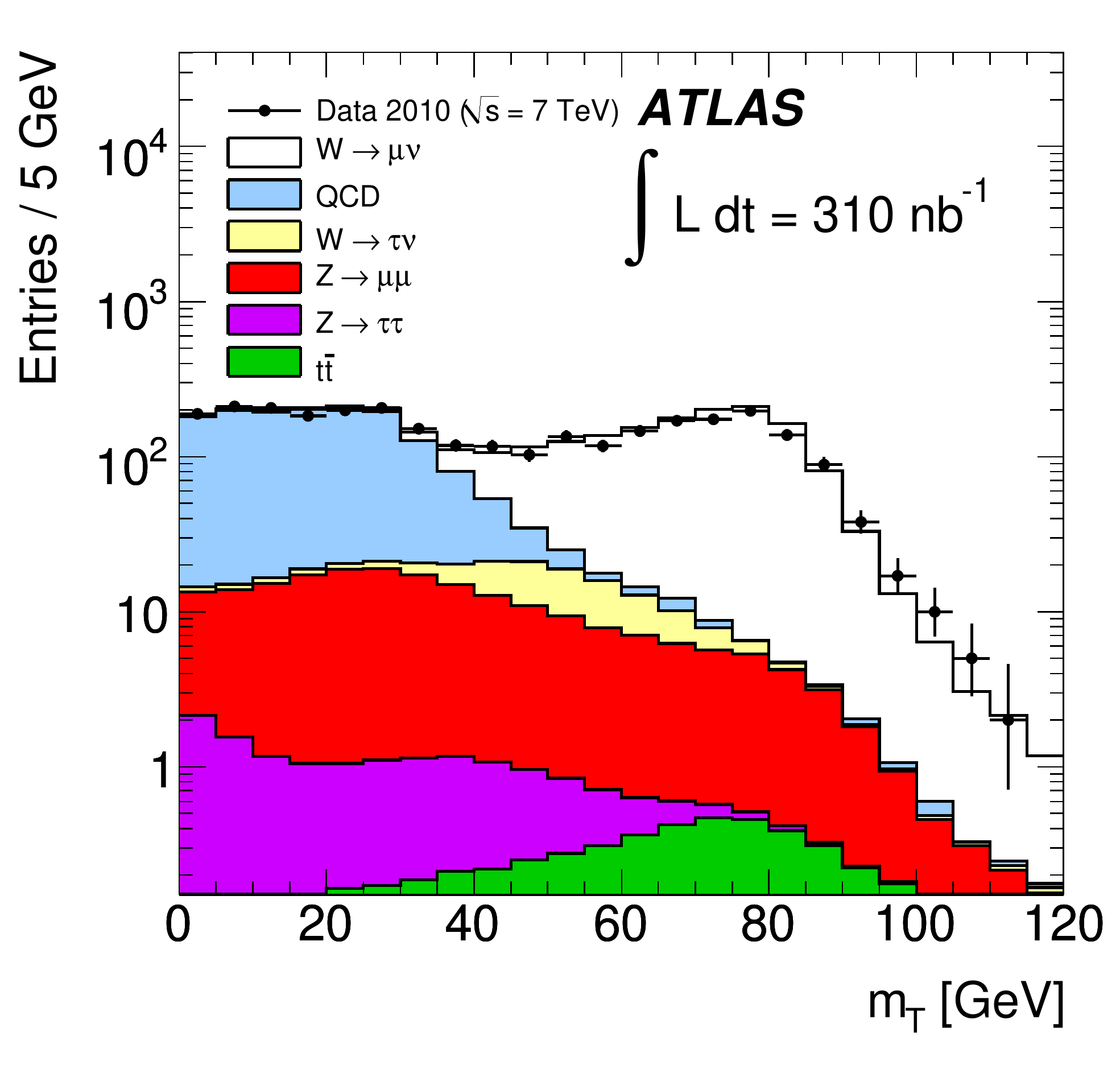}}
\caption{\it\small Distributions of the transverse mass, \MT, of the electron-\MET\ system~(a) and muon-\MET\ system~(b) without an \MET\  requirement. The data are compared to Monte-Carlo simulation, broken down into the signal and various background components.
\label{fig:mt_ele}}
\end{center}
\end{figure}

%Figure 4
\begin{figure}[!]
\begin{center}
\subfigure[]{\label{mt_w_ele}\includegraphics[width=0.47\textwidth]{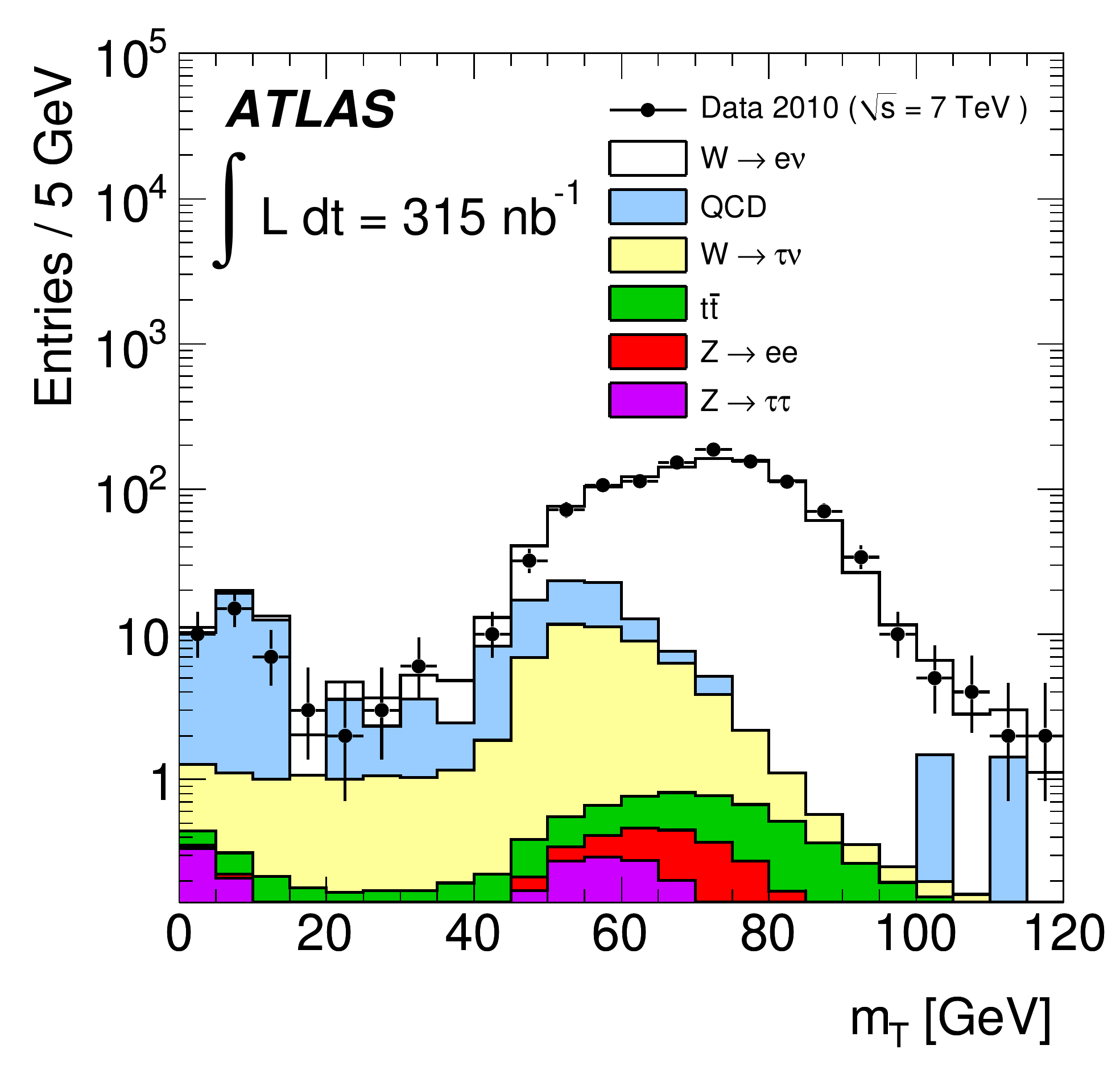}}
\subfigure[]{\label{mt_w_muon}\includegraphics[width=0.47\textwidth]{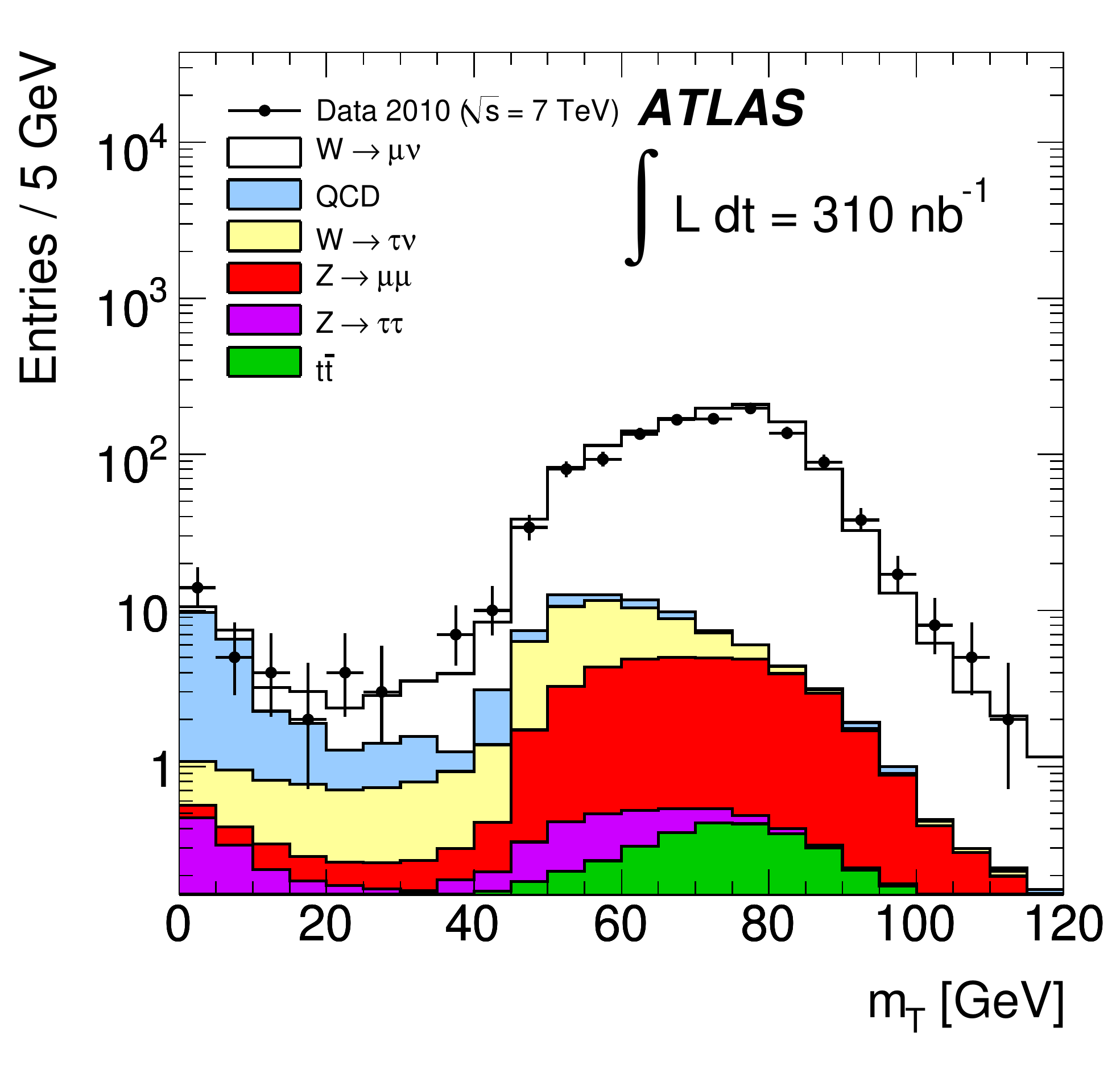}}
\caption{\it\small Distributions of the transverse mass, \MT, of the electron-\MET\ system~(a) and muon-\MET\ system (b) with a requirement of \MET~$>$~25~GeV. The data are compared to Monte-Carlo simulation, broken down into the signal and various background components.
\label{fig:mt_muon}}
\end{center}
\end{figure}

Figure~\ref{fig:etmiss} shows the \MET\ distributions of electron and muon candidates passing the requirements described above, except the \MET~and \MT~criteria. Both distributions indicate that applying a minimum requirement on \MET~greatly enhances the \Wboson\ signal over the expected background. True  $\ensuremath{W}\rightarrow \ell \nu$ events in the Monte Carlo are predominantly at high \MET\ due to the escaping neutrino in the event. Although some of the QCD background events may also have neutrinos in their final state, they mostly populate the regions of small \MET. Figures~\ref{fig:mt_ele} and~\ref{fig:mt_muon} show  the \MT~distributions without and with the requirement of \MET~$>$~25~GeV.

\subsection{\Wboson~and \Zboson~candidates after final selection}

\begin{figure}[t]
\begin{center}
\subfigure[]{\label{et_ele}\includegraphics[width=0.47\textwidth]{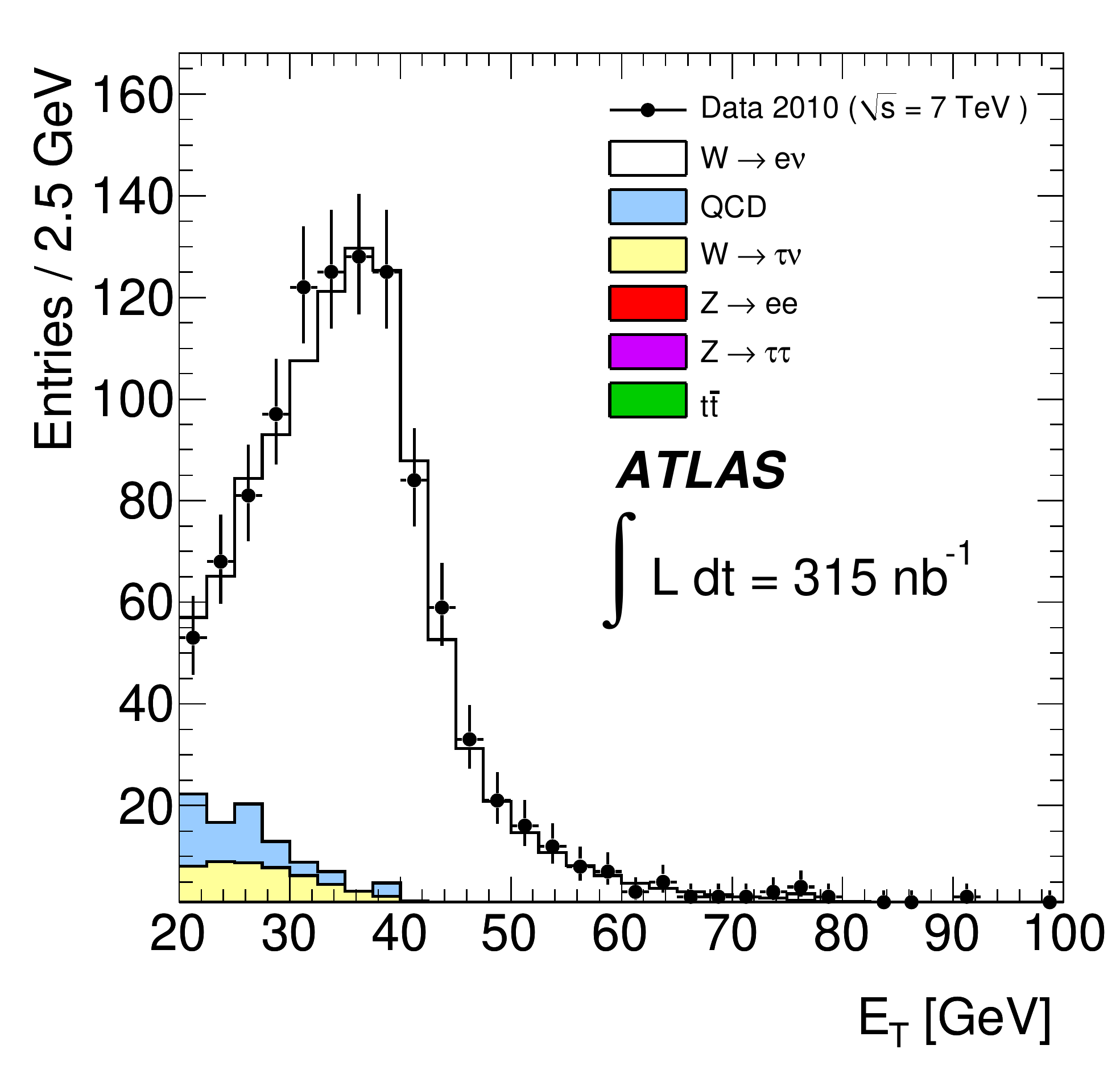}}
\subfigure[]{\label{pt_muon}\includegraphics[width=0.47\textwidth]{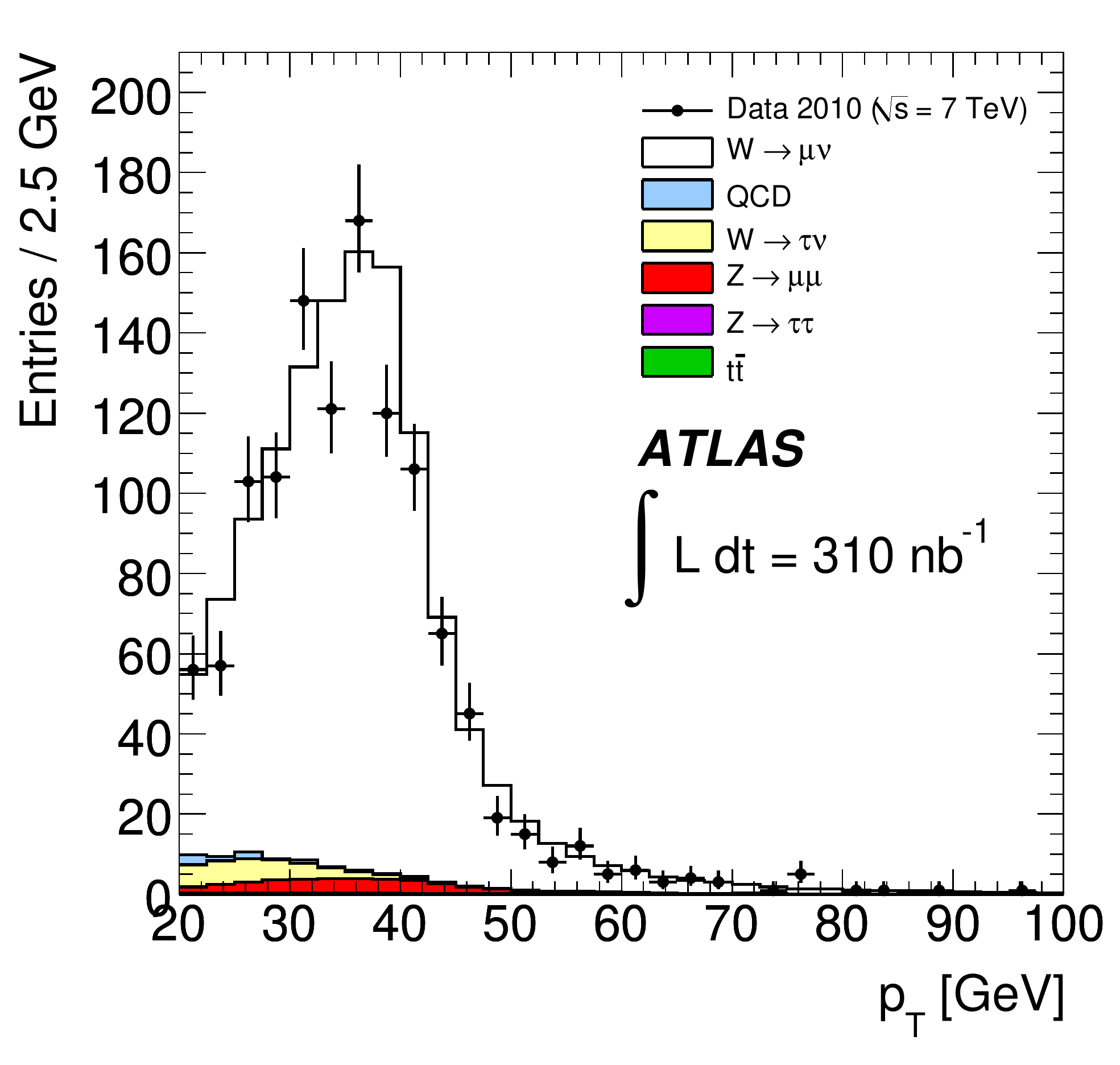}} %YES
\caption{\it\small Distributions of the electron cluster $E_T$~(a) and muon \pT~~(b) of the W candidates after final selection.  The requirements of \MET~$>$~25~GeV and \MT~$>$~40~GeV are applied.  The data are compared to Monte-Carlo simulation, broken down into the signal and various background components.
\label{fig:etW}}
\end{center}
\end{figure}

\begin{figure}[!]
\begin{center}
\subfigure[]{\label{ptw_ele}\includegraphics[width=0.47\textwidth]{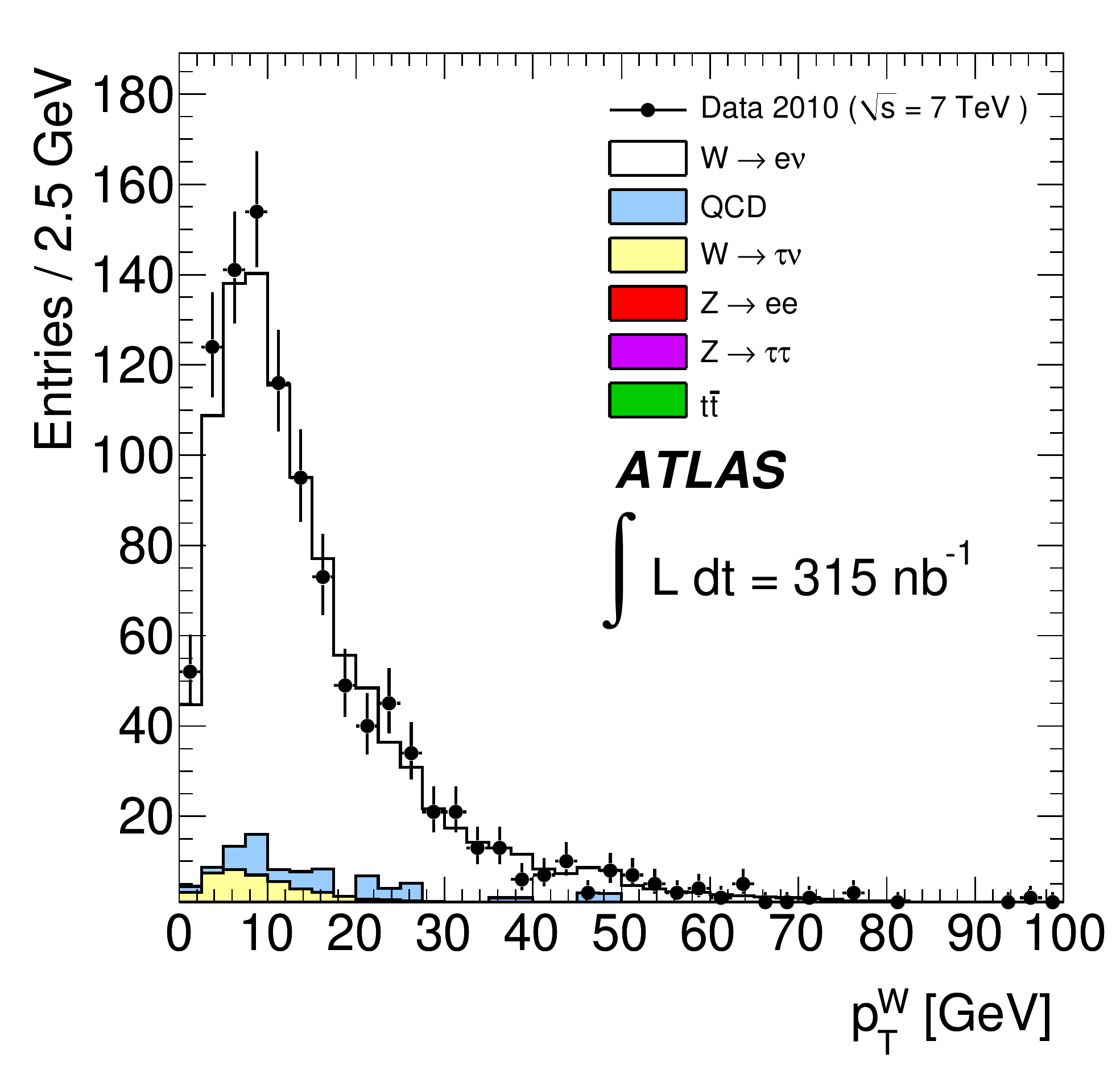}}
\subfigure[]{\label{ptw_muon}\includegraphics[width=0.47\textwidth]{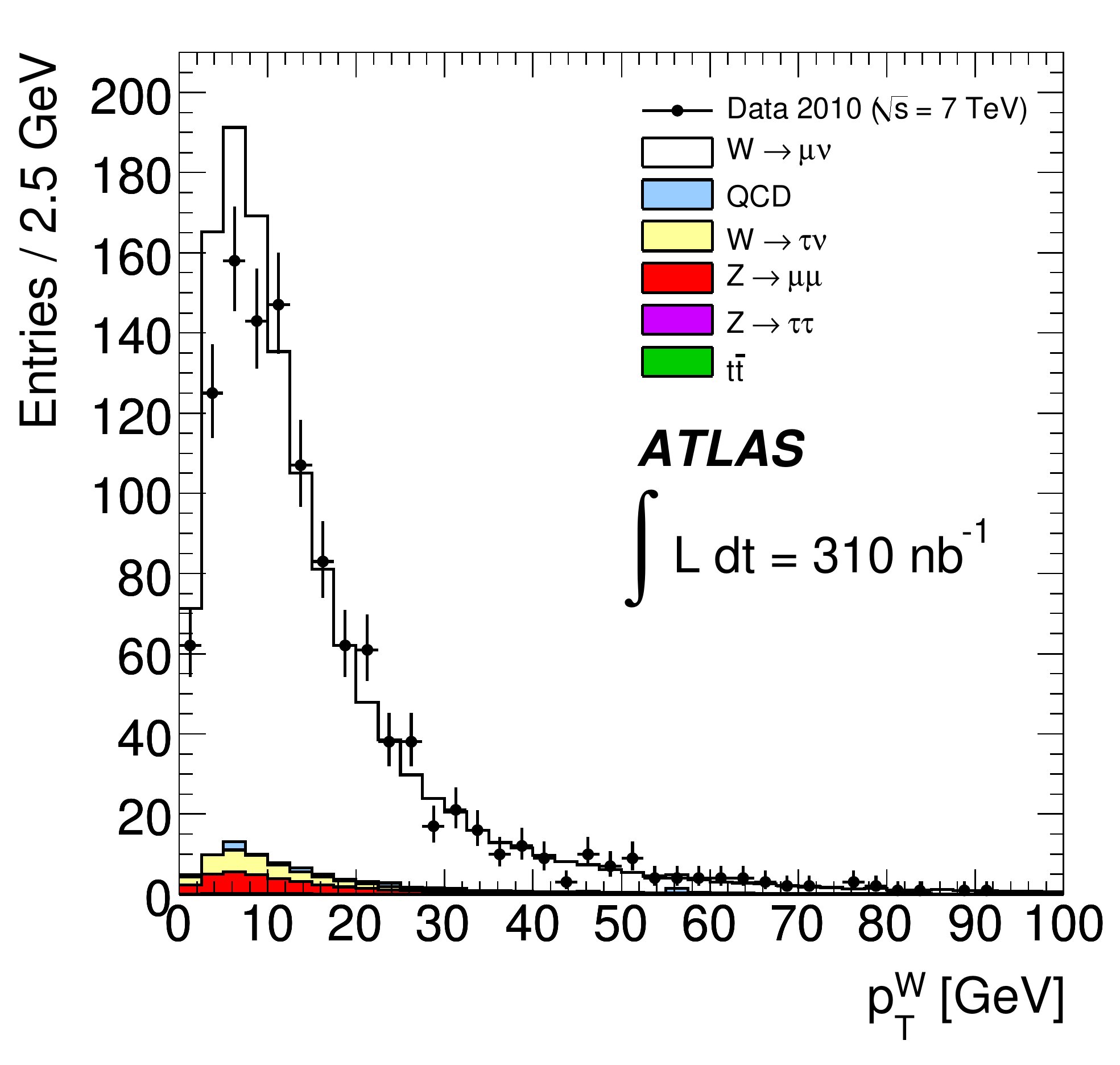}} %YES
\caption{\it\small Distributions of the transverse momentum \pT\ of the W candidates in the electron channel~(a) and muon channel~(b) after final selection. The requirements of \MET~$>$~25~GeV and \MT~$>$~40~GeV are applied. The data are compared to Monte-Carlo simulation, broken down into the signal and various background components.
\label{fig:ptw}}
\end{center}
\end{figure}

Table~\ref{tab:wecandidates} summarises the number of $\ensuremath{W}\rightarrow \ell \nu$ candidates remaining after each major requirement in the respective analyses. A total of \ntotWe~candidates (\ntotWeplus~$e^+$ and \ntotWeminus~$e^-$) pass all requirements in the electron channel and \ntotWmu~candidates (\ntotWmuplus~$\mu^+$ and \ntotWmuminus~$\mu^-$) in the muon channel. Figure~\ref{fig:etW} shows the electron cluster \eT~and muon combined \pT\ of the lepton candidates, while Fig.~\ref{fig:ptw} shows the \pT\ spectrum of the $\ensuremath{W}\rightarrow \ell \nu$ candidates. Both channels demonstrate a clear \Wboson\ signal over an almost negligible background.

\begin{table}[t]
\centering
\begin{tabular}{l|c|c}
\hline \hline
Requirement &  \multicolumn{2}{|c}{Number of candidates} \\
            &  $W \to e \nu$  &   $ W \to \mu \nu$ \\ \hline \hline
\raisebox{-0.4ex}{Trigger} & \raisebox{-0.4ex}{\ntrigWZe}  & {$\ntrigWZmu$}   \\
Lepton:
$e$ with \eT~$>$~20~GeV  or $\mu$ with $\pT>$ 20 GeV
 & \ntotTightWe & \ntotpTWmu\   \\
 \raisebox{-0.4ex}{Muon isolation: $\sum p_T^{\mathrm{ID}}/p_T<0.2$ }               &  --  &  \raisebox{-0.4ex}{\ntotIsolWmu}        \\*[0.2cm]
\MET~$>$~25~GeV                                                                     & \ntotMETWe   & \ntotMETWmu\  \\
\MT~$>$~40~GeV                                                                      & \ntotWe      & \ntotWmu      \\
\hline
\hline
\end{tabular}
\caption{Number of $\ensuremath{W}\rightarrow e \nu$ and $W \to \mu \nu$ candidates in data, remaining after each major requirement.}
\label{tab:wecandidates}
\end{table}

%Figure 05

%Figure 06

%Figure 08 (Electron ET and muon Pt for final Z candidates)
\begin{figure}[h]
\begin{center}
\subfigure[]{\label{et_ele_z}\includegraphics[width=0.46\textwidth]{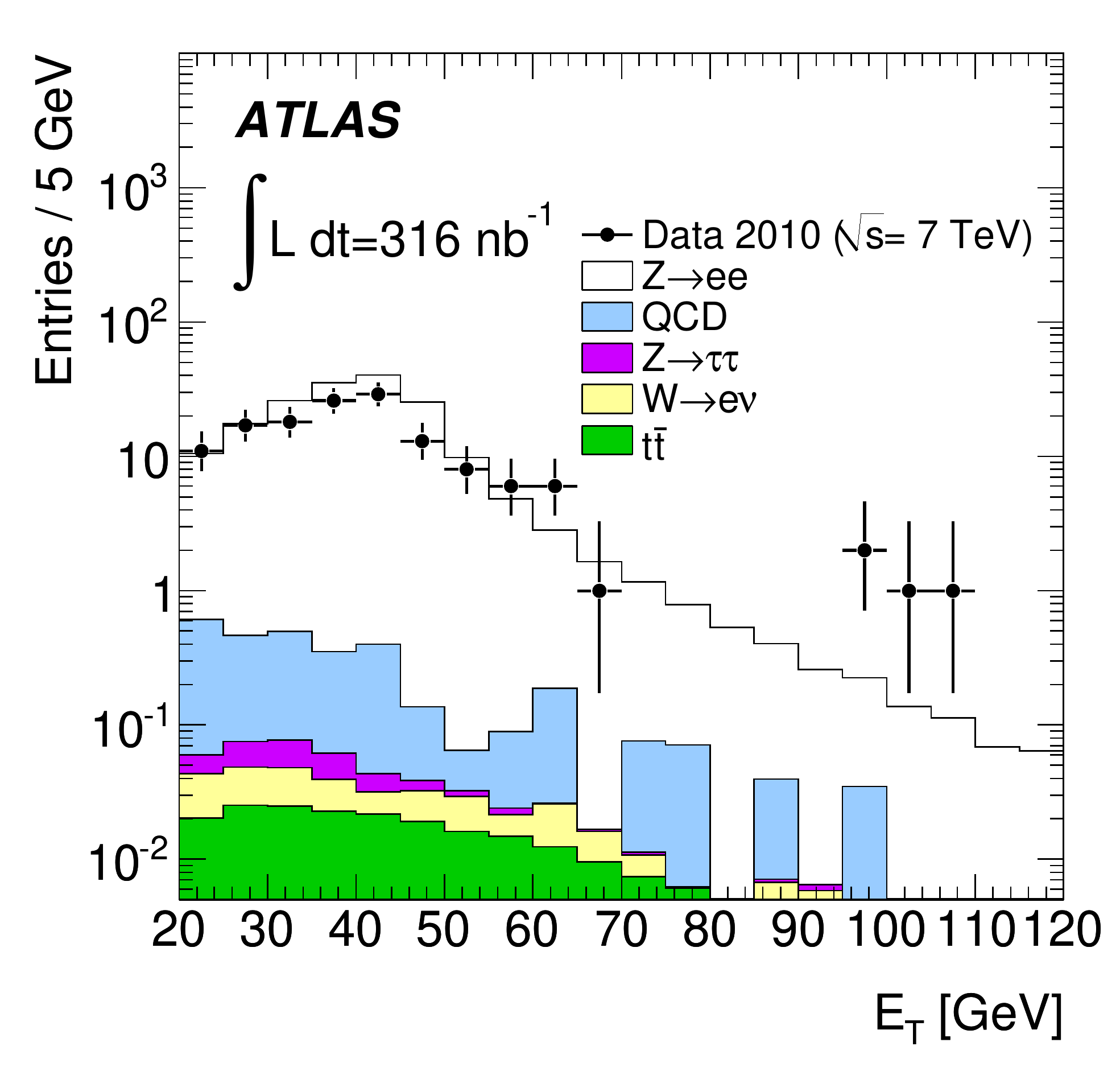}}
\subfigure[]{\label{pt_muon_z}\includegraphics[width=0.45\textwidth]{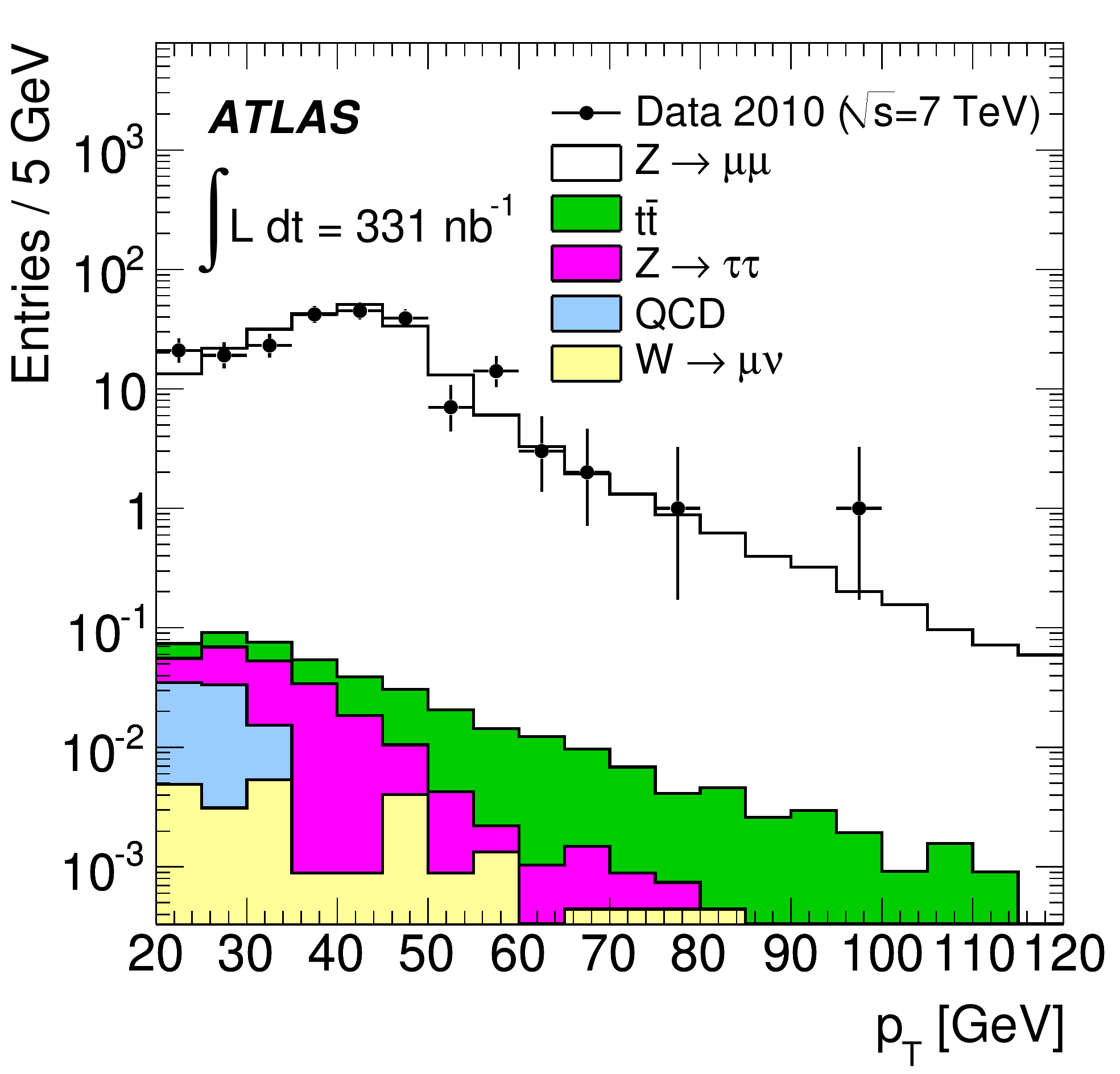}} %YES
\caption{\it\small Distributions of the electron cluster \eT~(a) and muon \pT~(b) of the \Zboson~candidate leptons after final selection. The data are compared to Monte-Carlo simulation, broken down into the signal and various background components.
\label{fig:etZ}}
\end{center}
\end{figure}

\begin{table}[h,t]
\centering
\begin{tabular}{l|c|c}
\hline \hline
Requirement &  \multicolumn{2}{|c}{Number of candidates} \\
            &  $Z \to e e$  &   $ Z \to \mu \mu$ \\
\hline \hline
\raisebox{-0.4ex}{Trigger}  & \raisebox{-0.4ex}{\ntrigWZe}& \raisebox{-0.4ex}{\ntrigWZmu}\\
Two leptons ($ee$ or $\mu \mu$ with $\et (\pt) >$20 GeV) & 83  & 144 \\
Muon isolation: $\sum p_T^{\mathrm{ID}}/p_T<0.2$   &  --   & 117 \\
Opposite charge  $ee$ or $\mu \mu$ pair:        & \ntotZenowindow             & \ntotZmunowindow \\
$66< m_{\ell \ell} < 116$~GeV                   & \ntotZe                     & \ntotZmu\\
\hline
\hline
\end{tabular}
\caption{Number of $\ensuremath{Z}\rightarrow e e$ and $\ensuremath{Z}\rightarrow \mu \mu $
candidates in data, remaining after each major requirement.}
\label{tab:zemucandidates}
\end{table}

%Figure 09 (Pt(Z) for electrons and muons)
\begin{figure}[!]
\begin{center}
\subfigure[]{\label{ptz_ele}\includegraphics[width=0.47\textwidth]{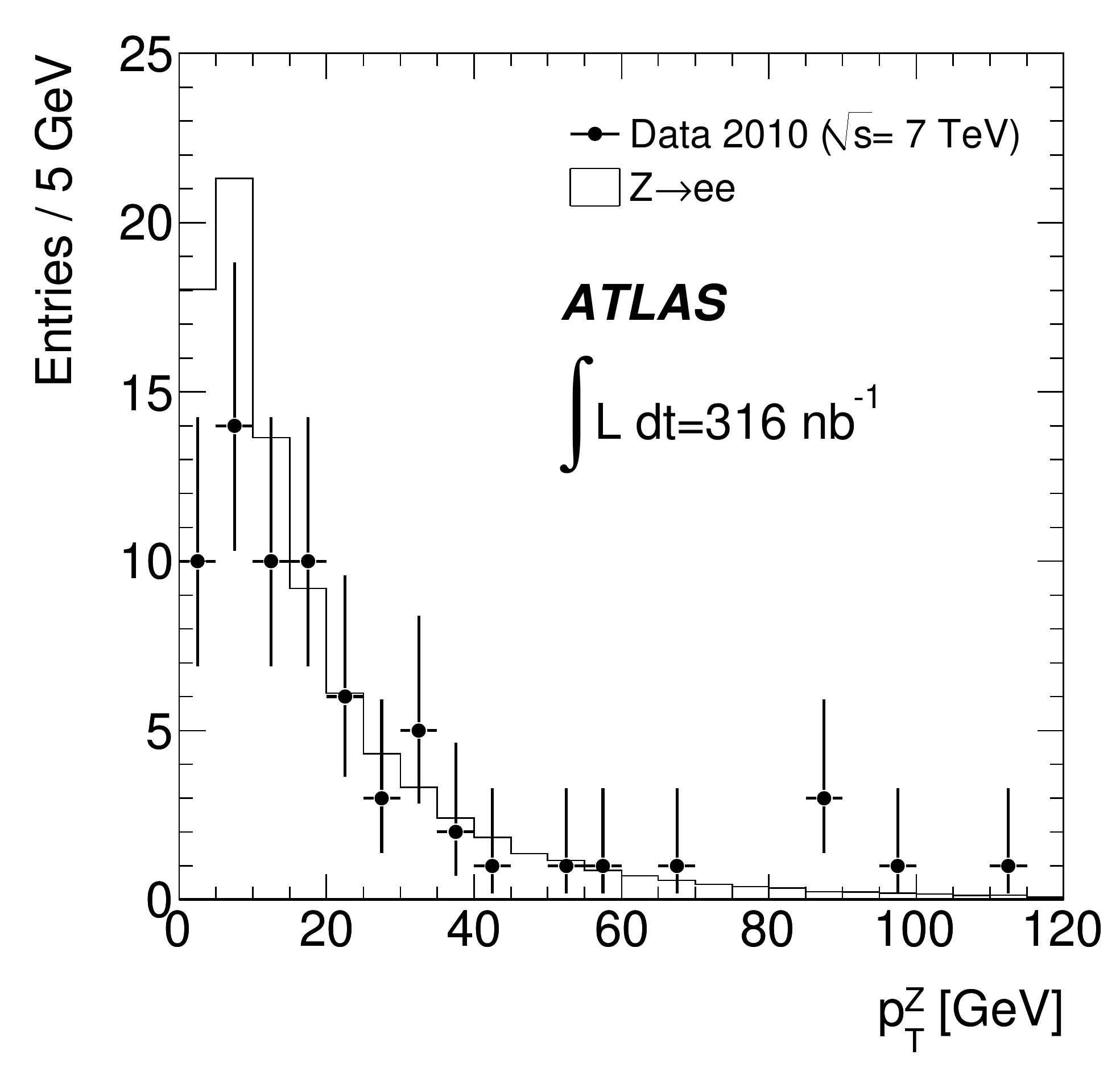}}
\subfigure[]{\label{ptz_muon}\includegraphics[width=0.47\textwidth]{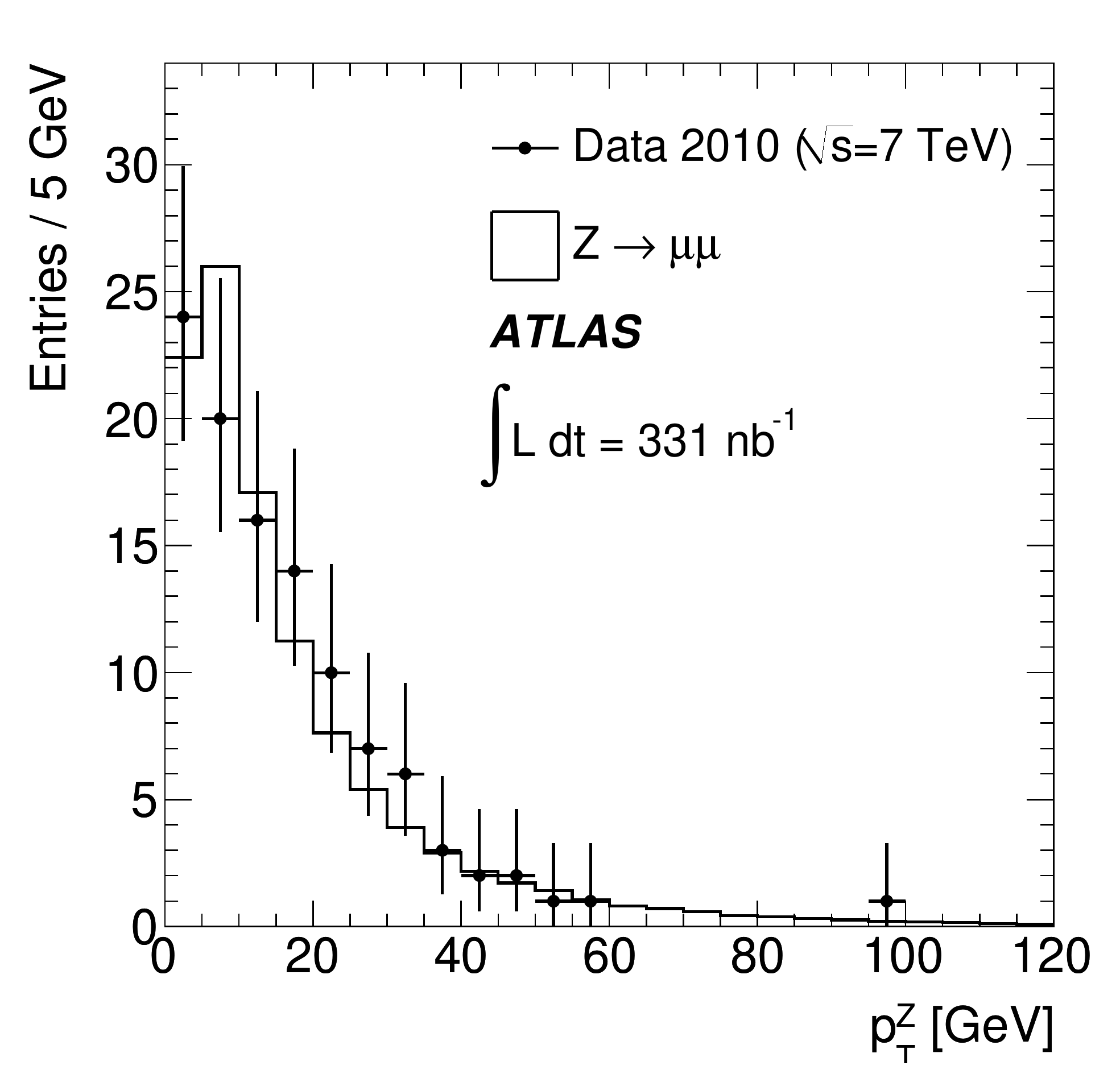}} %YES
\caption{\it\small Distributions of the transverse momentum $\pT$ of the \Zboson~candidates in the electron channel~(a) and muon channel~(b) after final selection. The data are compared to the expectations from Monte-Carlo simulation.
\label{fig:ptZ}}
\end{center}
\end{figure}

\begin{figure}[t]
\begin{center}
\subfigure[]{\label{mee_eleos2}\includegraphics[width=0.47\textwidth]{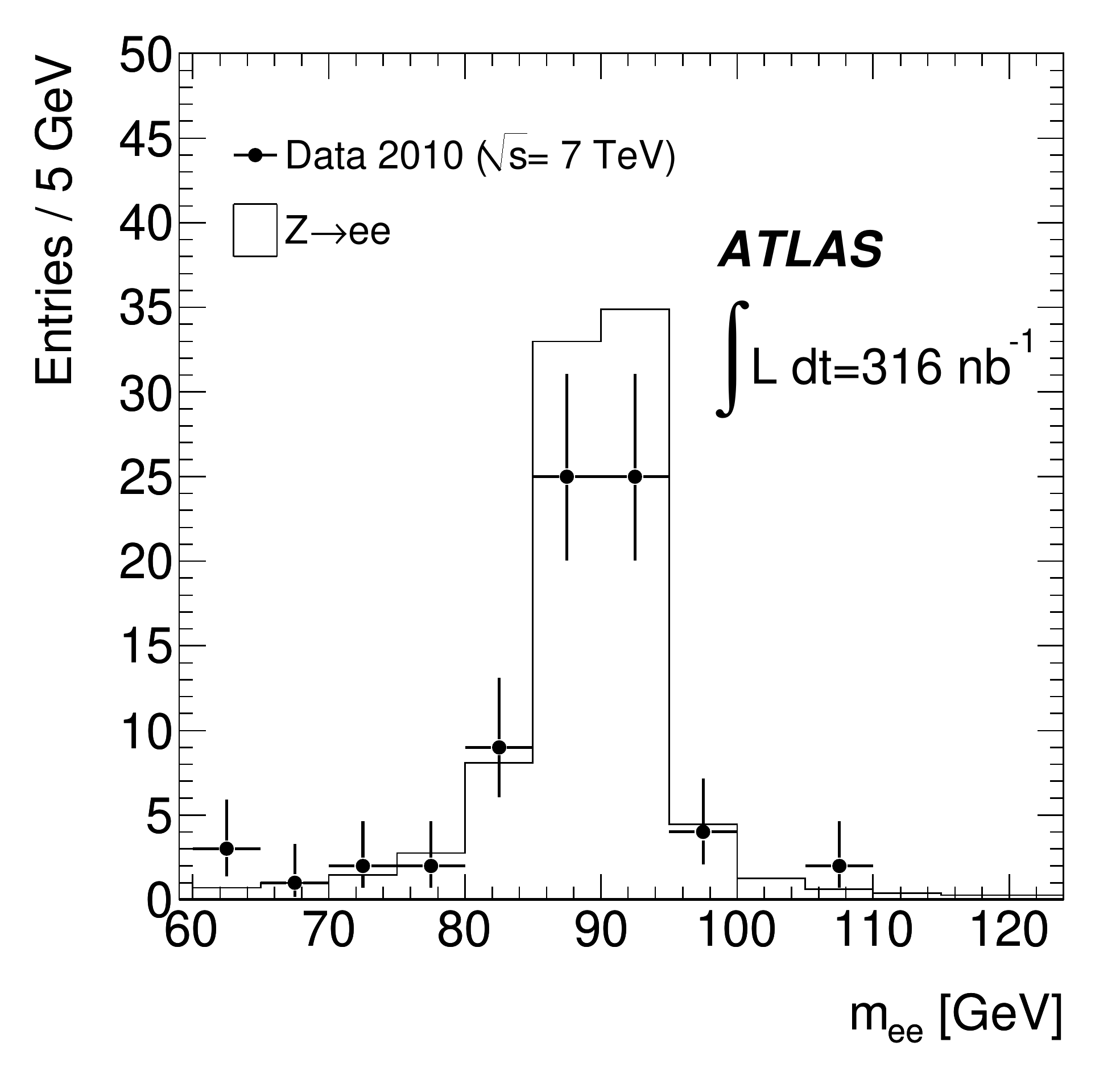}}
\subfigure[]{\label{mmumu_muonos2}\includegraphics[width=0.47\textwidth]{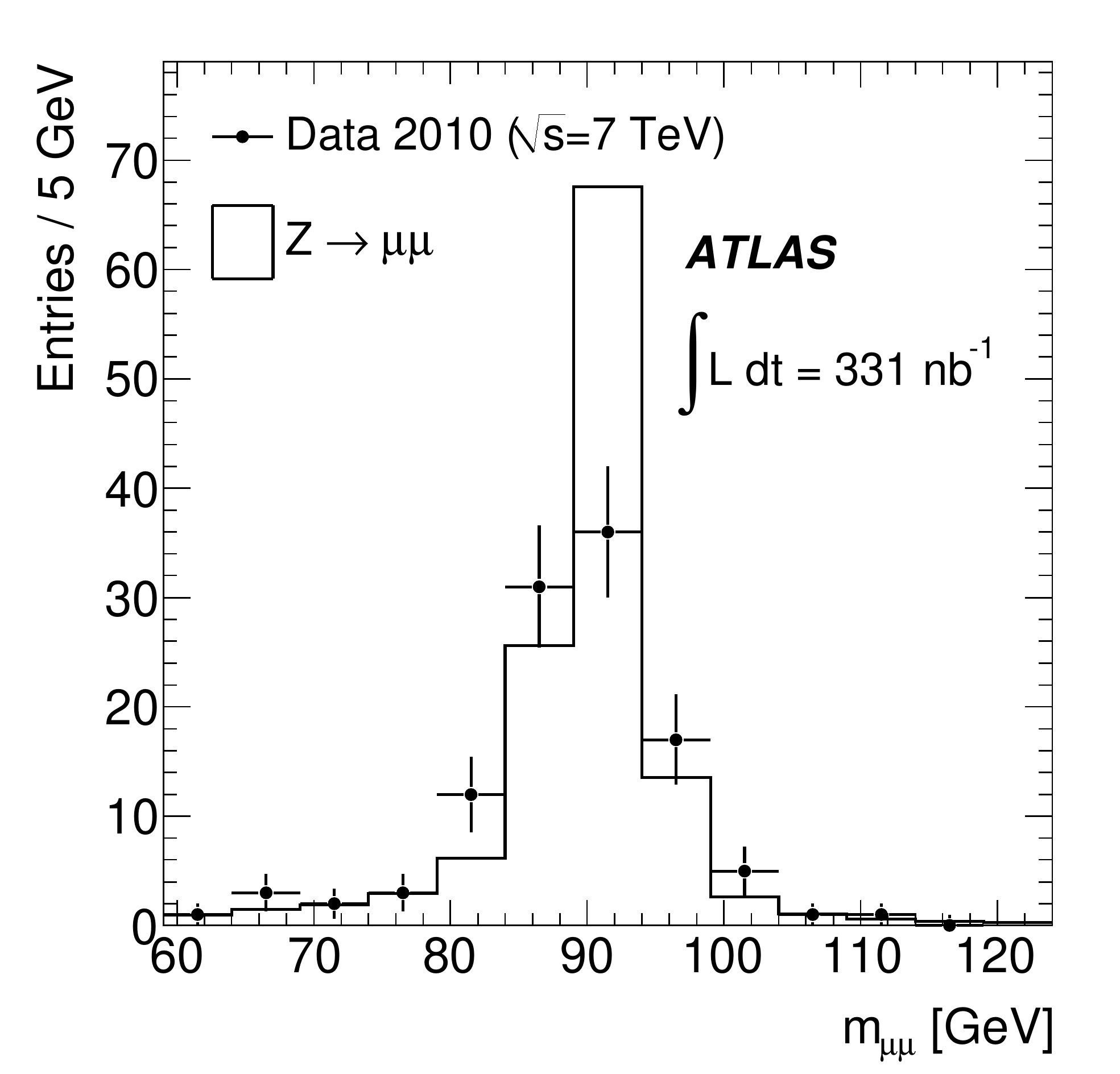}}
\caption{\it\small Distributions of the invariant mass $m_{\ell\ell}$ of  \Zboson\ candidates in the electron~(a) and muon~(b) channels. The data are compared to the expectations from Monte-Carlo simulation.
\label{fig:Z_mass_fit}}
\end{center}
\end{figure}

Table~\ref{tab:zemucandidates} summarises the number of $\ensuremath{Z}\rightarrow \ell \ell$ candidates remaining after each major requirement has been imposed. A total of \ntotZe~candidates  pass all requirements in the electron channel and \ntotZmu~candidates in the muon channel, within the invariant mass window
66 $< m_{\ell\ell} <$ 116~GeV.  Figure~\ref{fig:etZ} shows the electron cluster \eT~and muon combined \pT\ of the lepton candidates. The breakdown of the various background contributions are also shown in this figure.
Due to the small size of the backgrounds in both channels, backgrounds are not shown in the subsequent distributions for the $Z$~analysis. Figure~\ref{fig:ptZ} shows the \pT\ spectrum of the $\ensuremath{Z}\rightarrow \ell \ell$ candidates. The invariant mass distribution of the lepton pairs is presented in Fig.~\ref{fig:Z_mass_fit}. The observed resolution degradation in the muon data compared to design expectations is currently under investigation. It has been taken into account in the systematic uncertainties of the cross-section measurement.

% Figure X

\clearpage
%-------------------------------------------
%  Signal and Background rates
%

\section{$W$ and $Z$ boson signals and backgrounds}
\label{sec:wbkg}

In this section, estimates of the various background components in the $W$ and $Z$-candidate samples,
and background-subtracted signal numbers, are presented. Except for the $Z \to \mu \mu$ final state,
the QCD components of the backgrounds were estimated from the data. The electroweak and $t\overline{t}$ components were obtained for all channels from Monte-Carlo simulation.

\subsection{Background estimate for the $W \to e \nu$ channel}
\label{Z selection}

The expected contributions from the $\ensuremath{W}\rightarrow \tau \nu$, $\ensuremath{Z} \rightarrow ee$ and $Z \to \tau \tau$ processes were estimated to be \nWtaunuWe, \nZeeWe, and
\nZtautauWe~events, respectively, while from $t\overline{t}$ production \nttbarWe~events are expected. 

\begin{figure}[t]
  \begin{center}
\includegraphics[width=0.44\textwidth]{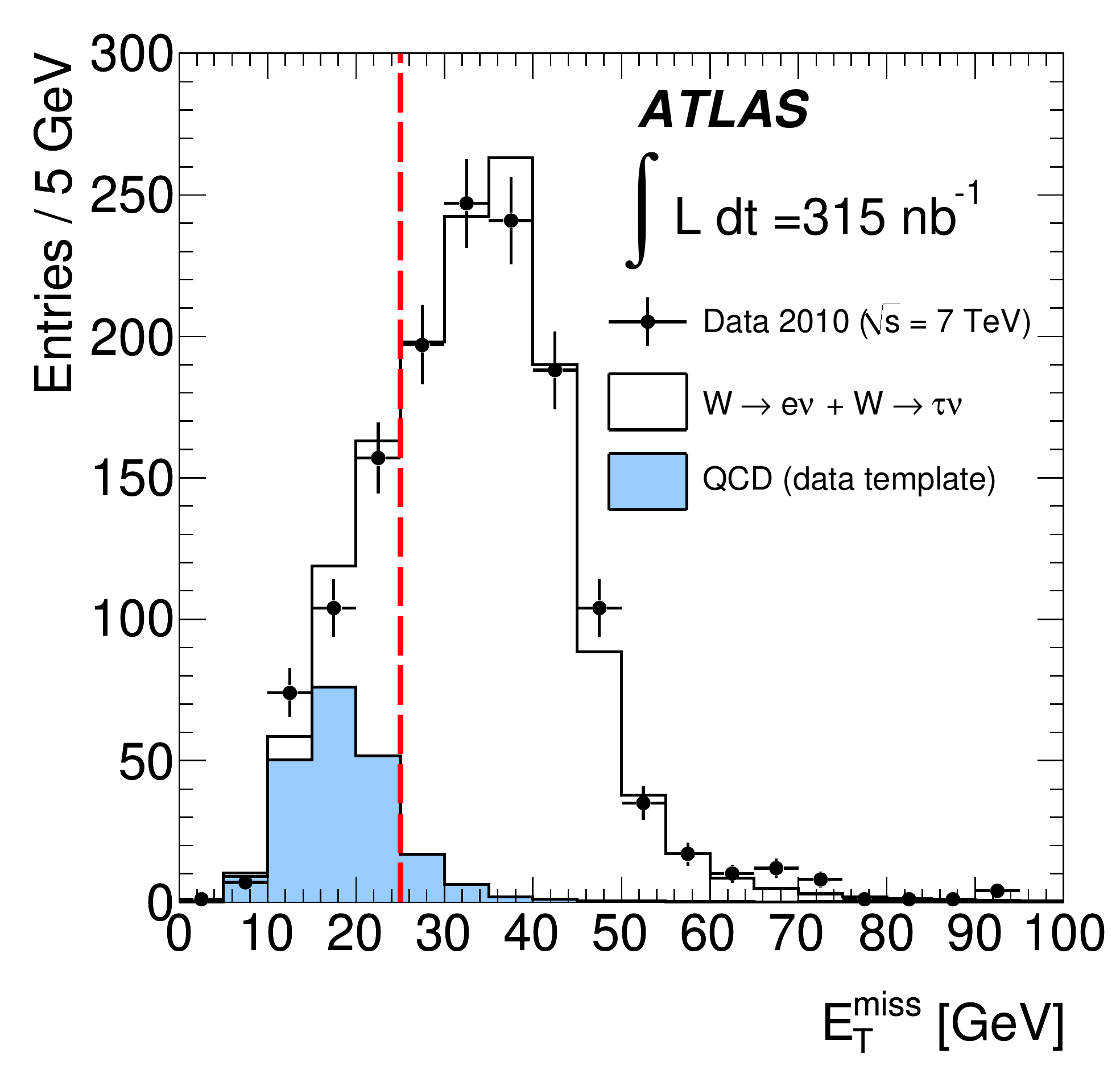}
    \caption{\it\small \label{fig:CaloIsolationPreselection} The distribution of \met\ after applying all 
$W$ selection cuts, except the \met\ cut. The data are shown together with the results of a template fit 
for signal (including the dominant $W \to \tau \nu$ electroweak background contribution) 
and the QCD background. 
The dashed line indicates the cut on \met, as applied in the $W$ analysis.}
\label{iso_2}
      \end{center}
\end{figure}

The QCD background was estimated using the distribution of the 
missing transverse energy \MET\ as measured in data. Events were selected by applying all cuts used in the 
$W$ selection, except the \MET\ cut at 25 GeV. The resulting distribution is displayed in Fig.~\ref{iso_2}. 
The signal and background components in this sample were obtained from a binned maximum likelihood 
template fit. The shapes of the $W \to e \nu$ signal and of the  dominant $W \to \tau \nu$ background were 
taken from Monte-Carlo simulation, whereas the shape of the QCD background was determined from data.

The background template was obtained by using the \Wboson~selection, but not applying all electron identification requirements and reversing
some of the requirements in the ``tight'' electron identification. In order to suppress the residual contribution from $W \to e \nu$ signal events and to obtain an essentially signal-free sample, isolated candidates were rejected. Using a high-statistics QCD-dijet Monte-Carlo sample, it was verified that these requirements produce a background template similar in shape to the background expected from the $W$ selection. The result of the fit to the data is shown in Fig.~\ref{iso_2}. It provides a background estimate in the signal region ($\MET >$~25 GeV) of $N_{\mathrm{QCD}} =$~\nQCDbkgWestat~events, where the uncertainty contains the statistical uncertainty of the data and of the templates. This estimate is used in the extraction of the cross section in Section~\ref{WXsection}.

To estimate the systematic uncertainty, the shape of the background template was varied by applying different
event selection criteria, in particular by varying isolation cuts. In addition, two extreme 
ranges in \MET~(0--25~GeV and 15--100~GeV) were considered as fit ranges.
Based on these studies, the systematic uncertainty on the QCD background was estimated to be $\pm$ 10 events.

As an alternative estimate, the calorimeter isolation variable, 
$\sum^{\Delta R < 0.3}_i E_{\rm{T}}^i / E_{\rm{T}}$, as defined in Section~\ref{sec:results_isol}, was used
as discriminating variable. Due to the limited statistics and the few background events, the fit was performed
after applying the ``loose''  instead of the ``tight'' electron identification, while the requirements 
on \MET\ and \MT\ were kept.
Using this method, the number of QCD background events was estimated to be \nQCDbkgTightWe.
The large error results from the large uncertainty on the
estimation of the jet rejection factor for the ``tight'' requirement with respect to the ``loose'' requirement.
As a further cross-check the background was also estimated from the dijet Monte-Carlo simulation, including the normalisation factor
discussed in Section~\ref{sec:results_presel}, and was found  to be
\nQCDWe~events, which is in agreement with the estimates presented above.

%Figure 13

\subsection{Background estimate for the $W \to \mu \nu$ channel}
\label{sec:mubkg}

For the muon channel,  the expected contributions from $\ensuremath{Z}\rightarrow \mu \mu$,  $\ensuremath{W}\rightarrow \tau \nu$, and $\ensuremath{Z}\rightarrow \tau \tau$ decays are \nZmumuWmu, \nWtaunuWmu, and \nZtautauWmu~events, respectively, while  the $t\overline{t}$ contribution is expected to be \nttbarWmu~events.

The QCD background is primarily composed of heavy-quark decays, with smaller contributions from pion and kaon decays and hadrons faking muons.
Given the large uncertainty in the dijet cross section and the difficulty to properly simulate fake prompt muons,
the QCD background has been derived
from data using the two methods described in the following.

In the baseline method, the QCD background was estimated from a comparison of the number of events seen in data ($N_{\mathrm{iso}}$) after the full $W$ selection,
to the number of events observed ($N_{\mathrm{loose}}$) if the muon isolation requirement is not applied. The number of events in the two samples can be expressed as:
\begin{eqnarray}
N_{\mathrm{loose}} &=& N_{\mathrm{nonQCD}} +  N_{\mathrm{QCD}} \nonumber \\
N_{\mathrm{iso}} &=& \epsilon^{\mathrm{iso}}_{\mathrm{nonQCD}} N_{\mathrm{nonQCD}} + \epsilon^{\mathrm{iso}}_{\mathrm{QCD}} N_{\mathrm{QCD}},
\end{eqnarray}
where $N_{\rm{nonQCD}}$ includes the $W$ signal and the background from the other, non-QCD, physics processes and $\epsilon^{\mathrm{iso}}_{\mathrm{nonQCD}}$ and
$\epsilon^{\mathrm{iso}}_{\mathrm{QCD}}$ denote the corresponding efficiencies of the muon isolation requirement for the two event classes.
If these efficiencies are known, the equations can be solved for $N_{QCD}$.
The muon isolation efficiency for non-QCD events was measured in the data $\Zmm$ sample, while the efficiency for QCD events was estimated from a sample of muons
with transverse momenta in the range of 15 - 20~GeV, which is dominated by dijet events
(see Fig.~\ref{presel_pt_muon}). The efficiency factor was extrapolated to higher $\pt$ values
relevant for the $W$-signal selection using Monte-Carlo simulation.
This method yields a background estimate in the $W$ signal region of  \nQCDWmulabel~events.
The systematic uncertainty is dominated by the uncertainty on the isolation efficiency for QCD events.

This estimate was cross-checked using a method where a similarity relationship in the  plane of \met\ versus lepton isolation
was exploited~\cite{Lohwasser:1265829}.
The plane was divided into four separate regions and the number of background events in the signal region (high \MET\ and low values of the isolation variable)
was estimated from non-isolated events at high \MET\ by applying the corresponding scale factor observed at low \MET.
The calculation was corrected for the contributions from the signal and the electroweak backgrounds and takes into account the
correlation between the two variables, as predicted by Monte-Carlo simulation.
This method yields a background estimate
of \nQCDWmuAlternateTwo~events, in agreement with the baseline estimate.

As a further cross-check the background was also estimated from the dijet Monte-Carlo simulation, after applying the normalisation factor
discussed in Section~\ref{sec:results_presel}, and was found to be \nQCDWmuAlternate,
which is in agreement with the estimates presented above.

The muon channel is also subject to background contamination from cosmic-ray muons
that overlap in time with a collision event. Looking at cosmic-ray muons from non-collision bunches
and events that pass the full $W$ selection but fail the primary vertex selection,
this background component was estimated to be~\nCosmic\ events.

\subsection{Background estimate for the $\ensuremath{Z}\rightarrow  e e $ channel}
\label{sec:zbkg}

Within the invariant mass window $66<m_{ee}<116$~GeV, the contributions from the $\ensuremath{W}\rightarrow e \nu$, $\ensuremath{Z}\rightarrow \tau\tau$, and $t\overline{t}$ processes were determined to be \nWenuZe, \nZtautauZe, and \nttbarZe~events,  respectively, from the Monte-Carlo simulation.

A data-driven estimate of the QCD background was made. The lepton requirement was relaxed from ``medium'' to ``loose'' (as described in Section~\ref{sec:electrons}) and the invariant-mass distribution of the resulting electron-positron pairs was used as a template. A fit consisting of a Breit-Wigner convolved with a Gaussian function, to model the signal, and a second-order polynomial, to model the background, was made to the mass distribution within the mass window $50<m_{ee}<130$~GeV. The number of loose electron background candidate events within the mass window $66 < m_{ee} < 116$~GeV was estimated to be 48.5~$\pm$~6.0(stat) events.
A data-derived ``loose'' to ``medium'' rejection factor for the leptons was determined to be \rejectZe~and was then used to estimate the expected number of lepton pairs which both pass the nominal $\ensuremath{Z}\rightarrow ee$ requirements.  By applying this data-derived rejection factor to each lepton in this ``loose'' pair, a QCD-background estimate totalling \nQCDbkgZelabel~events  in the opposite-charge distribution within the $Z$-mass window was derived. This same procedure was applied to the $\ensuremath{Z}\rightarrow ee$ and corresponding background Monte-Carlo samples, resulting in an estimated QCD background of \nQCDbkgZeMClabel~events, in agreement with the data-derived result.

A systematic uncertainty on the number of background events within the invariant mass window was assessed  by selecting pairs of candidate leptons of varying levels of electron identification, e.g. pairs of lepton candidates before the ``loose'' selection, one ``loose'' and one ``medium'' lepton candidate, and using the corresponding rejection factors measured from the data. The fit stability was verified by changing the bin size of the invariant mass distribution and replacing the second-order polynomial function by a first-order one. The systematic uncertainties of the rejection factors were evaluated by exploring their kinematic dependencies as well as the background composition and signal contamination of the samples used to derive these factors at the various levels of electron identification. The total estimated QCD background within the invariant mass window $66<m_{ee}<116$~GeV is \nQCDbkgZesyslabel.

The number of same-charge lepton pairs that otherwise satisfy all other requirements is a good indicator of the level of background in the selection. In the electron channel, three same-charge lepton pairs satisfy all 
$Z$-boson selection requirements within the invariant mass window.
This is in agreement with the expectation based on Monte-Carlo simulation (see Section~\ref{sec:electrons})
from which 2.3 same-charge lepton pairs are expected
from $Z \to ee$ decays. In addition 0.9 events are expected from QCD background.

\subsection{Background estimate for the $Z \to \mu \mu$ channel}

Within the invariant mass window $66<m_{\mu\mu}<116$~GeV, the contributions from $t\overline{t}$, $Z \rightarrow \tau\tau$, and
$\ensuremath{W}\rightarrow \mu \nu$ are expected to be \nttbarZmu, \nZtautauZmu, and \nWmunuZmu~events, respectively.

For this channel, also the QCD background was determined from Monte-Carlo simulation and 0.04 events are predicted from a simulation of
$b\bar{b}$ production. Given the large uncertainty on the prediction of absolute rates from Monte-Carlo simulation, a 100\% uncertainty is assigned to this estimate. This is considered to be a conservative assumption
because the QCD background to both the single and di-muon samples was found to be overestimated by the same factor of about 1.6.
An estimate of the QCD background from data is still limited by the statistical uncertainty, e.g. no same-charge muon pair was found to satisfy all $Z$-boson selection requirements in the invariant mass window considered and only few events are observed reversing the isolation requirement.
Compared to the contributions described above, all other background sources are negligible.

\subsection{Background-subtracted \Wboson~and \Zboson~candidate events}

\begin{table}[t]
\centering
\small
\begin{tabular}{ c||c||c|c||c }
\hline
\hline
$\ell$ & Observed & Background & Background & Background-subtracted \\
\raisebox{-0.4ex}{}       & \raisebox{-0.4ex}{candidates} & \raisebox{-0.4ex}{(EW$+t\overline{t}$)}     & \raisebox{-0.4ex}{(QCD)}    & \raisebox{-0.4ex}{signal $ N_W^{sig}$}\\*[0.2cm] \hline \hline
\raisebox{-0.4ex}{$e^+$} & \raisebox{-0.4ex}{\ntotWeplus} &
\raisebox{-0.4ex}{\nEWttbarbkgWeplus} &
\raisebox{-0.4ex}{\nQCDbkgWeplus} &
\raisebox{-0.4ex}{\ntotsignalWeplus}
\\*[0.2cm] \hline
\raisebox{-0.4ex}{$e^-$} & \raisebox{-0.4ex}{\ntotWeminus} &
\raisebox{-0.4ex}{\nEWttbarbkgWeminus} &
\raisebox{-0.4ex}{\nQCDbkgWeminus} &
\raisebox{-0.4ex}{\ntotsignalWeminus}
\\*[0.2cm] \hline
\raisebox{-0.4ex}{$e^{\pm}$} & \raisebox{-0.4ex}{\ntotWe} &
\raisebox{-0.4ex}{\nEWttbarbkgWe} &
\raisebox{-0.4ex}{\nQCDbkgWe} &
\raisebox{-0.4ex}{\ntotsignalWe}
\\*[0.2cm] \hline \hline
%muons
\raisebox{-0.4ex}{$\mu^+$}  & \raisebox{-0.4ex}{\ntotWmuplus} &
\raisebox{-0.4ex}{\nEWttbarbkgWmuplus}&
\raisebox{-0.4ex}{\nnonEWWmuplus}&
\raisebox{-0.4ex}{\ntotsignalWmuplus}
\\*[0.2cm] \hline
\raisebox{-0.4ex}{$\mu^-$}  &
\raisebox{-0.4ex}{\ntotWmuminus} &
\raisebox{-0.4ex}{\nEWttbarbkgWmuminus}&
\raisebox{-0.4ex}{\nnonEWWmuminus}&
\raisebox{-0.4ex}{\ntotsignalWmuminus}
\\*[0.2cm]  \hline
\raisebox{-0.4ex}{$\mu^{\pm}$} & \raisebox{-0.4ex}{\ntotWmu} &
\raisebox{-0.4ex}{\nEWttbarbkgWmu} &
\raisebox{-0.4ex}{\nnonEWWmu} &
\raisebox{-0.4ex}{\ntotsignalWmu}
\\*[0.2cm] \hline
\hline
\end{tabular}
\caption{Numbers of observed candidate events for the $\ensuremath{W}\rightarrow \ell \nu$ channel, electroweak ($\ensuremath{W}\rightarrow \tau \nu$, $\ensuremath{Z} \to \ell \ell$,  $Z \to \tau \tau$) plus $t\overline{t}$, and QCD background events, as well as background-subtracted signal events. For the muon channel, the QCD background also contains a small cosmic-ray component. For the electron channel, the QCD background is assumed to be charge independent. The background fits were also performed separately for $W^+$ and $W^-$ production and were found to agree  within uncertainties. The first uncertainty is statistical. The second uncertainty represents the systematics (as described in the text). In addition to what is quoted in this table, a $\pm$11\% uncertainty on the luminosity determination is applicable to the electroweak plus $t\overline{t}$ backgrounds. }
\label{tab:wtable}
\end{table}

\begin{table}[t]
\centering
\small
\begin{tabular}{ c||c||c|c||c }
\hline
\hline
$\ell$ & Observed & Background & Background & Background-subtracted \\
\raisebox{-0.4ex}{}       & \raisebox{-0.4ex}{candidates} & \raisebox{-0.4ex}{(EW$+t\overline{t}$)}     & \raisebox{-0.4ex}{(QCD)}    & \raisebox{-0.4ex}{signal $ N_Z^{sig}$}\\*[0.2cm] \hline \hline
\raisebox{-0.4ex}{$e^{\pm}$} & \raisebox{-0.4ex}{\ntotZe} &
\raisebox{-0.4ex}{\nEWttbarbkgZe} &
\raisebox{-0.4ex}{\nQCDbkgZe} &
\raisebox{-0.4ex}{\ntotsignalZe}
\\*[0.2cm] \hline \hline
%muons
\raisebox{-0.4ex}{$\mu^{\pm}$} & \raisebox{-0.4ex}{\ntotZmu} &
\raisebox{-0.4ex}{\nEWttbarbkgZmu} &
\raisebox{-0.4ex}{\nQCDbkgZmu} &
\raisebox{-0.4ex}{\ntotsignalZmu}
\\*[0.2cm] \hline
\hline
\end{tabular}
\caption{Numbers of observed candidate events for the $\ensuremath{Z}\rightarrow \ell \ell$ channel, electroweak ($\ensuremath{W} \to \ell \nu$, $\ensuremath{Z} \to \tau\tau$) plus $t\overline{t}$, and  QCD background events, as well as background-subtracted signal events. The first uncertainty is statistical. The second uncertainty represents the systematics (as described in the text). In addition to what is quoted in this table, a $\pm$11\% uncertainty on the luminosity determination is applicable to the electroweak plus $t\overline{t}$ backgrounds. }
\label{tab:ztable}
\end{table}

The numbers of observed candidate events for the $\ensuremath{W}\rightarrow \ell \nu$ and $\ensuremath{Z}\rightarrow \ell \ell$ channels, the estimated background events from both the QCD processes and electroweak plus $t\overline{t}$  processes and the number of background-subtracted signal events are summarised in Tables~\ref{tab:wtable} and~\ref{tab:ztable} together with their statistical and systematic uncertainties.
The systematic uncertainties receive contributions from experimental systematic uncertainties
(see Section~\ref{WXsection_effic}), from theoretical uncertainties on the predicted cross sections for
$W$, $Z$ and $t \bar{t}$ production (see Sections 3 and \ref{WXsection_comp}), and from uncertainties on the
parton distribution functions (see Section \ref{WXsection_acc}). The statistical component of the background uncertainty
 is included in the systematic uncertainty of the background-subtracted signal.
The luminosity determination uncertainty of $\pm$11\% is used in all channels but is only applicable to the electroweak and $t\overline{t}$ backgrounds as they are determined from Monte-Carlo simulation. The resulting correlation of the luminosity systematic uncertainty is fully taken into account in the calculation of the cross sections in Section~\ref{WXsection}.

%-------------------------------------------

\clearpage
%---------------------------------------------------------------
%    Cross Section Chapter
%

\section{Cross-section measurements}
\label{WXsection}

\subsection{Methodology}
\label{WXsection_intro}

The production cross sections for the $W$ and $Z$ bosons times the branching ratios for decays into
leptons can be expressed as:

\begin{equation}
\stotWZ \cdot BR(W(Z) \to \ell \nu \ (\ell \ell)) = \frac{ N_{W(Z)}^{\rm{sig}}}
{A_{W(Z)} \cdot C_{W(Z)} \cdot L_{W(Z)}},
\label{EQ::Wxsec}
\end{equation}
where
\begin{itemize}
\item
$ N_W^{\rm{sig}}$ and $N_Z^{\rm{sig}}$ denote the numbers of background-subtracted signal events passing the selection criteria of the analyses in the \Wboson\ and
$Z$  channels, as defined in Section~\ref{sec:wkine}.
\item
$A_W$ and $A_Z$ denote the acceptances for the \Wboson~and \Zboson-boson decays under consideration, defined as the fraction of decays satisfying the geometrical and kinematical constraints at the generator level (fiducial acceptance).
These quantities can only be determined from Monte-Carlo simulations and are
defined here before the decay leptons emit photons via QED final state radiation.
\item
$C_W$ and $C_Z$ denote the ratios between the total number of generated events which pass the final selection requirements after reconstruction and the total number of generated events within the fiducial acceptance. These correction factors include the efficiencies for triggering, reconstructing, and identifying
the $W$ and \Zboson-boson decays falling within the acceptance.
\item
$L_W$ and $L_Z$ denote the integrated luminosities for the channels of interest.
\end{itemize}

The resulting cross sections, as defined by Eq.~(\ref{EQ::Wxsec}),
define measured total inclusive cross sections. For the $W$ boson they are measured separately for $W^+$, $W^-$ and $W$ production. The total
cross sections are denoted as $\stotWp$,  $\stotWm$ and $\stotW$. The corresponding $Z$ cross section
in the invariant mass range 66 $< m_{\ell \ell} <$ 116 GeV~is referred to as $\stotZ$.

These total cross sections are derived from the measurements of the cross sections in the fiducial region, which are denoted as
fiducial cross sections $\sfidWZ$. They are related to the total cross sections via
\begin{equation}
\sfidWZ \cdot BR(W(Z) \to \ell \nu \ (\ell \ell))    =  \stotWZ \cdot BR(W(Z) \to \ell \nu \ (\ell \ell)) \ \cdot A_{W(Z)} = \frac{ N_{W(Z)}^{\rm{sig}}}
{C_{W(Z)} \cdot L_{W(Z)}}. \label{eq:sig-fid}
\end{equation}
By definition, no acceptance correction factors are needed for the measurement of the fiducial cross sections. Therefore these cross sections are not affected by
significant theoretical uncertainties.
Hence, future improvements on the predictions of $A_W$ and $A_Z$ can be used to extract improved
total cross-section measurements. Cross-section results in this paper are presented below for both the electron and muon channels as well as for their combination.

\subsection{The correction factors $C_W$ and $C_Z$}
\label{WXsection_effic}

The central values of the correction factors \cw\ and \cz\ are computed using Monte-Carlo simulation. Only in the case of 
the trigger efficiencies for muons, corrections determined from data are applied. To assess the
uncertainties affecting these factors, the following decomposition is used:
\begin{equation}
C_W = \epsilon^W_{\rm{event}} \cdot \alpha^{W}_{\rm{reco}} \cdot \epsilon^{W}_{\rm{lep}} \cdot \epsilon^{W}_{\rm{trig}}
\label{eq:CW}
\end{equation}
\[
C_Z = \epsilon^Z_{\rm{event}} \cdot \alpha^Z_{\rm{reco}} \cdot (\epsilon^Z_{\rm{lep}})^2 \cdot  [ 1 - (1-\epsilon^Z_{\rm{trig}})^2],
\]
where the individual factors account for event selection efficiencies ($\epsilon_{\ \rm{event}}$), e.g. primary vertex requirements,
lepton reconstruction and identification efficiencies ($\epsilon_{\rm{lep}}$) and the trigger efficiency with respect to
selected lepton candidates ($\epsilon_{\rm{trig}}$).
The exact definitions of these terms for electron and muon final states are given below in the respective
subsections.
The factor $\alpha_{\rm{reco}}$ accounts for all differences observed between the efficiencies of applying
the kinematic and geometrical cuts at generator level and reconstruction level. It includes for example
effects due to the detector resolution on the lepton transverse momenta/energies and on the missing transverse
energy. This factor also includes basic reconstruction efficiencies.
The choice mentioned above of calculating the acceptance factors for leptons before they emit final-state radiation of photons
also affects this correction factor in a significant way, in particular for electron final states. Finally,
this factor includes migration and combinatorial effects and therefore may have values larger than unity.

\subsubsection{Electron final states}
For electrons, the efficiency of the L1 trigger with its nominal threshold of~10~GeV was measured to be close to~100\%, using minimum-bias data and samples obtained with lower-threshold electron triggers at lower luminosities.

The term $\epsilon_{\rm{lep}}$ refers to the
``tight'' and ``medium'' electron identification efficiencies for the $W$ and $Z$ selection, respectively. They
are defined with respect to all reconstructed electron candidates and were determined from Monte-Carlo simulation.
A strong~\et\ and~$\eta$ dependence is observed for these efficiencies, which arises mainly from material
interactions in the inner detector. It is a significant source of systematic uncertainty for $C_W$ and $C_Z$.

This uncertainty was evaluated by combining the results from dedicated simulations, including additional material
in the inner detector and in front of the electromagnetic calorimeter, with those obtained from direct measurements of the efficiencies from data. These measurements were performed with limited statistical precision using as probes unbiased electrons selected together with a well identified tag electron in \Zee\ candidate events, and with better accuracy using as probes unbiased electrons in selected \Wen\ candidate events with large and isolated~\met, as discussed in~Section~\ref{sec:electrons}. All these direct measurements are in agreement with the nominal values, within the estimated overall systematic uncertainties quoted in~Table~\ref{eff_cwz_ele} of~5.2\% and~4.2\%, for the ``tight'' and ``medium'' electron identification efficiencies, respectively.

The factor $\alpha_{\rm{reco}}$ includes in addition to the electron and $\met$ resolution effects
the basic reconstruction efficiency, e.g. the probability for an electron
that an electromagnetic cluster in the calorimeter is reconstructed in a fiducial region of the detector and
is loosely matched to a reconstructed track. This includes losses of leptons due to imperfect
regions of the detector within the geometrical acceptance.
In the case of \Zee\ candidates, the value of~$\alpha_{reco}$ is significantly lower than for \Wen\ candidates because both electrons must fall outside the imperfect regions of the detector and also because~3.3\% of the \Zee\
candidates fail the requirement of a pair of oppositely charged leptons, as discussed in~Section~\ref{sec:zbkg}.

The central values as well as the relative uncertainties of the efficiencies and of $\alpha_{\rm{reco}}$ are summarised for both
$W \to e \nu$ and $ Z \to ee$ final states in Table~\ref{eff_cwz_ele}.

\begin{table}[htb]
\centering
\begin{footnotesize}
\begin{tabular}{ l|c|c || c |c|| c|c || c | c}
\hline
\hline
% & \multicolumn{4}{|c||}{Electrons} &  \multicolumn{4}{|c}{Muons} \\
&  \multicolumn{2}{|c||}{$W \to e \nu$}    &  \multicolumn{2}{|c||}{$Z \to ee $}
         &  \multicolumn{2}{|c||}{$W \to \mu \nu$}    &  \multicolumn{2}{|c}{$Z \to \mu \mu $} \\
       & Central 	& Relative & Central  & Relative     & Central 	& Relative & Central & Relative             \\
       &    value	& uncertainty & value & uncertainty  & value	& uncertainty & value           & uncertainty           \\
\hline
\hline
$\epsilon_{\rm{event}}$      & 1.000  &  $<$ 0.2\%  & 1.000   & $<$ 0.2\%  & 0.998 & $<$0.2\% & 0.998 & $<$0.2\% \\
$\epsilon_{\rm{lep}}  $      & 0.749   & 	5.2\%   & 0.943   & 	4.2\%      & 0.886 & 2.7\%     & 0.894 & 2.7\%    \\
$\epsilon_{\rm{trig}} $      & 0.998   &$<$ 0.2\%   & 0.998   & $<$ 0.2\%  & 0.815 & 1.9\%     & 0.811 & 1.9\%    \\
\hline
$\alpha_{\rm{reco}}$       & 0.882      &    3.9\%   & 0.732   & 3.2\%       & 1.051 & 2.3 \%     & 1.007 & 0.7 \%   \\
\hline
\hline
$C_W, C_Z$                    & 0.659      &    7.0\%   & 0.651   & 9.4\%       & 0.758 & 4.0\%      & 0.773  & 5.5\%    \\
\hline
\hline
\end{tabular}
\caption{Efficiency factors per lepton and $\alpha_{\rm{reco}}$ as well as their relative uncertainties which enter the calculation of the correction factors $C_W$ and $C_Z$ for both lepton channels.
The trigger efficiencies were measured from data.
The other efficiencies and their uncertainties were determined from Monte-Carlo simulation and have been validated with data,
 as described in the text.
It should be noted that for $Z$ bosons the trigger and identification efficiencies are given per lepton,
according to the definition given in Eq.~(\ref{eq:CW}).
\label{eff_cwz_ele}}
\end{footnotesize}
\end{table}

\subsubsection{Muon final states}

For the muon channels, the trigger efficiency was measured in data relative to reconstructed muons, using a control
sample selected with an independent jet trigger.
Combined reconstructed muons above 20~GeV are selected by applying the same criteria as adopted for the $W$ selection.
Tracks are then extrapolated to the trigger chamber planes and the efficiency is measured by looking at 
associated trigger signals in the barrel or end-cap regions separately.
The ratio of the event trigger efficiency measured in data and predicted by Monte-Carlo simulation
is 0.929 $\pm$ 0.010~(stat) $\pm$ 0.015~(syst)
in the $W$ channel and 0.981 $\pm$ 0.003~(stat) $\pm$ 0.006~(syst) in the $Z$ channel.
These values are significantly different from 1 and
therefore a correction is applied to the central values of  $C_W$ and $C_Z$.
The systematic uncertainty is derived from changing the tolerance on the association between tracks and 
trigger signals, by checking the stability of the 20~GeV threshold in the plateau region and by comparing measurements obtained using different muon reconstruction algorithms.

The term $\epsilon_{\rm{lep}}$ includes the combined-muon reconstruction efficiency relative to the inner detector track
($\epsilon$ = 0.924~$\pm$~0.023) and the efficiencies of quality ($\epsilon$ = 0.966~$\pm$~0.001) and isolation 
($\epsilon$ = 0.993~$\pm$~0.010) requirements.
The combined-muon reconstruction efficiency was determined by Monte-Carlo simulation and cross-checked to be in agreement with
data, as explained in
Section~\ref{sec:muons}. The isolation efficiency was measured in data using a sample of muons from $Z$ decays and found to be in agreement with
Monte-Carlo simulation within $\pm$1\%. This difference was assigned as a systematic uncertainty.

The factor $\alpha_{\rm{reco}}$ includes in addition to the muon and $\met$ resolution effects the efficiency for
the reconstruction of a track in the inner detector ($\epsilon$~=~0.989~$\pm$~0.010).
The dominant systematic uncertainties on $\alpha_{\rm{reco}}$ result from the uncertainties on the muon momentum scale and resolution
(as derived in Section~\ref{sec:muons}) and —for the $W$ analysis— from uncertainties on the $\met$ scale and resolution.

The central values, as well as the relative uncertainties, of the efficiencies and of $\alpha_{\rm{reco}}$ are also summarised for both
$W \to \mu \nu$ and $ Z \to \mu \mu$ final states in Table~\ref{eff_cwz_ele}.

\subsubsection{$C_W$ and $C_Z$ and their  uncertainties}
The central values of the correction factors $C_W$ and $C_Z$, shown in Table~\ref{eff_cwz_ele},  were
determined to a large extent using Monte-Carlo simulation. For muons, data-driven corrections for the trigger efficiencies are included.
The uncertainties of $C_W$ and $C_Z$ receive contributions from
the uncertainties of the efficiencies discussed above and
from uncertainties on the correction factor $\alpha_{\rm{reco}}$. A breakdown of the various components is given in
Tables~\ref{t:cw-errors-elec} and \ref{t:cw-errors-muon} for electron and muon final states, respectively.
The decomposition was made in such a way that the correlations between the different contributions are negligible.
 For electrons, the main contributions result from
uncertainties on the electron reconstruction efficiency and from material effects in the inner detector as well as
from uncertainties on the electron energy scale and resolution.
For muons, the uncertainties on the reconstruction efficiency and on the \MET\ scale and resolution are dominant.

The uncertainties on $C_W$ linked to uncertainties on the scale of the missing transverse energy were
determined from a variation of the response of cells in topological clusters within the range given in Section~\ref{sec:met}.
These changes propagate to an uncertainty of $\pm$1.5\% on the
number of accepted $W \to \ell \nu$ events. Other sources of uncertainty, namely the imperfect modelling of the
overall \MET\ response (low energy hadrons) and resolution, of the underlying event and pile-up effects,
lead to acceptance changes at the level of $\pm$1\%, resulting in a total uncertainty
of $\pm$ 2\% on $C_W$.

\begin{table}[t]
\begin{center}
\small
\begin{tabular}{l| c | c}
\hline
\hline
Parameter                                              &    $\delta C_W / C_W $(\%)  &    $\delta C_Z / C_Z $(\%)  \\
\hline %----------------------------------------------------------------------------------------------------------------
Trigger efficiency                                     &   $<$0.2 & $<$0.2  \\
%------------------------------------------------------------------------------------------------------------------------
Material effects, reconstruction and identification    &   5.6    & 8.8 \\
%------------------------------------------------------------------------------------------------------------------------
Energy scale and resolution                            &   3.3    & 1.9 \\
%------------------------------------------------------------------------------------------------------------------------
$\met$ scale and resolution                             &   2.0    & - \\
%------------------------------------------------------------------------------------------------------------------------
Problematic regions in the calorimeter                 &   1.4    & 2.7 \\
%------------------------------------------------------------------------------------------------------------------------
Pile-up                                                &   0.5    & 0.2 \\
Charge misidentification                               &   0.5    & 0.5 \\
%------------------------------------------------------------------------------------------------------------------------
FSR modelling                                          &   0.3    & 0.3 \\
\hline
Theoretical uncertainty (PDFs)                         &   0.3    & 0.3    \\
\hline
\hline %---------------------------------------------------------------------------------------------------------------------
Total uncertainty                                      &   7.0    & 9.4     \\
\hline
\hline
\end{tabular}
\caption{Summary of the different terms contributing to the uncertainty on $C_W$ and $C_Z$ for electron final states.
The decomposition has been made such that correlations between the various contributions are negligible.
\label{t:cw-errors-elec}}
\end{center}
\end{table}

\begin{table}[t]
\begin{center}
\small
\begin{tabular}{l|c |c}
\hline
\hline %----------------------------------------------------------------------------------------------------------------
Parameter                                                    &    $\delta C_W / C_W$(\%)  &    $\delta C_Z / C_Z $(\%)  \\
\hline %----------------------------------------------------------------------------------------------------------------
Trigger efficiency                                           &   1.9   & 0.7  \\
%------------------------------------------------------------------------------------------------------------------------
Reconstruction efficiency                                    &   2.5    & 5.0 \\
%------------------------------------------------------------------------------------------------------------------------
Momentum scale                                               &   1.2   & 0.5 \\
%------------------------------------------------------------------------------------------------------------------------
Momentum resolution                                          &   0.2   & 0.5 \\
%------------------------------------------------------------------------------------------------------------------------
$\met$ scale and resolution                                   &   2.0  & -  \\
%------------------------------------------------------------------------------------------------------------------------
Isolation efficiency                                         &   1.0  & 2.0  \\
\hline
Theoretical uncertainty (PDFs)                               &   0.3  & 0.3  \\
\hline
\hline %---------------------------------------------------------------------------------------------------------------------
Total  uncertainty                                           &    4.0    & 5.5     \\
\hline %---------------------------------------------------------------------------------------------------------------------
\hline %---------------------------------------------------------------------------------------------------------------------
\end{tabular}
\caption{Summary of the different terms contributing to the uncertainty on $C_W$ and $C_Z$ for muon final states.
The decomposition has been made such that correlations between the various contributions are negligible.
\label{t:cw-errors-muon}}
\end{center}
\end{table}

In addition uncertainties arising from QED final-state radiation and
theoretical uncertainties, resulting predominantly from structure function parametrisations, have been considered.
The purely theoretical uncertainty on the QED final-state radiation emission is very small, typically smaller
than~0.2\%~\cite{Nanava:2009vg,Golonka:2005pn}. It can be neglected compared to the other uncertainties
discussed in~Section~\ref{WXsection_acc}. In the case of electrons and collinear emission of QED photons, however,
there is an experimental uncertainty arising from the transport of low-energy photons through the detector material
and the response of the electromagnetic calorimeter which was estimated to be~$<$~0.3\% on $C_W$ and $C_Z$.
%from dedicated studies assuming that collinear photons with transverse energies below~1~GeV were not measured
%by the electromagnetic calorimeter.
Finally, using the prescription described in Section~\ref{WXsection_acc},
the relative uncertainties on $C_W$ and $C_Z$ resulting from structure function parametrisations were estimated to
be small, at the level of $\pm$0.3\%.

As can be seen from the numbers given in Tables~\ref{t:cw-errors-elec} and \ref{t:cw-errors-muon}, the total uncertainties
on $C_W$ and $C_Z$
are larger for electrons than for muons. This is mainly due to the higher sensitivity of electrons to material effects
in the inner detector and the current knowledge of the electron energy scale compared to the muon momentum scale.

\subsection{Measured fiducial cross sections}

According to Eq.(\ref{eq:sig-fid}), the correction factors $C_W$ and $C_Z$, the number of observed events,
and the integrated luminosity are the elements for the extraction of the fiducial cross sections.
All relevant numbers are summarised, separated for  \Wplus, \Wminus, $W$ and $Z$~production and decay in the electron and muon channels in
Tables~\ref{FiducialXSections} and \ref{FiducialZXSections}, respectively. Using these numbers, the fiducial cross sections
reported in Table~\ref{t:xs_fiducial} are obtained.

\begin{table}[t]
\begin{center}
\footnotesize
\begin{tabular}{ l||r|r|r|r||r|r|r|r||r|r|r|r}
\hline
\hline
%\cline{1-13}
   & \multicolumn{4}{ c||}{$W^+$} & \multicolumn{4}{|c||}{$W^-$}  & \multicolumn{4}{|c }{$W$}  \\
\hline
%\multicolumn{13}{ c }{}   \\
\multicolumn{13}{ c }{\bf Electron channel}   \\ \hline
 & value & stat & syst & lumi & value & stat & syst & lumi & value & stat & syst & lumi \\ \hline \hline
$N^{\rm{sig}}_W$ & 604.2  & 25.2 & 7.6 & 2.0 & 403.2  & 20.8 & 7.5 & 1.5 & 1007.5 & 32.7 & 10.8 & 3.5 \\
$L_W$  [nb$^{-1}$] &
\lumWenolabel & - & - & 35   &\lumWenolabel & - & - & 35   &\lumWenolabel & - & - & 35   \\
$C_W$ & \CWeplus  & - & 0.046 & -   & \CWeminus & - & 0.046 & -   & \CWe      & - & 0.046 & -      \\
$A_W$       & 0.466  & -  & 0.014  & -   & 0.457  & -  & 0.014  & -   & 0.462  & -  & 0.014  & -        \\
\hline
\hline
% \multicolumn{13}{ c }{}   \\
\multicolumn{13}{ c }{\bf Muon channel}   \\
\hline
& value & stat & syst & lumi & value & stat & syst & lumi & value & stat & syst & lumi \\ \hline \hline
$N^{\rm{sig}}_W$  & 655.6 & 26.6 & 6.2 & 4.7 & 425.0 & 21.7 & 5.4 & 3.9 & 1080.6 & 34.4 & 11.2 & 8.5        \\
$L_{W}$ [nb$^{-1}$]&
\lumWmunolabel & - & - & 34   & \lumWmunolabel & - & - & 34   & \lumWmunolabel & - & - & 34   \\ 
$C_W$ & \CWmuplus  & - & 0.031 & -   &
\CWmuminus & - & 0.030 & -   & \CWmu      & - & 0.030 & -      \\
$A_W$ & 0.484 & - & 0.015 & -  & 0.475 & - & 0.014 & -  & 0.480 & - & 0.014 & -       \\

\hline
\hline
\end{tabular}
\caption{ Summary of input quantities for the calculation of the $W^+$, $W^-$ and $W$ boson production cross sections. 
For each channel, the observed numbers of signal events after background subtraction, the correction factors~$C_W$, the acceptance factors $A_W$
and the integrated luminosities are given, with their statistical, systematic, and luminosity uncertainties.
\label{FiducialXSections}}
\end{center}
\end{table}

\begin{table}[t]
\begin{center}
\footnotesize
\begin{tabular}{ l||r|r|r|r||r|r|r|r}
\hline
\hline
%\cline{1-9}
   & \multicolumn{8}{|c}{$\Zg$}  \\
\hline \hline
& \multicolumn{4}{|c||}{\bf Electron channel} &  \multicolumn{4}{|c}{\bf Muon channel} \\ \hline
& value & stat & syst & lumi  & value & stat & syst & lumi \\ \hline \hline
$N^{\rm{sig}}_Z$  & 68.8 & 8.4 & 0.4 & 0.0  & 108.8 & 10.4 & 0.0 & 0.0 \\ 
$L_{Z}$ [nb$^{-1}$] & \lumZenolabel & - & - & 35  & \lumZmunolabel & - & - & 35   \\ 
$C_Z$   & \CZe & - & 0.061 & -  & \CZmu & - & 0.043 & -    \\ 
$A_Z$   & 0.446  & -  & 0.018  & -  & 0.486 & - & 0.019 & -       \\
\hline
\hline
\end{tabular}
\caption{
Summary of input quantities for the calculation of the $\Zg$ boson production cross section. 
For the electron and muon channels, the observed numbers of signal events after background subtraction, 
the correction factors~$C_Z$, the acceptance factors $A_Z$
and the integrated luminosities are given, with their statistical, systematic, and luminosity uncertainties.
\label{FiducialZXSections}}
\end{center}
\end{table}

\begin{table}[tbqh]
\small
\begin{center}
\begin{tabular}{l | c | c }
\hline
\hline
 & & \\
 & {\small $\sfidWpm \cdot$ BR($W \to e \nu$) \ \ [nb]}  & {\small $\sfidWpm \cdot$ BR($W \to \mu \nu$) \ \ [nb]} \\
\hline
 & & \\
${W^+} $      &   $~\sigfidWeplus$  & $\sigfidWmuplus$   \\
 & & \\
${W^-} $      &   $~\sigfidWeminus$ & $\sigfidWmuminus$  \\
  & & \\
$ W $  &   $~\sigfidWe$      & $\sigfidWmu$       \\
 &   \\
\hline
\hline
 & & \\
 & {\small $\sfidZg \cdot$ BR($Z/\gamma^* \to e e$)\ [nb],}  & {\small $\sfidZg \cdot$ BR($Z/\gamma^* \to \mu \mu$) \ [nb],}
 \\
&  {\small $66<m_{ee}<116$~GeV}   &  {\small $66<m_{\mu \mu}<116$~GeV }\\
\hline
& & \\
$Z/\gamma^*$      &   $~\sigfidZe$  & $\sigfidZmu$   \\
\hline
\hline
\end{tabular}
\caption{Measured fiducial cross sections times leptonic branching ratios for $W^+$, $W^-$, $W$ and $\Zg$ production in the
electron and muon final states.}
\label{t:xs_fiducial}
\end{center}
\end{table}

Even with the rather low integrated luminosity of about 320~nb$^{-1},$ these $W$ cross-section measurements are already dominated by
systematic uncertainties, most prominently by the luminosity uncertainty of $\pm$11\% and to a lesser degree by the experimental
uncertainties discussed in the previous section.
As already mentioned, these cross sections are only very weakly affected by theoretical uncertainties related to the calculation of
acceptance corrections. The calculation of these correction factors and of the related uncertainties is discussed in the next
section.

\subsection{Acceptances and uncertainties}\label{WXsection_acc}

The total cross sections are derived from the measured fiducial cross sections by applying the factors
$A_W$ and $A_Z$ for the phase-space requirements applied in the analysis:
\begin{itemize}
\item $W\rightarrow e\nu$ : $\ET^e>20$~GeV, $|\eta|<2.47$, excluding $1.37<| \eta | <1.52$, $\pT^\nu>25$~GeV, $m_T>40$~GeV;
\item $W\rightarrow \mu\nu$ : $\pT^\mu>20$~GeV, $|\eta|<2.4$, $\pT^\nu>25$~GeV, $m_T>40$~GeV;

\item $Z\rightarrow ee$ : $\ET^e>20$~GeV, $|\eta|<2.47$, excluding $1.37< | \eta  | <1.52$, \   66$<m_{ee}<116$~GeV;
\item $Z\rightarrow \mu\mu$ : $\pT^\mu>20$~GeV, $|\eta|<2.4$, \ 66$< m_{\mu\mu} < 116$~GeV.
\end{itemize}

The calculation of $A_W$ and $A_Z$ is based on Monte-Carlo simulation.
Losses due to QED final-state radiation~\cite{Nanava:2009vg,Golonka:2005pn} are
included in the correction factors $C_W$ and $C_Z$, evaluated with a full simulation of the detector response.

The acceptances are calculated using the PYTHIA Monte-Carlo generator with the modified leading order parton distribution function set MRST LO*~\cite{mrst} and the corresponding ATLAS MC09 tune~\cite{MC09tune}. The central values of the acceptances are provided in Table~\ref{table_Acc_all}, separated for $W^+$, $W^-$,
$W$ and $\Zg$ production. In addition, the ratio $A_W/A_Z$ is
given, which is relevant for the measurement of the cross-section ratio (see Section~\ref{s:xs-ratio}).
The  statistical uncertainties resulting from the Monte-Carlo samples are negligible.

The systematic uncertainties on the acceptances are dominated by the limited knowledge of the proton PDFs and the modelling of the $W$ and $Z$ boson production at the LHC. These uncertainties therefore were derived by combining three different components:

\begin{itemize}
\item The uncertainties within one PDF set were derived using the CTEQ 6.6 PDF~\cite{Nadolsky:2008zw} error eigenvector sets at the 90\% C.L. limit, in combination with the
MC@NLO acceptance calculation. The relative uncertainties on the acceptances were found to be $\pm$1.0\% for $W^+$,  $\pm$1.8\% for $W^-$,
and $\pm$1.6\% for $\Zg$-boson production.
\item
Larger uncertainties were found between different PDF sets. They  have been estimated using PYTHIA, based on the maximal difference between the
MRST LO*,  CTEQ 6.6 and HERAPDF 1.0~\cite{:2009wt} sets. The relative uncertainties on the acceptances were found to be $\pm$2.7\% for $W^+$,  $\pm$0.9\% for $W^-$,
and $\pm$2.0\% for $\Zg$-boson production.

\item  The uncertainties due to the modelling of $W$ and $Z$ production were derived from the difference obtained between the PYTHIA and
MC@NLO simulations, using the same PDF set, CTEQ~6.6. In this case the relative uncertainties on the acceptances were found to be $\pm$0.4\% for $W^+$,  $\pm$1.4\% for $W^-$, and $\pm$2.3\% for $\Zg$-boson production.

\end{itemize}

Adding these components in quadrature results in systematic uncertainties on the acceptance values
for $\Wplus$, $\Wminus$ and
$\Zg$ production of $\pm 3.2$\%, $\pm 2.7$\% and $\pm 3.8$\%, respectively. Approximate numbers of
$\pm$3\% and $\pm$4\% are used in the following
as the overall relative systematic uncertainties for the PYTHIA
acceptance values $A_W$ and $A_Z$, respectively.

\begin{table}[t]
\centering
\begin{small}
\begin{tabular}{l|l|l|l||l|| l }
\hline
\hline
MC  & $A_{W^+}$  &  $A_{W^-}$ & $A_W$ & $A_Z$ & $A_W / A_Z$ \\
 & $\ensuremath{W}^+\rightarrow e^+ \nu$
 & $\ensuremath{W}^-\rightarrow e^- \nu$
 & $\ensuremath{W}  \rightarrow e \nu$
 & $\ensuremath{\Zg}  \rightarrow e^+e^-$  &
\\ \hline \hline
PYTHIA MRST LO*    & 0.466 & 0.457 & 0.462 & 0.446 & 1.036 \\
\hline
PYTHIA CTEQ6.6   & 0.479 & 0.458 & 0.471 & 0.455 & 1.035  \\
PYTHIA HERAPDF1.0 & 0.477 & 0.461 & 0.470 & 0.451 & 1.042  \\
MC@NLO HERAPDF1.0 & 0.475 & 0.454 & 0.465 & 0.440 & 1.057 \\
MC@NLO CTEQ6.6    & 0.478 & 0.452 & 0.465 & 0.445 & 1.045 \\
\hline
\hline
 & $A_{W^+}$  &  $A_{W^-}$ & $A_W$ & $A_Z$  & $A_W / A_Z$   \\
 & $\ensuremath{W}^+\rightarrow \mu^+ \nu$
 & $\ensuremath{W}^-\rightarrow \mu^- \nu$
 & $\ensuremath{W}  \rightarrow \mu \nu$
 & $\ensuremath{\Zg}  \rightarrow \mu^+\mu^-$  &
\\ \hline
\hline
PYTHIA MRSTLO*    & 0.484  & 0.475  & 0.480  & 0.486 & 0.988 \\
\hline
PYTHIA CTEQ6.6   & 0.499 & 0.477 & 0.490 & 0.496 &  0.987  \\
PYTHIA HERAPDF1.0 & 0.496 & 0.479 & 0.489 & 0.492 &  0.994   \\
MC@NLO HERAPDF1.0 & 0.494 & 0.472  & 0.483  & 0.479  & 1.008 \\
MC@NLO CTEQ6.6    & 0.496 & 0.470  & 0.483  & 0.485  & 0.996 \\
\hline
\hline
\end{tabular}
\caption{Summary of acceptance values  $A_{W}$ for $\ensuremath{W}\rightarrow e \nu$ and  $\ensuremath{W}\rightarrow \mu \nu$ (separated for charges and combined) and $A_Z$ for $\Zg \to ee$ and  $\Zg \to \mu \mu$ as well as the ratio $A_W / A_Z$
using various Monte-Carlo simulations. \label{table_Acc_all}}
\end{small}
\end{table}

%--------------------------------------
%  W / Z cross section input quantities
%--------------------------------------

The uncertainties on the ratios of acceptances cannot be naively calculated via error propagation since the theoretical uncertainties exhibit significant correlations and the PDF uncertainties are expected to cancel partially. Using the same combination of the three sources of uncertainties, as discussed for the individual acceptances, the uncertaintiess on the ratios are estimated to be
$\pm$3.0\% for $A_{W^+} / A_Z$,
$\pm$2.5\% for $A_{W^-} / A_Z$,
and $\pm$1.5\% for the charge combined ratio  $A_{W} / A_Z$.

\subsection{Measured total cross sections}
\label{WXsection_totboson}

The total cross sections are obtained by dividing the
measured fiducial cross sections by the acceptance factors.
The results are summarized in Table~\ref{t:xs_total},
separated for the electron and muon final states.

\begin{table}[tbqh]
%\footnotesize
\small
\begin{center}
\begin{tabular}{l | c | c }
\hline
\hline
 & & \\
 & {\small $\stotWpm  \cdot$ BR($W \to e \nu$) \ \   [nb]} & {\small $\stotWpm  \cdot$ BR($W \to \mu \nu$) \ \   [nb]} \\
\hline
 & & \\
${W^+} $      &   $~\sigWeplus$  & $\sigWmuplus$   \\
 & & \\
${W^-} $      &   $~\sigWeminus$ & $\sigWmuminus$  \\
  & & \\
$ W $  &   $~\sigWe$      & $\sigWmu$       \\
 &   \\
\hline
\hline
 & & \\
 & {\small $\stotZg \cdot$ BR($Z/\gamma^* \to e e$) \ [nb],}   & {\small $\stotZg \cdot$ BR($Z/\gamma^* \to \mu \mu$) \ [nb],}\\
  &  \small{$66<m_{ee}<116$~GeV}   &  { \small $66<m_{\mu \mu}<116$~GeV }\\
\hline
& & \\
$Z/\gamma^*$      &   $~\sigZe$  & $\sigZmu$   \\
\hline
\hline
\end{tabular}
\caption{Measured total cross sections times leptonic branching ratios for $W^+$, $W^-$, $W$ and $\Zg$ production in the
electron and muon final states.}
\label{t:xs_total}
\end{center}
\end{table}

Assuming lepton universality, the measured total cross sections in the two lepton final states can be combined
to decrease the statistical uncertainty. For the combination, it is assumed that the uncertainties on the integrated
luminosity, on the acceptance factors
$A_W$ and $A_Z$ and the uncertainty resulting from the hadronic part of the \met\ measurement are fully correlated
between the electron and muon channels. All other uncertainties are assumed to be uncorrelated.
For the $W$ production cross sections the following results are obtained:
\[
\stotWp  \cdot  \rm{BR}(W \to \ell \nu) \ = \ \sigWplus \ \rm{nb},     \nonumber
\]
\[
\stotWm  \cdot  \rm{BR}(W \to \ell \nu) \ = \ \sigWminus \ \rm{nb},     \nonumber
\]
\[
\stotW   \cdot  \rm{BR}(W \to \ell \nu) \ = \ \sigW \ \rm{nb}.    \nonumber
\]
For the $\Zg$ production cross section, measured in the mass range $ 66<m_{\ell \ell}<116$~GeV, the combined result is:
\[
\stotZg   \cdot  \rm{BR}(\Zg \to \ell \ell) \ = \ \sigZnb \ \rm{nb}  \hspace*{1.0cm}  (66<m_{\ell \ell}<116~\rm{GeV}).  \nonumber
\]

It should be noted that the pure $Z$-boson cross section is expected to be 2\% lower in the mass range considered.

Due to the additional uncertainties on $A_W$ and $A_Z$ the relative systematic uncertainties have slightly increased, as compared
to the fiducial cross sections. For the total $W$ production cross section, the relative uncertainties are
$\pm$3.3\%(stat), $\pm$7.7\%(syst) and $\pm$11\%(lumi). For the $\Zg$ production cross section the statistical uncertainty
is still larger than the experimental systematic uncertainty. The relative uncertainties are
$\pm$7.2\%(stat), $\pm$4.8\%(syst) and $\pm$11\%(lumi).

\subsection{Comparison to theoretical calculations}
\label{WXsection_comp}

\begin{figure}[t,h]
  \begin{center}
{\includegraphics[width=0.48\textwidth]{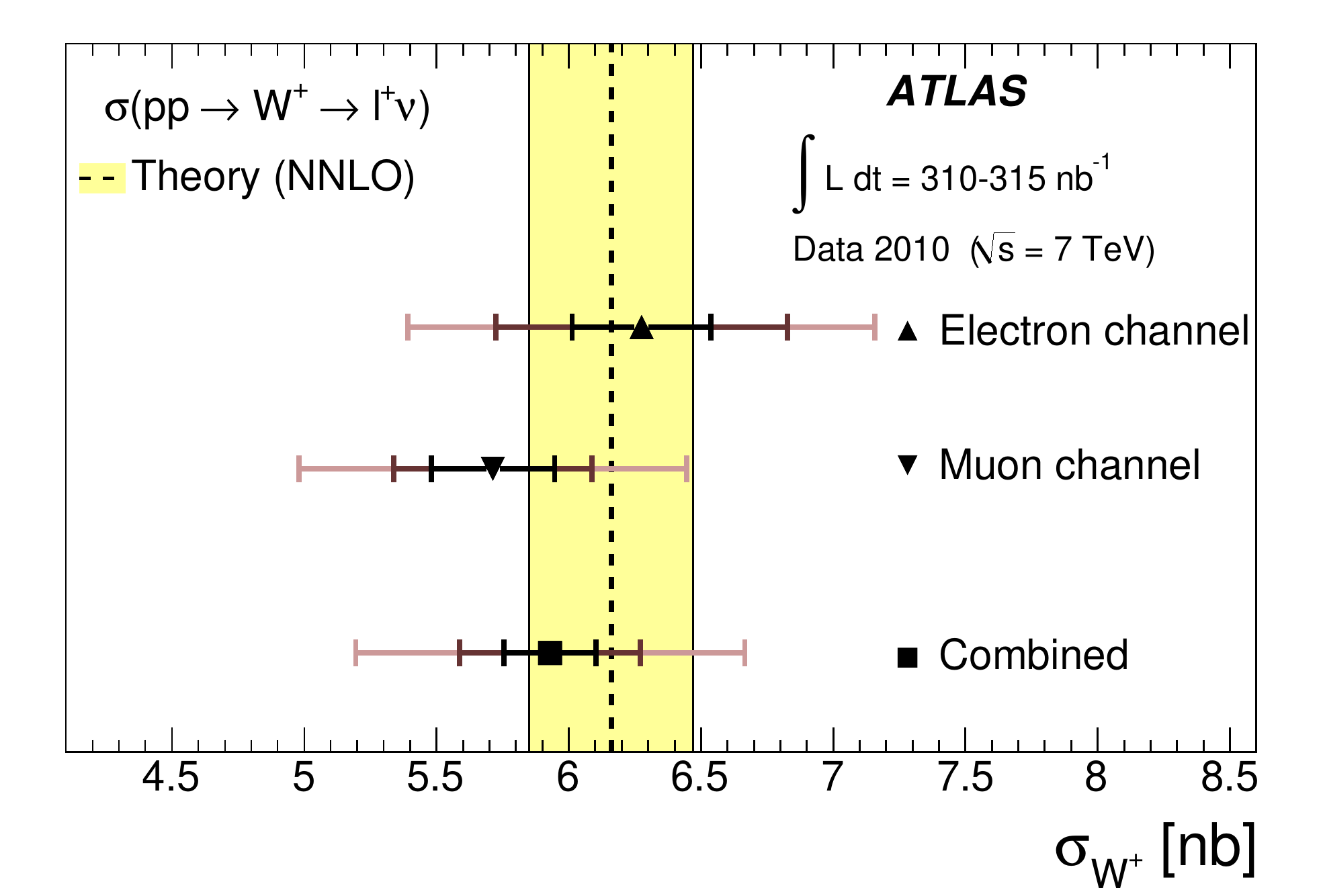}}
{\includegraphics[width=0.48\textwidth]{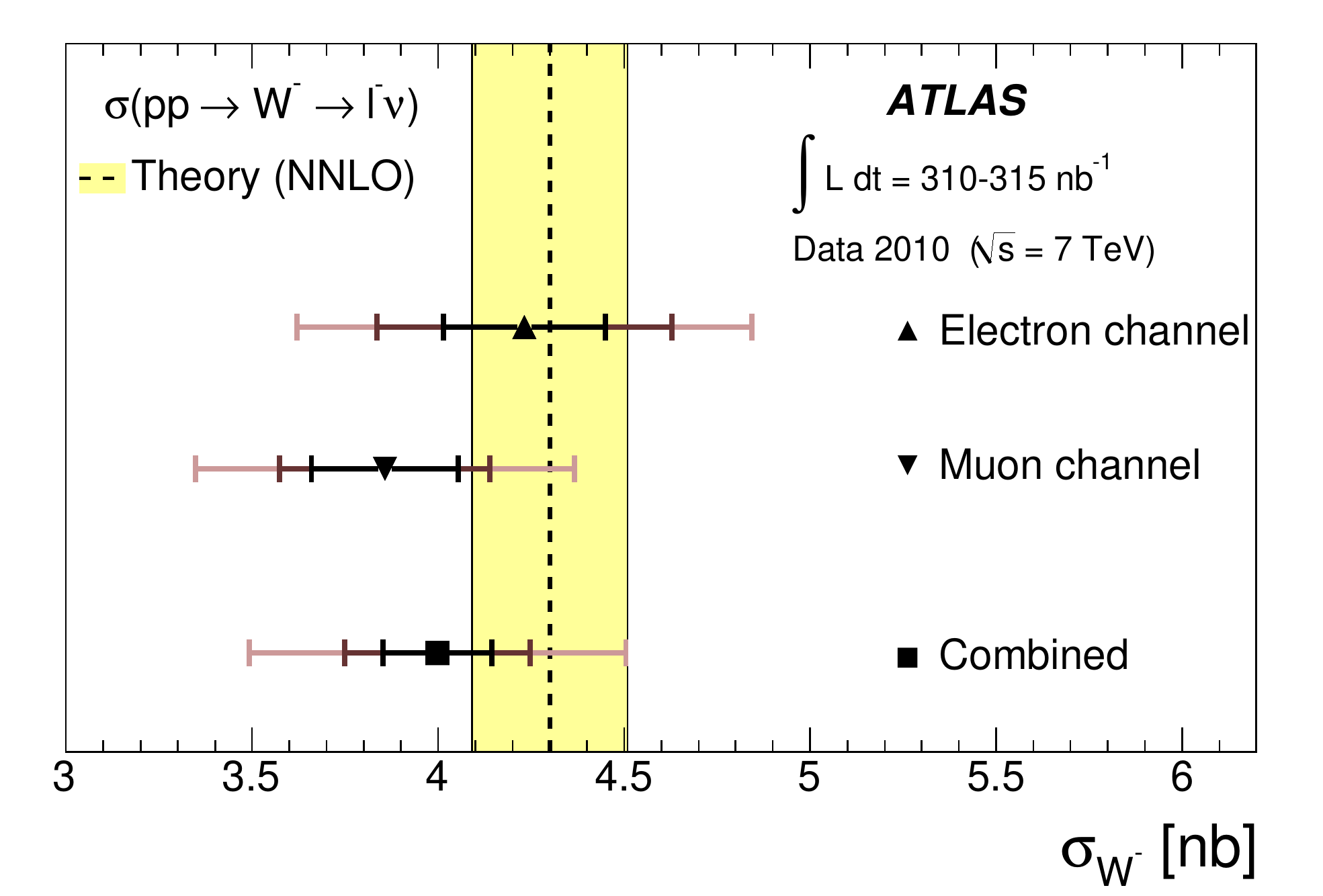}}
{\includegraphics[width=0.48\textwidth]{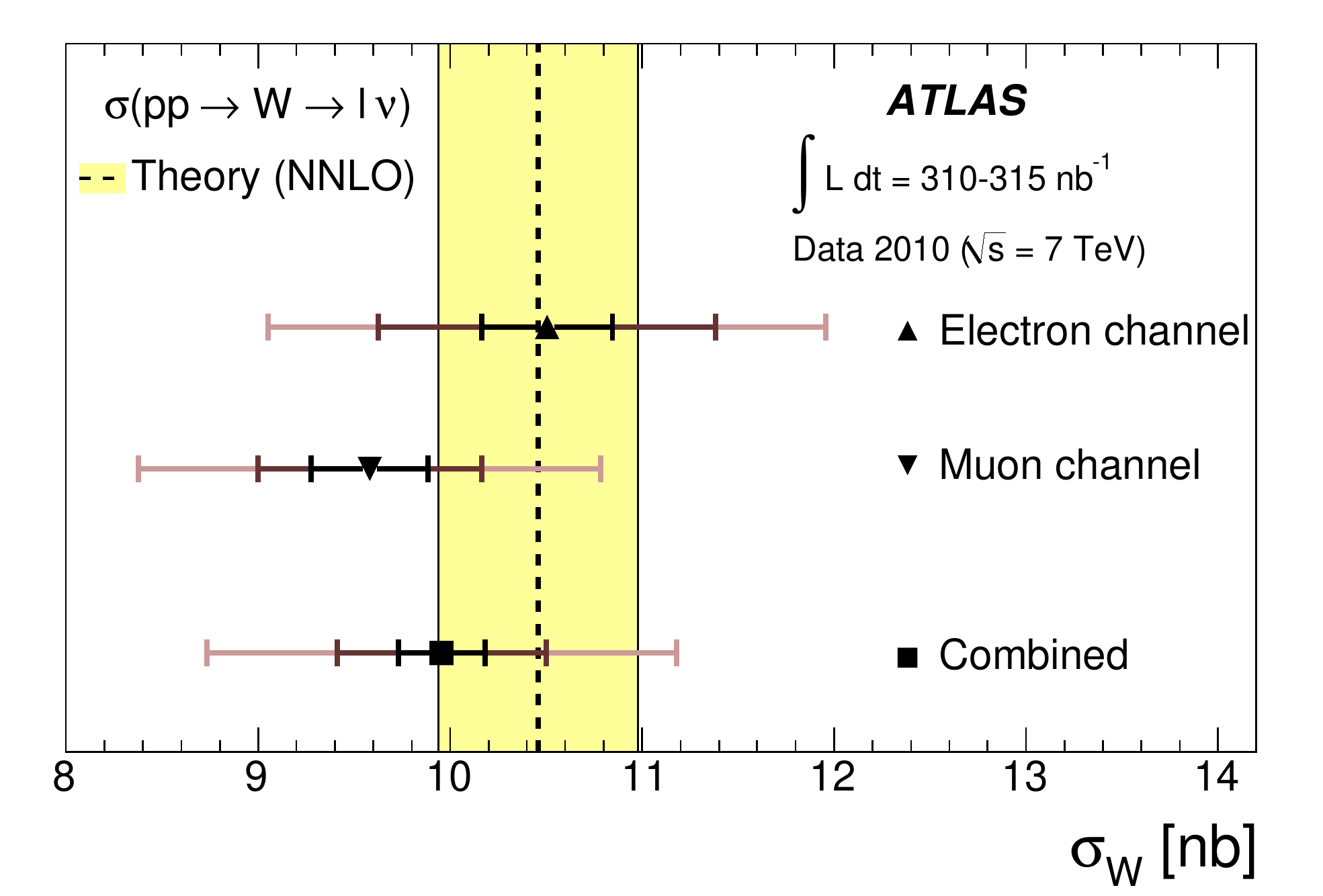}}
{\includegraphics[width=0.48\textwidth]{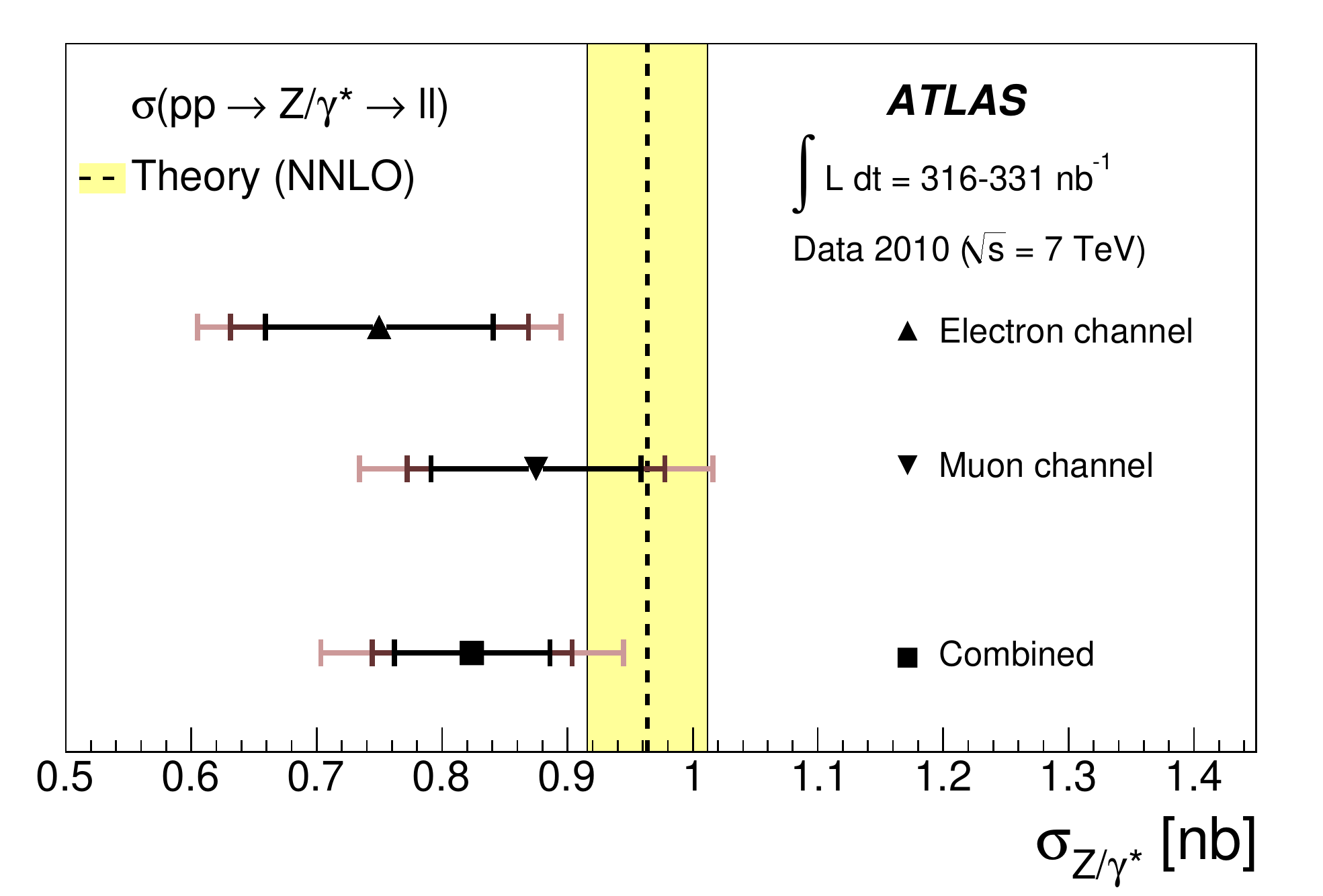}}
    \caption{\it\small  The measured values of $\sigma_W \cdot \mathrm{BR}$ ($\ensuremath{W}\rightarrow \ell \nu)$ for \Wplus, \Wminus~and for their sum
and of $\sigma_{\ensuremath{Z}/\gamma^*} \cdot \mathrm{BR}$ ($\Zg \to \ell \ell)$
compared to the theoretical predictions based on NNLO QCD calculations (see text). Results are shown for the electron and muon final states as well as for their
combination. The error bars represent successively the statistical, the statistical plus systematic
and the total uncertainties (statistical, systematic and luminosity). All uncertainties are added in quadrature. 
\label{fig:WZcross_compare}}
\end{center}
\end{figure}

A comparison of the measured cross-section values for $W$ and $Z$ production to theoretical predictions including next-to-next-to-leading order QCD corrections are shown in Fig.~\ref{fig:WZcross_compare}. The calculations were performed using the programs
 FEWZ~\cite{Anastasiou:2003ds} and ZWPROD~\cite{Hamberg:1990np,vanNeerven:1991gh} with the MSTW 08 NNLO structure function parameterisation~\cite{Martin:2009iq}. The following results were obtained: \\
\ \\
\hspace*{1.0cm}
$\sigma^{NNLO}_{\ensuremath{W^+} \rightarrow \ell^+ \nu}$ = 6.16$\pm$0.31~nb, \hspace*{0.5cm}
$\sigma^{NNLO}_{\ensuremath{W^-} \rightarrow \ell^- \nu}$ = 4.30$\pm$0.21~nb, \hspace*{0.5cm}
$\sigma^{NNLO}_{\ensuremath{W} \rightarrow \ell \nu}$ = 10.46$\pm$0.52~nb  \ \ and   \\
\ \\
\hspace*{1.0cm}
 $\sigma^{NNLO}_{\ensuremath{Z/\gamma^*} \rightarrow \ell^+ \ell^-}$ = 0.96$\pm$0.05~nb, \hspace*{0.2cm}
for  66 $< m_{\ell \ell} <$ 116 GeV.\\
\ \\
An overall uncertainty of the NNLO $W$ and $Z$-boson cross sections of $\pm$5\% was estimated using the MSTW 08
NNLO PDF error eigenvectors at the 90\% C.L. limit, variations of $\alpha_s$ in the range 0.1145 -- 0.1176, 
and variations of the renormalisation and factorisation scales by factors of two around the nominal scales $\mu_R = \mu_F = m_{W/Z}$.
Within the uncertainties, the calculations for $W$ production agree well  
with the measured cross sections. In particular, the expected
asymmetry between the $W^+$ and $W^-$ cross sections is confirmed. For the $Z$ cross section, the present measurements are below the theoretical predictions, but are still consistent within uncertainties.

\begin{figure}[h,t]
  \begin{center}
{\includegraphics[width=0.75\textwidth]{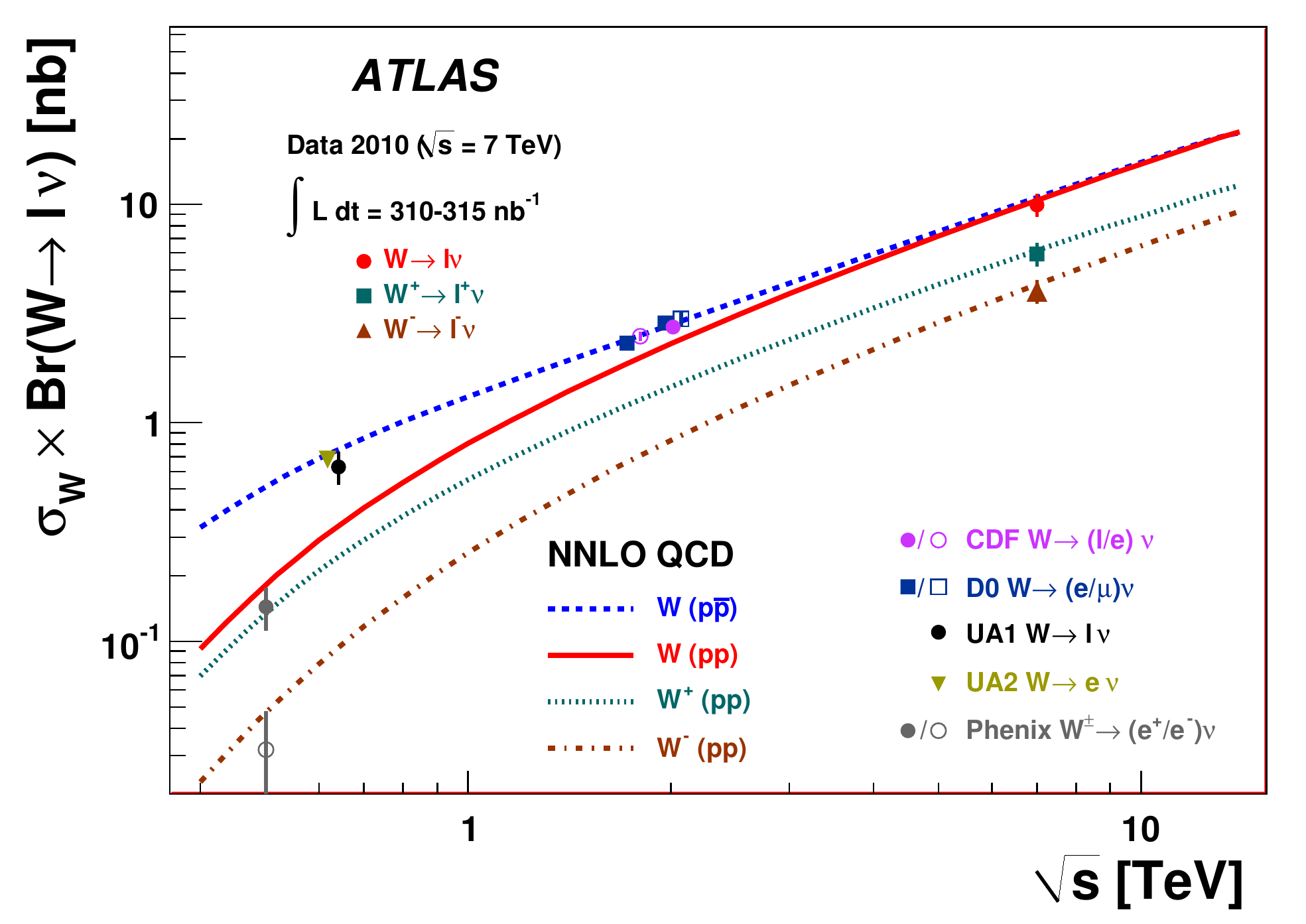}}
    \caption{\it\small  The measured values of $\sigma_W \cdot \mathrm{BR}$ ($\ensuremath{W}\rightarrow \ell \nu)$ for \Wplus, \Wminus~and for their sum compared to the theoretical predictions based on NNLO QCD calculations (see text). Results are shown for the combined electron-muon results. The predictions are shown for both proton-proton (\Wplus, \Wminus and their sum) and proton-antiproton colliders (\Wboson) as a function of $\sqrt{s}$. In addition, previous measurements at proton-antiproton and proton-proton colliders are shown.
The data points at the various energies are staggered to improve readability.
The CDF and D0 measurements are shown for both Tevatron 
collider energies, $\sqrt{s}$ = 1.8 TeV and $\sqrt{s}$ = 1.96 TeV.
All data points are displayed with their total uncertainty. The theoretical uncertainties are not shown.
\label{fig:Wcross_compare2}}
\end{center}
\end{figure}

\begin{figure}[h,t]
  \begin{center}
{\includegraphics[width=0.75\textwidth]{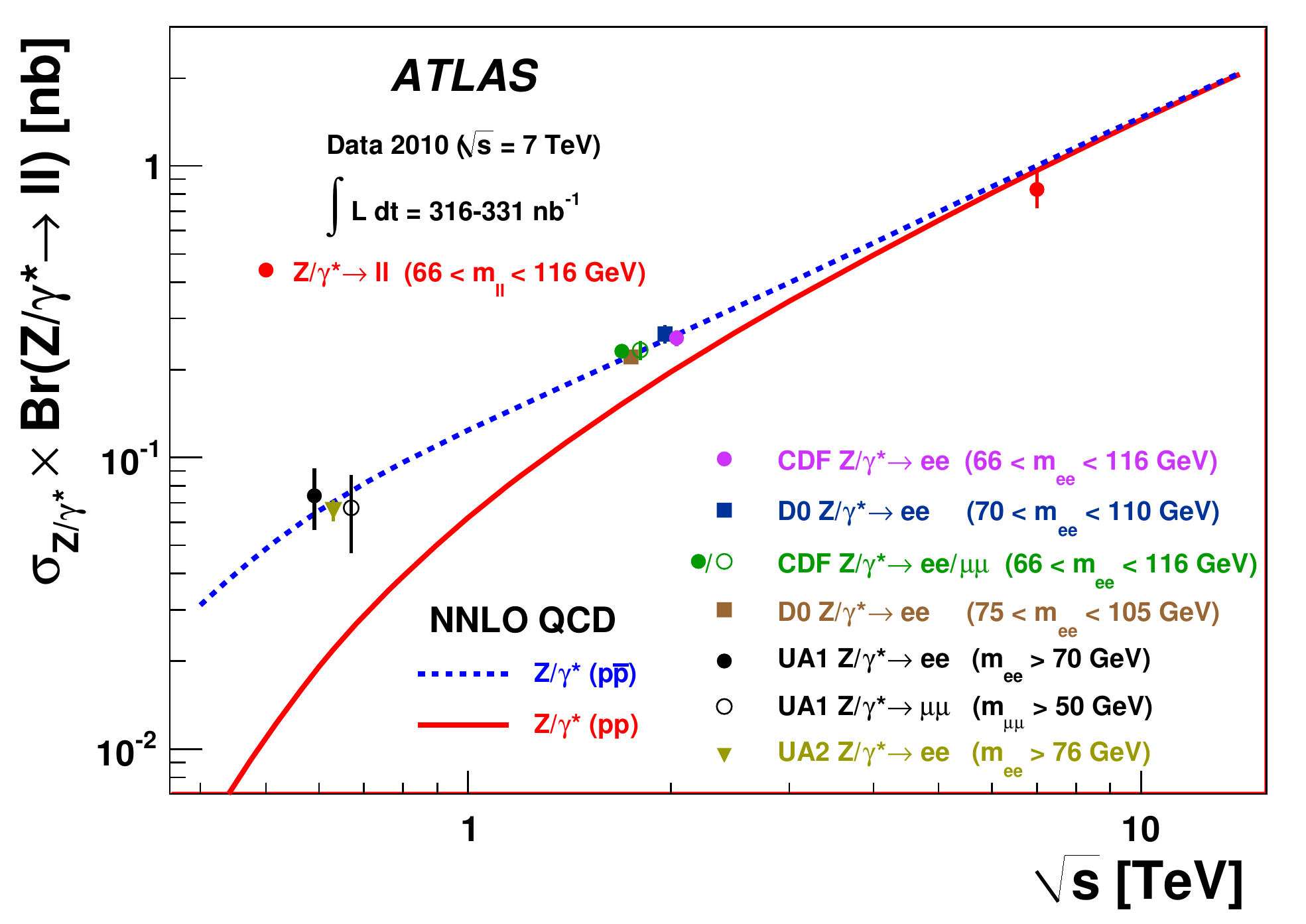}}
    \caption{\it\small  The measured value of $\sigma_{\ensuremath{Z}/\gamma^*} \times \mathrm{BR}$ ($\ensuremath{Z}/\gamma^*\rightarrow \ell \ell$)  where the electron and muon channels have been combined, compared to the theoretical predictions based on NNLO QCD calculations (see text). The predictions are shown for both proton-proton  and proton-antiproton colliders as a function of $\sqrt{s}$. In addition, previous measurements at proton-antiproton colliders are shown.
The data points at the various energies are staggered to improve readability.
The CDF and D0 measurements are shown for both Tevatron
collider energies, $\sqrt{s}$ = 1.8 TeV and $\sqrt{s}$ = 1.96 TeV.
All data points are displayed with their total uncertainty. The theoretical uncertainties are not shown.
\label{fig:Zcross_compare2}}
\end{center}
\end{figure}

In Figures~\ref{fig:Wcross_compare2} and~\ref{fig:Zcross_compare2}, the combined electron and muon measurements at $\sqrt{s}$ = 7 TeV are compared
to the theoretical predictions and to previous measurements  of the total $W$ and $Z$-production cross sections
by the UA1~\cite{UA1W2} and UA2~\cite{UA2W} experiments at $\sqrt{s} = 0.63$~TeV at the CERN Sp$\overline{\mathrm{p}}$S
and
by the CDF~\cite{CDFW} and D0~\cite{D0W} experiments at $\sqrt{s} = 1.8$~TeV and $\sqrt{s} = 1.96$~TeV at the Fermilab Tevatron colliders and to
the recent $W$ production cross-section measurement by the PHENIX \cite{Adare:2010xa} experiment in proton-proton collisions
at $\sqrt{s} = 0.5$~TeV at the RHIC collider.
The theoretical predictions are in good agreement with all measurements. 
The energy dependence of the total $W$ and $Z$ production cross sections is well described.

% Figure X

% Figure X

\subsection{The ratio of the $W$ to $Z$ cross sections}
\label{s:xs-ratio}

The measurement of the ratio of the $W$ to $Z$ cross sections times branching ratios,
\begin{equation}
R  \ = \ \frac{\sigma_W \cdot BR (W \to \ell \nu)}
{\sigma_Z  \cdot BR (Z  \to \ell \ell)},
\label{eq:xs-ratio}
\end{equation}
constitutes an important test of the Standard Model. It can be measured with a higher relative precision than the
individual cross sections since both experimental and theoretical uncertainties partially cancel.
In addition, it is sensitive to new physics processes which change the $W$ or $Z$ production rates or
the $W \to \ell \nu$ branching ratio.

Based on the theoretical cross-section calculations presented in Section~\ref{WXsection_comp} the ratios
of the $W^+, W^-, W$ to the $Z/\gamma^*$ cross sections are predicted to be:\\
\ \\
\hspace*{1.0cm}
$ R_{W^+/Z}^{NNLO}$ \ = \ 6.387$^{+0.077}_{-0.057}$, \hspace*{0,5cm}
$ R_{W^-/Z}^{NNLO}$  \ = \ 4.445$^{+0.036}_{-0.054}$, \hspace*{0,2cm} and \hspace*{0,2cm}
$R_{W/Z}^{NNLO}$  \ = \ 10.840 $\pm$ 0.054.\\
\ \\
\noindent
In terms of the experimental quantities defined in the previous sections, the ratio $R$ can be written as
\begin{equation}
R = \frac{N_W^{sig}}{N_{Z}^{sig}} \cdot \frac{A_Z}{A_W} \cdot \frac{C_Z}{C_W}.
\end{equation}

\noindent
In particular, the integrated luminosity and the related uncertainty cancel.
The uncertainties on the ratio
of the acceptance factors have already been discussed in Section~\ref{WXsection_acc}.
The uncertainty on the ratio of the
correction factors $C_Z / C_W$ was evaluated separately for the electron and the muon channels. For
both electrons and muons, the correlation between the uncertainties on $\alpha_{\rm{reco}}^W$ and
$\alpha_{\rm{reco}}^Z$ was taken to be one for the contribution of the lepton energy scale and resolution and
zero for the uncertainties resulting from the \MET\ scale (hadronic recoil), which affects only
$\alpha_{\rm{reco}}^W$. In addition, in the case of electrons, a correlation
between the ``tight" (applied in the $W$ analysis)  and ``medium" (applied in the $Z$ analysis)  electron
identification criteria is relevant and was taken into account. The
total uncertainty on $C_W/C_Z$ was estimated to be $\pm$6.0\% for the electron channel and $\pm$3.8\%
for the muon channel.

\begin{figure}[t]
  \begin{center}
{\includegraphics[width=0.48\textwidth]{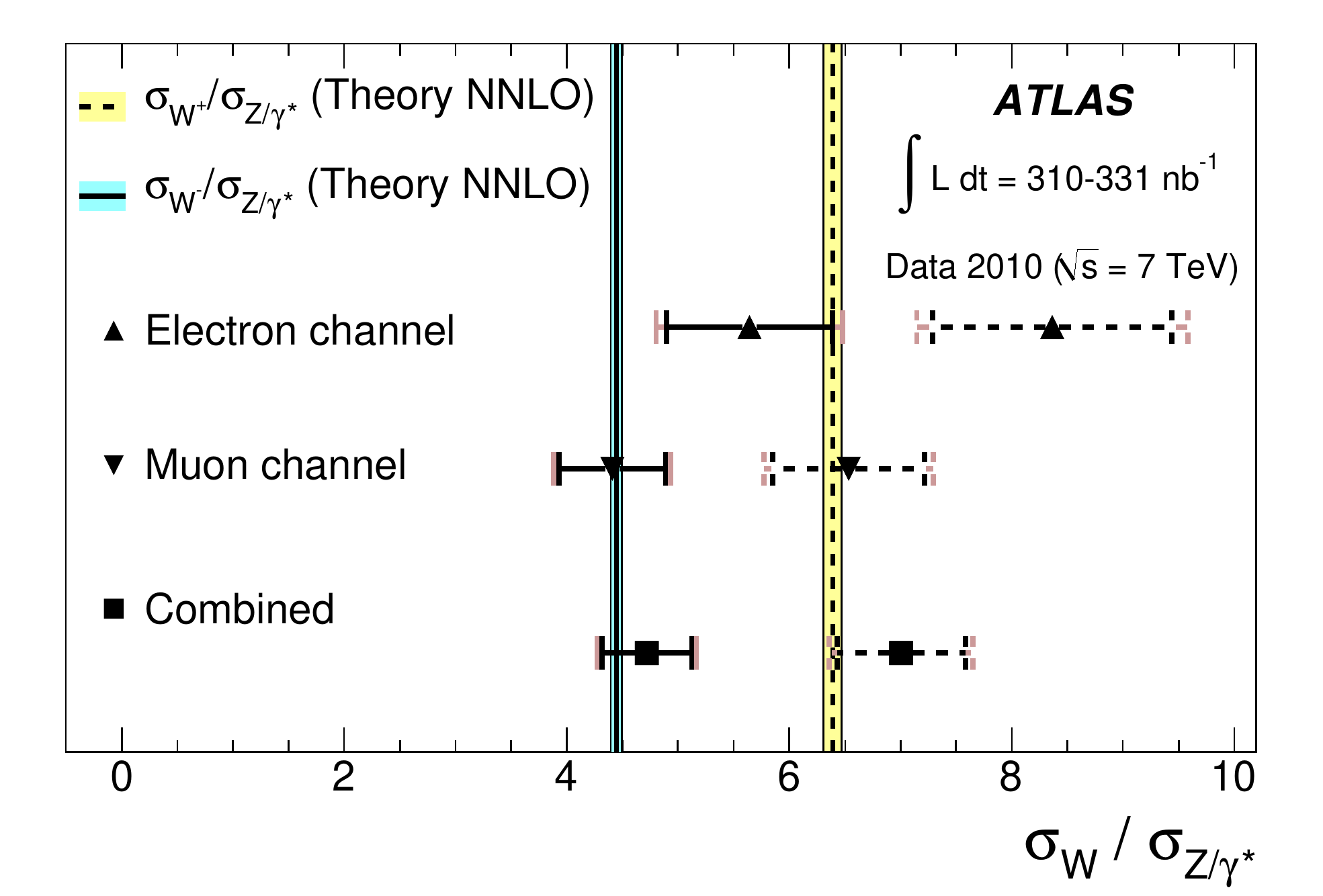}}
{\includegraphics[width=0.48\textwidth]{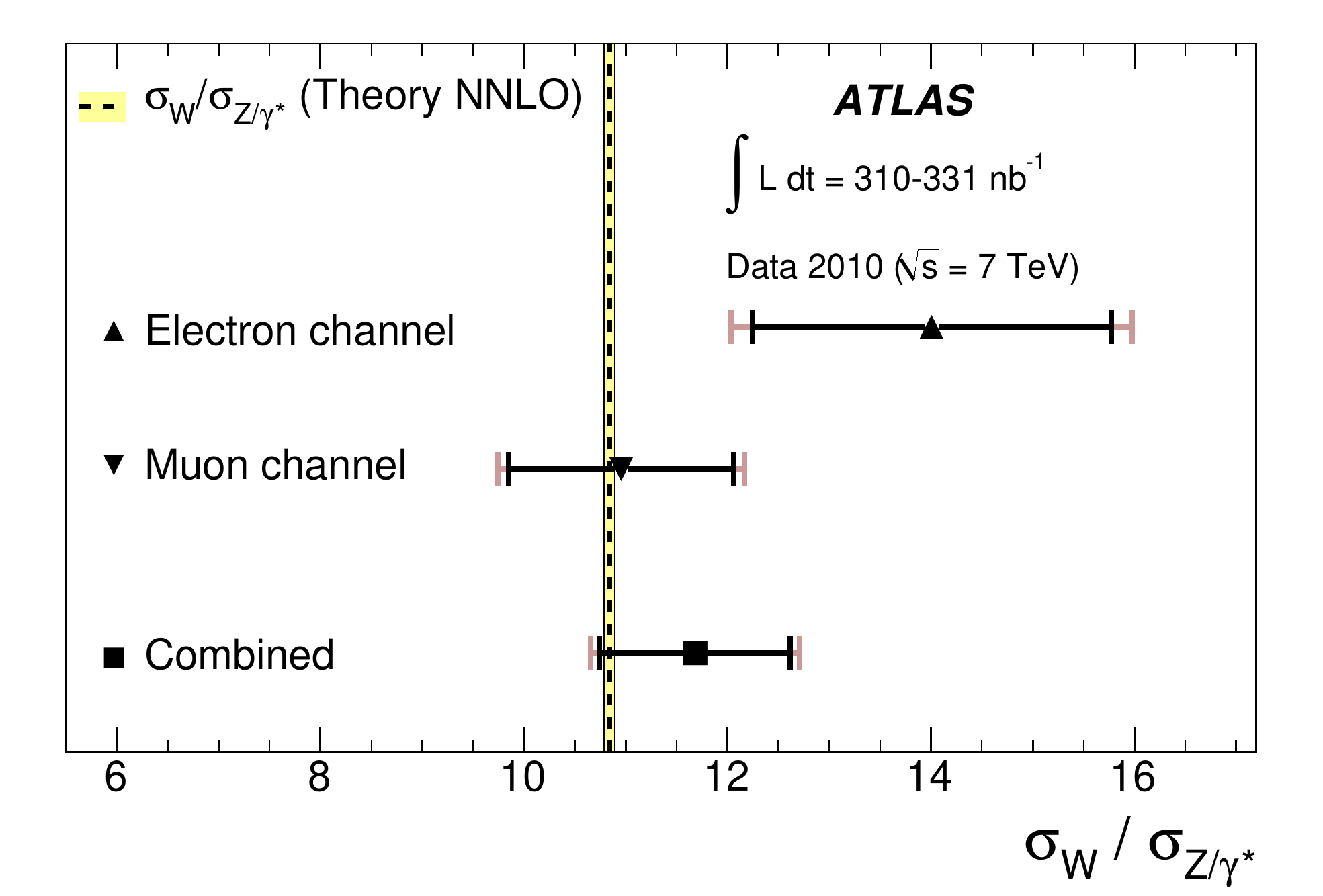}}
    \caption{\it\small  The measured ratios between the $W^+$ and $W^-$ and the $\Zg$ cross section (left)
in the electron and muon decay channels as well as the combined result (right)
compared to the theoretical predictions based on NNLO QCD calculations (see text).
The error bars represent successively the statistical, the statistical plus systematic and the total
uncertainties (statistical, systematic and luminosity). All uncertainties are added in quadrature. 
\label{f:R_WZ}}
\end{center}
\end{figure}

Using the measured cross-section values presented in Section~\ref{WXsection_totboson}
the results given in Table~\ref{t:xs_ratios} are obtained for the cross-section ratios  for the  electron and
muon channels.
\begin{table}[tb]
% \small
\begin{center}
\begin{tabular}{l | c | c }
\hline
\hline
% & & \\
 & $ R^{e}_{W^{(\pm)}/Z}$    &  $ R^{\mu}_{W^{(\pm)}/Z}$ \\
\hline
% & & \\
${W^+} $     & 8.4 $\pm$ 1.1 (stat) $\pm$ 0.6 (syst) &  6.5 $\pm$ 0.7 (stat) $\pm$ 0.3 (syst) \\
% & & \\
${W^-} $     & 5.7 $\pm$ 0.7 (stat) $\pm$ 0.4 (syst) &  4.4 $\pm$ 0.5 (stat) $\pm$ 0.2 (syst) \\
%  & & \\
$ W $        & 14.0 $\pm$ 1.8 (stat) $\pm$ 0.9 (syst) & 11.0 $\pm$ 1.1 (stat) $\pm$ 0.5 (syst) \\
% & &  \\
\hline
\hline
\end{tabular}
\caption{Measured cross-section ratios $R_{W^+/Z}^{e,\mu}$, $R_{W^-/Z}^{e,\mu}$ and $R_{W/Z}^{e,\mu}$
in the electron and muon final states.}
\label{t:xs_ratios}
\end{center}
\end{table}
\noindent
The combination of the two lepton flavours leads to:
\begin{center}
 $ R^{\ell}_{W^+/Z}$   = 7.0 $\pm$ 0.6 (stat) $\pm$ 0.3 (syst),  \\
\  \\
 $ R^{\ell}_{W^-/Z}$   = 4.7 $\pm$ 0.4 (stat) $\pm$ 0.2 (syst),    \\
\  \\
 $ R^{\ell}_{W/Z}$    = 11.7 $\pm$ 0.9 (stat) $\pm$ 0.4 (syst).  
\end{center}
\noindent
The results are shown in Fig.~\ref{f:R_WZ} and compared to the theoretical predictions. Within
the large uncertainties, which are still dominated by the statistical uncertainties, the theoretical 
predictions agree with the measured ratios. Due to the low value of the measured $Z \to ee$ cross section,
the ratios in the electron channel are above the theoretical expectations. However, it should be noted
that the three ratio measurements are correlated via the common low $Z \to ee$ cross-section value and
are still compatible within uncertainties with the theory value.

Updated measurements using larger data samples will provide
interesting constraints on $\Gamma_W$ and allow for a precise test of the Standard Model predictions.
For such measurements the ratios would have to be normalised to the pure $Z$ boson contribution and
electroweak corrections would need to be addressed more carefully.

%------------------------------
% Figure W-Z coss section ratio
%------------------------------

%---------------------------------------------------------------

%---------------------------------------------------------------
%    Asymmetry Chapter
%
\section{Measurement of the $W \to \ell \nu$ charge asymmetry}
 \label{Wasymmetry}

The measurement of the charge asymmetry of the \Wboson-bosons produced at
hadron colliders provides important information about parton distribution functions. Inclusive
measurements have been performed at the Tevatron~\cite{Abe:1998rv,Lai:2010vv} and the data have been included
in global fits of parton distributions~\cite{Martin:2009iq,Pumplin:2002vw}.

The \Wboson-boson charge asymmetry is obtained from the charge of the decay leptons.
The lepton charge asymmetry measured in this paper is defined via the fiducial cross sections,
$\sfidWp$ and $\sfidWm$ (see Section~\ref{WXsection_intro} for the definition):
\begin{equation}
  A_\ell =
  \frac{\sfidWp -\sfidWm} {\sfidWp + \sfidWm}.
  \label{eqn:asymmetry}
\end{equation}
This implies that the asymmetry is measured for leptons, satisfying the geometrical
and kinematic constraints at generator level, as defined in Section~\ref{sec:wkine}, with all detector effects corrected for.

Given the difference in the cross-section measurements for $W^+$ and $W^-$ presented in the previous section,
the overall asymmetry is different from zero. This reflects the different content of $u$ and $d$ 
valence quarks in the proton. In addition, the asymmetry is expected to depend on the
lepton pseudorapidity. This dependence on $\eta$ provides valuable constraints on the
parton distribution functions of the proton, since different $\eta$ bins probe different
average values of the momentum fractions $x$ of the partons producing the $W$ boson.
As the $W$ lepton asymmetry is mainly sensitive to valence quark distributions~\cite{Berger:1988tu}, it provides complementary
information to that obtained from measurements of structure functions in deep inelastic scattering at
HERA~\cite{Chekanov:2008aa,Chekanov:2009gm,Adloff:2003uh,:2009wt},
which do not strongly constrain the ratio between $u$ and $d$ quarks
in the kinematic regime probed at the LHC.

\begin{figure}[t,b,h]
\begin{center}
\subfigure[]{\label{asy_eleca}\includegraphics[width=0.40\textwidth]{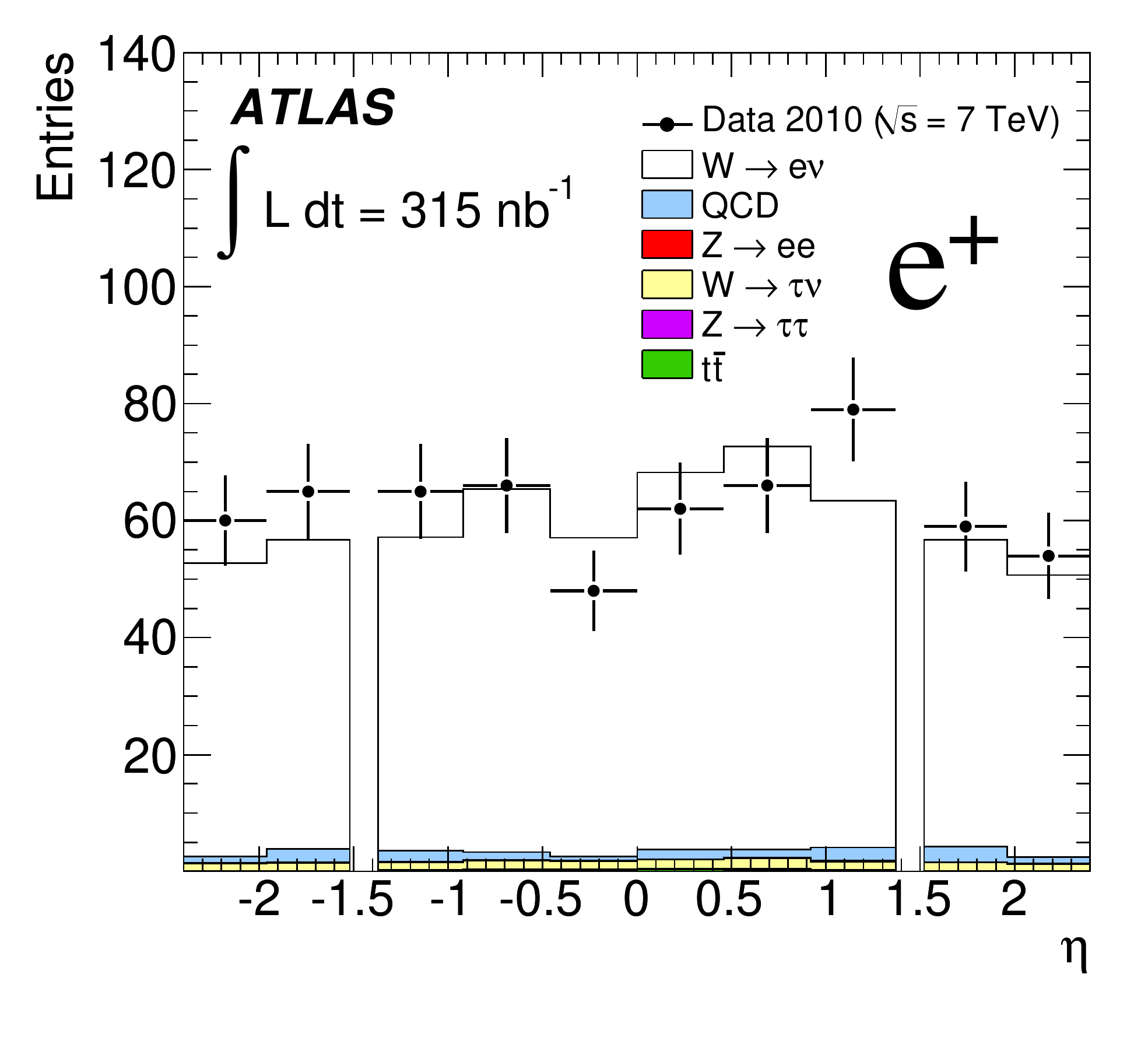}}
\subfigure[]{\label{asy_muona}\includegraphics[width=0.40\textwidth]{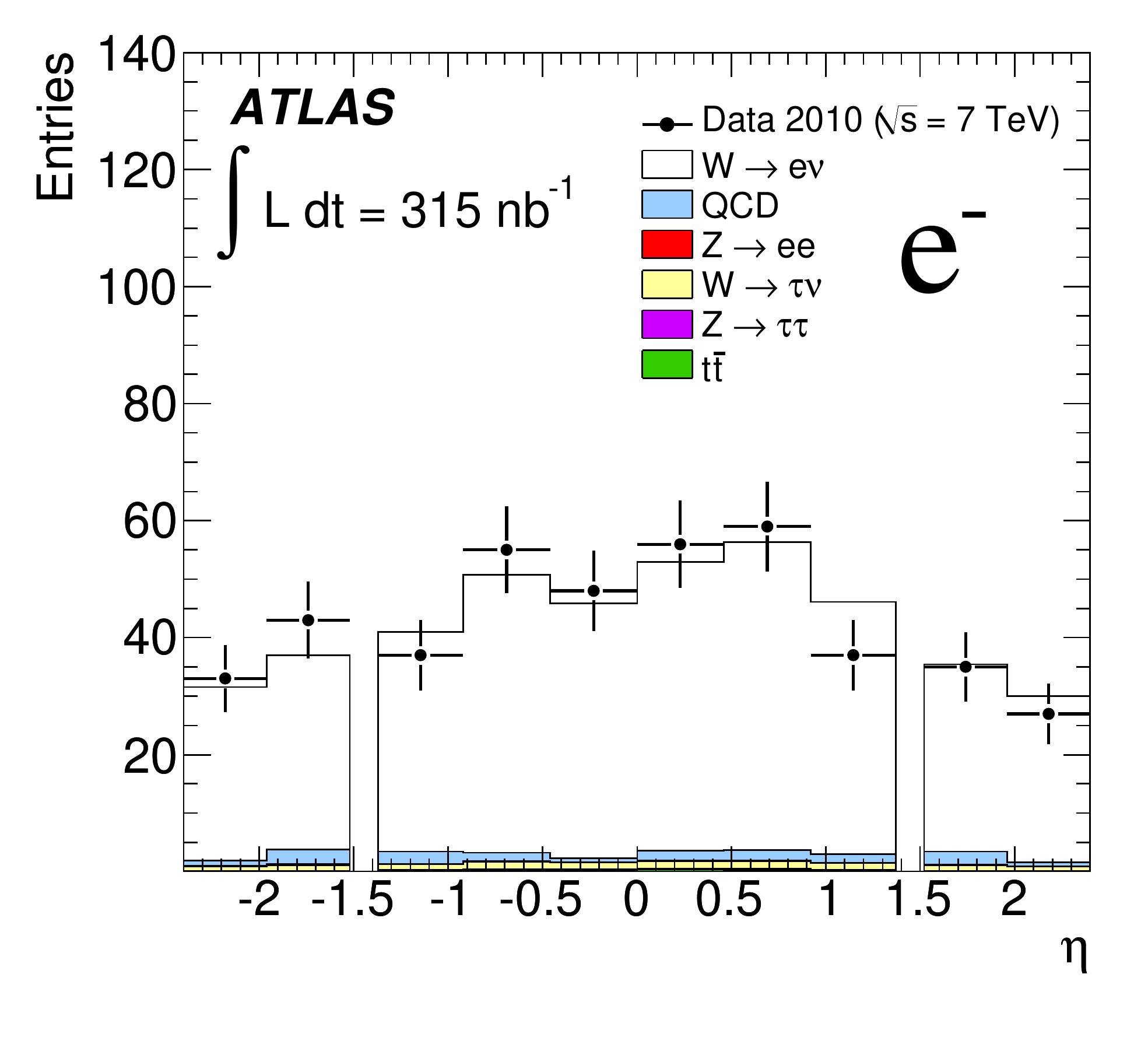}}
\subfigure[]{\label{asy_elecb}\includegraphics[width=0.40\textwidth]{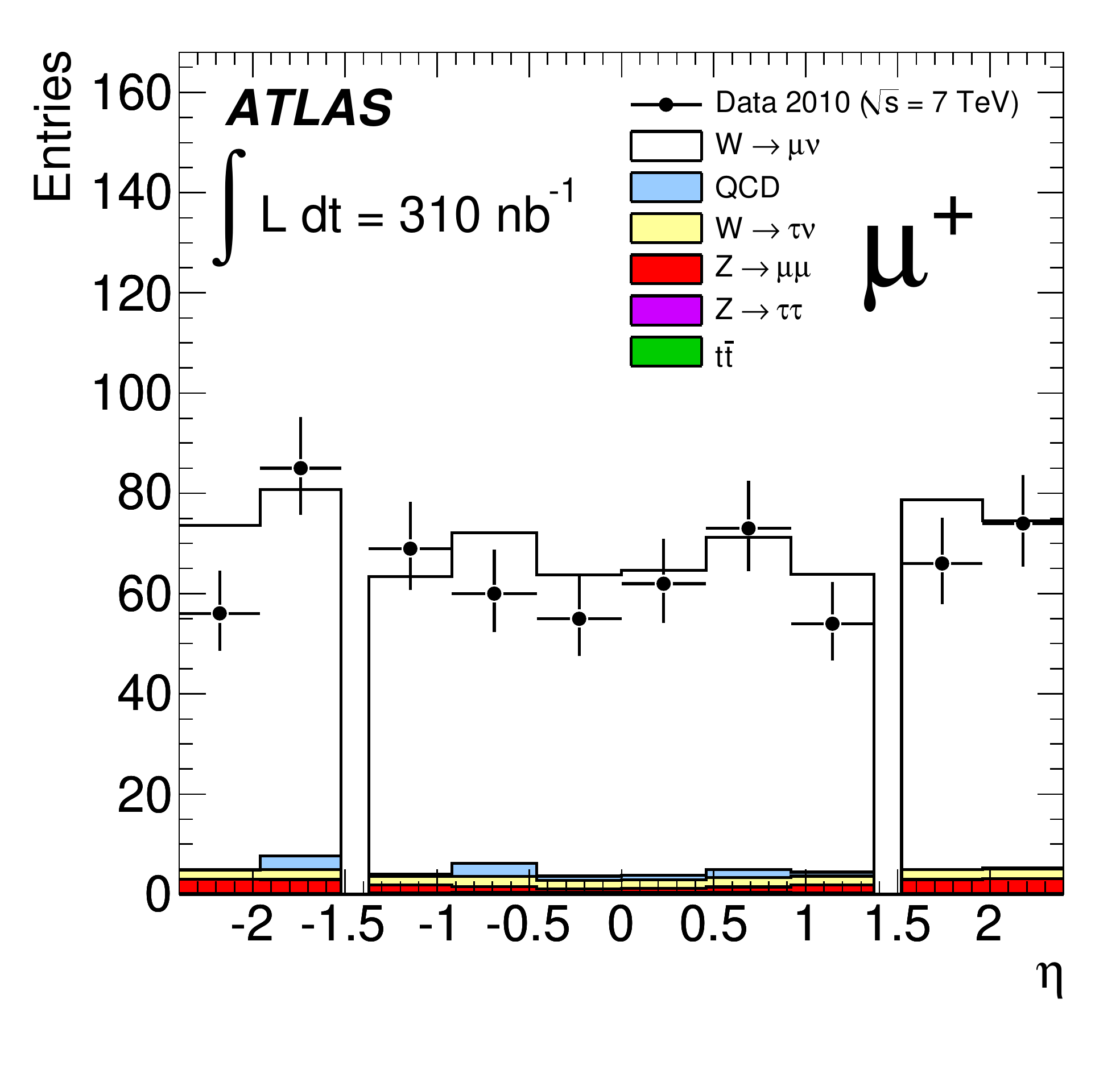}}
\subfigure[]{\label{asy_muonb}\includegraphics[width=0.40\textwidth]{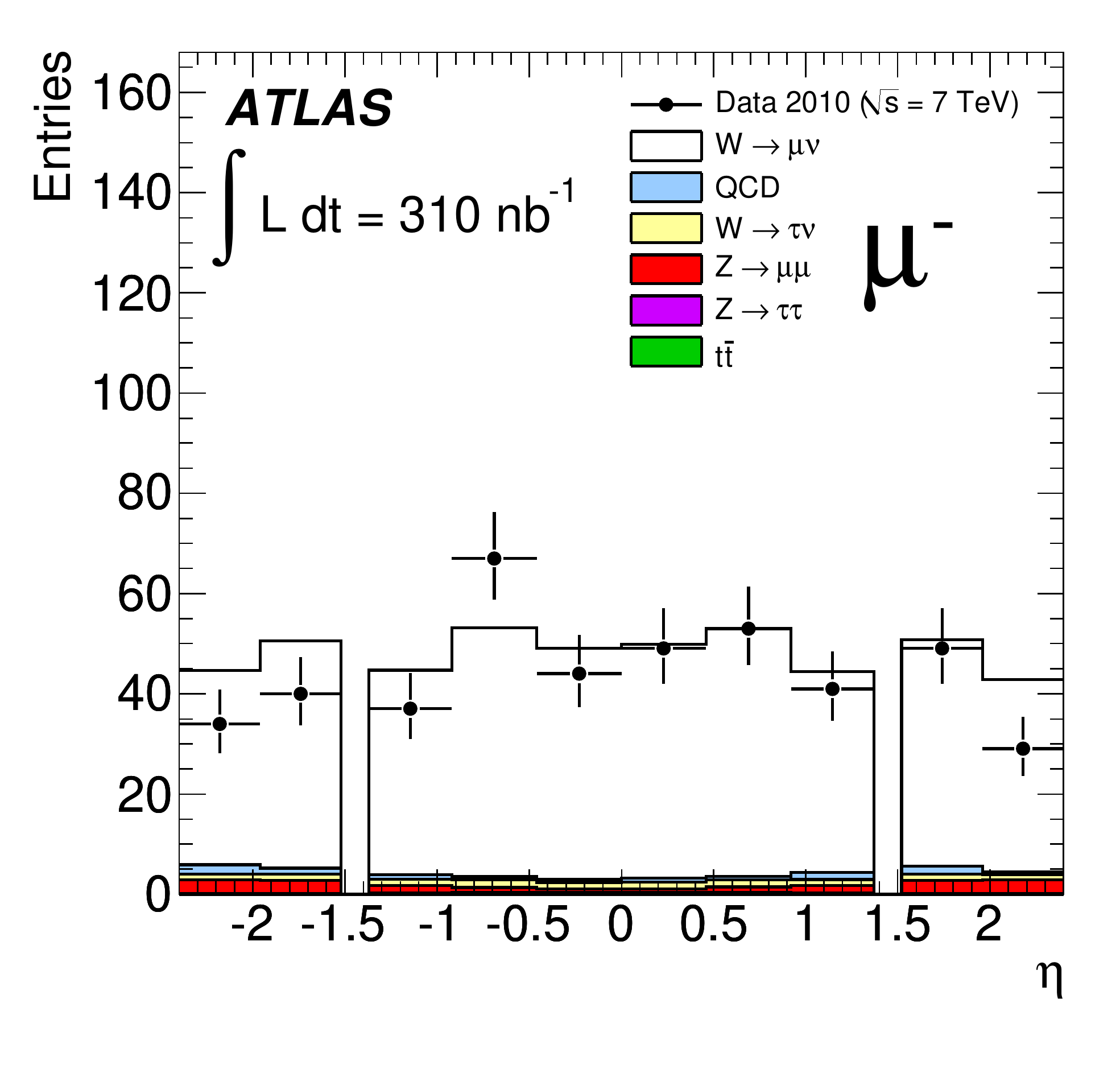}}
\caption{\it\small Pseudorapidity distributions of $e^+$ (a), $e^-$ (b), $\mu^+$ (c) and $\mu^-$ (d) candidates satisfying all
$W$ requirements (see Section~\ref{sec:wkine}). The data are compared to the Monte Carlo simulation, 
broken down into the signal and various background components. The Monte-Carlo distributions are normalised to 
the integrated luminosity of the data, as described in Section~\ref{sec:results_presel}. 
\label{f:eta-leptons}}
\end{center}
\end{figure}

The $\eta$ distributions of reconstructed electrons and muons after the final $W$ selection cuts (see Section~\ref{sec:wkine}) are shown in Fig.~\ref{f:eta-leptons}.
It should be noted that common $\eta$ acceptance cuts for electrons and muons are used and the asymmetry is measured
over the pseudorapidity range  0 $< | \eta | <$ 1.37 and 1.52 $< | \eta | <$ 2.4, which allows for a combination
of the results of the two lepton flavours. The Monte Carlo simulation is found to be in good agreement with the 
measured $\eta$ distributions.

The lepton charge asymmetry is measured in two bins of
pseudorapidity. For the calculation of the asymmetry, the correction factors $C_W$ were calculated separately for the two charges and for each of the  $ |\eta|$ bins and all background contributions are subtracted.
For the ratio defined in Eq.~(\ref{eqn:asymmetry}),
the luminosity uncertainty cancels and $C_W$-related uncertainties appear to be dominant. Also for some of those,
e.g. efficiency uncertainties, cancellations appear as long as they affect positive and negative charged leptons in a
symmetric way. Given the different production rates between the two lepton charges, the charge misidentification might
lead to a bias in the result. For electrons it is of the order of 0.1\% for the barrel and 1.3\% for the end-cap regions
and has been implicitly taken into account in the $C_W$ corrections applied. For muons, the charge misidentification
is found to be negligible.

The results obtained for the different $\eta$ bins as well as after integration over the full pseudorapidity interval
are listed in Table~\ref{tab:asymmetryint} together with their statistical and systematic uncertainties.
Consistent results are obtained for the two lepton channels. The precision of the measurements is
limited for both channels by the statistical uncertainties. 

For the electron channel, major contributions to the systematic uncertainties result from uncertainties on the electron identification and charge misidentification ($\pm 2.0$\%), electron energy scale ($\pm 1.0$\%), and on the QCD ($\pm 1.5$\%) and electroweak backgrounds ($\pm 0.5$\%). The systematic uncertainty on the QCD background is much larger than that on the electroweak background because of the larger relative uncertainty on the QCD background. In addition, potential distortions of the asymmetry are much larger from the QCD background than from the electroweak background, which predominantly consists of $W \rightarrow \tau \nu$ decays exhibiting an asymmetry similar to that expected in $W \rightarrow e \nu$ decays. The systematic uncertainty on the electron identification and misidentification is determined by comparing the variation of the asymmetry as a function of identification requirements in data to the same variation as predicted by Monte Carlo simulation.

For the muon channel, the systematic uncertainty is derived from uncertainties on the muon momentum scale and resolution ($\pm 5.0$\%), from  uncertainties on the trigger efficiency ($\pm 2.7$\%), and on the QCD ($\pm 0.8$\%) and electroweak backgrounds ($\pm 0.5$\%). 
The  muon momentum scale and resolution may depend significantly on charge. Scale and resolution uncertainties on the muon momentum measurement are considered to be anti-correlated, since they could affect in opposite directions the bending of tracks of opposite sign.

\begin{table}[!]
\begin{center}
\small
%\footnotesize
\begin{tabular}{l||c|c|c}
\hline  \hline
$\eta$ range   & Electron channel   &  Muon channel   & Combination \\
               &    $A_{e}$          &  $ A_{\mu}$      & $A_{\ell}$    \\
\hline
$|\eta| < 1.37$       & $ 0.15 \pm 0.04 \pm 0.00$ & $ 0.12 \pm 0.04 \pm 0.01 $ & $ 0.14 \pm 0.03 \pm 0.01 $ \\
$1.52 < |\eta| < 2.4$ & $ 0.29 \pm 0.05 \pm 0.02$ & $ 0.32 \pm 0.05 \pm 0.02 $ & $ 0.31 \pm 0.04 \pm 0.01 $ \\ 
\hline
$|\eta| < 1.37$ and $1.52 < |\eta| < 2.4$ & $ 0.21 \pm 0.03 \pm 0.01$ & $ 0.19 \pm 0.03 \pm 0.01 $ & $ 0.20 \pm 0.02 \pm 0.01 $ \\ 
\hline
\hline
  \end{tabular}
  \caption{The measured lepton asymmetries integrated over the full pseudorapidity range, as well as separately for the barrel and end-cap regions. The quoted uncertainties are statistical (first) and systematic (second). 
    \label{tab:asymmetryint}}
\end{center}
\end{table}

\begin{figure}[t,b]
\begin{center}
\subfigure[]{\label{asy_elec}\includegraphics[width=0.48\textwidth]{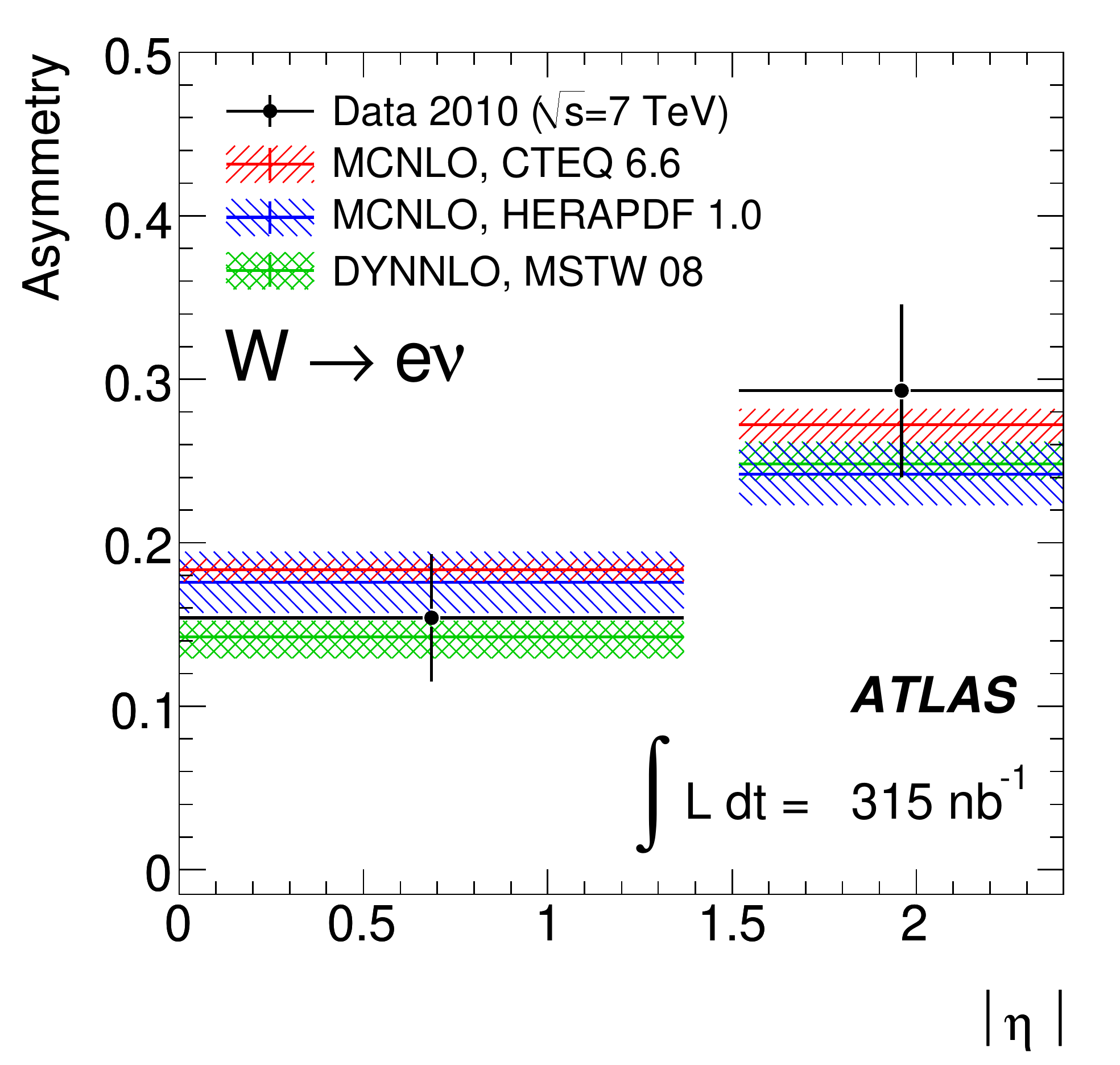}}
\subfigure[]{\label{asy_muon}\includegraphics[width=0.48\textwidth]{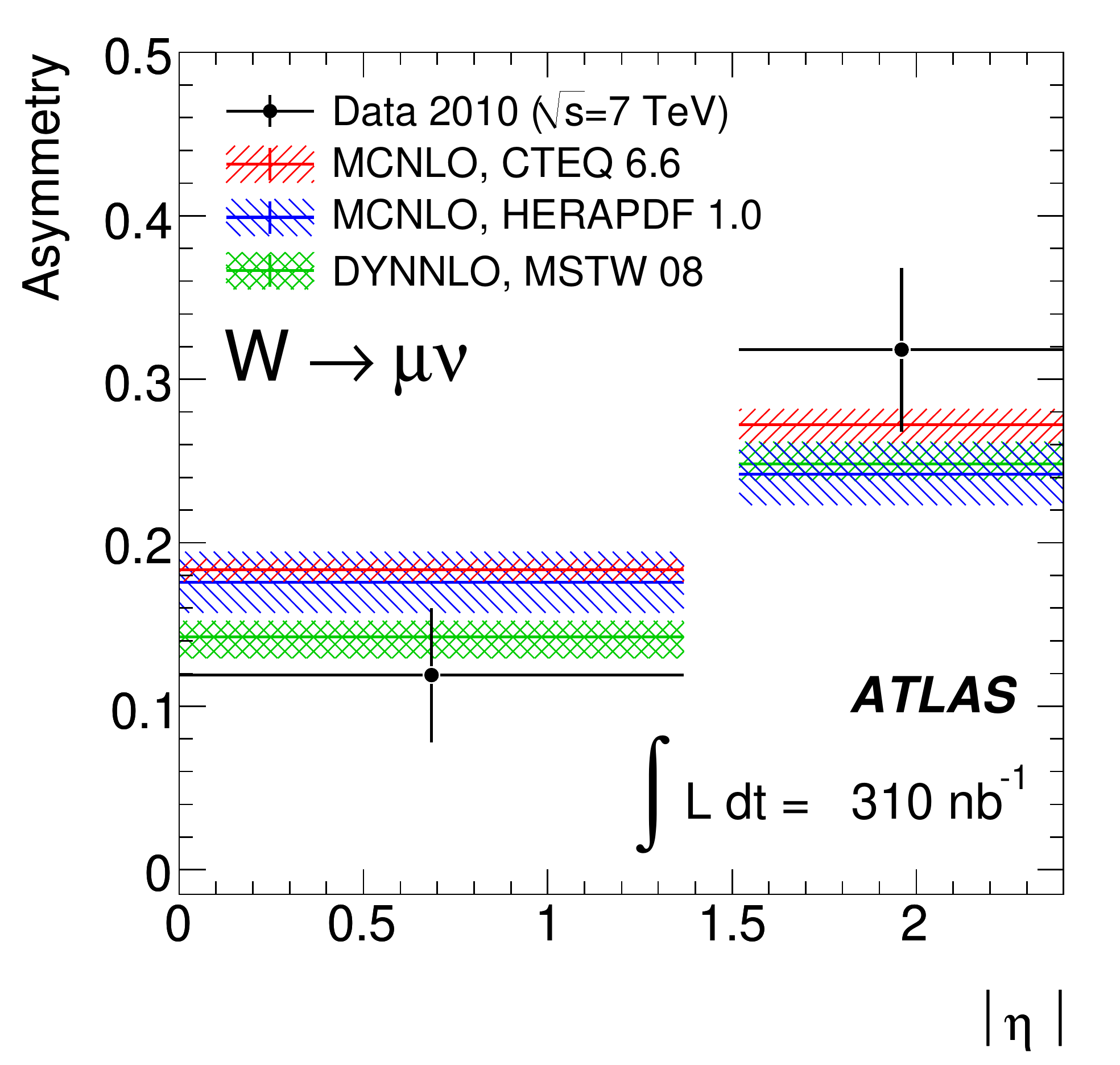}}
\caption{\it\small Lepton charge asymmetries for the electron~(a) and muon~(b) channels. Superimposed are several theoretical predictions (see text).
The bands show the uncertainties extracted from a variation of the error eigenvector sets of the PDFs
at the 90\% C.L. limit. 
\label{fig:AsymmetriePlots}}
\end{center}
\end{figure}

The measured lepton asymmetries are displayed in Fig.~\ref{fig:AsymmetriePlots} as a function of $ | \eta |$ and
compared to theoretical predictions obtained with NLO calculations, namely MC@NLO~\cite{mcatnloDYpatch} and DYNNLO~\cite{Catani:2007vq} which have been interfaced to various PDF parameterisations of the respective order. The parton distribution functions MSTW~08~\cite{Martin:2009iq}, CTEQ~6.6~\cite{Nadolsky:2008zw} and HERAPDF~1.0~\cite{:2009wt} were used. The predictions of these calculations for the integrated asymmetry ($ |\eta|<1.37$ and 1.52$ <|\eta|<$ 2.4) are
$0.218^{+0.008}_{-0.009}$ (MC@NLO, CTEQ 6.6), 0.202 $\pm$ 0.019 (MC@NLO, HERAPDF 1.0), and $0.184^{+0.011}_{-0.012}$ (DYNNLO, MSTW 08).
The bands shown for the theoretical predictions display the uncertainties extracted from a variation of  the error 
eigenvector sets of the PDFs at the 90\% C.L. limit.  
Within the large uncertainties, the theoretical predictions agree with the present 
measurements. However, the data do not provide sufficient separation power to
discriminate between various models.

%---------------------------------------------------------------

\section{Summary}
The ATLAS collaboration presents first measurements of the $\ensuremath{W}\rightarrow \ell \nu$ and $\ensuremath{Z}\rightarrow \ell \ell$ production cross sections in proton-proton collisions at $\sqrt{s}$ = 7 TeV. The results are based on data corresponding to an integrated luminosity of approximately \lumWZ. The total inclusive \Wboson-boson production cross sections times the leptonic branching ratios for the combined electron-muon channels are measured to be:
\[
\stotWp  \cdot  \rm{BR}(W \to \ell \nu) \ = \ \sigWplus \ \rm{nb},     \nonumber
\]
\[
\stotWm  \cdot  \rm{BR}(W \to \ell \nu) \ = \ \sigWminus \ \rm{nb},     \nonumber
\]
\[
\stotW   \cdot  \rm{BR}(W \to \ell \nu) \ = \ \sigW \ \rm{nb}.    \nonumber
\]
For the $\Zg$ production cross section, measured in the mass range $ 66<m_{\ell \ell}<116$~GeV, the result for the
combination of the electron and muon decay channels is:
\[
\stotZg   \cdot  \rm{BR}(\Zg \to \ell \ell) \ = \ \sigZnb \ \rm{nb}.   \nonumber
\]

\noindent
The ratio of the $W$ to $Z$-boson cross sections is measured to be
\[
R_{W/Z} \ = \ 11.7  \pm 0.9 \rm{(stat)} \pm 0.4 \rm{(syst)}.
\]
\noindent
Theoretical predictions, based on NNLO QCD calculations, are in good agreement with all measurements.

In addition, a measurement of the charge asymmetry of $W$-boson production is presented for the first time in proton-proton collisions at $\sqrt{s} = 7$ TeV. The charge asymmetry, defined via the fiducial cross sections, integrated over the acceptance
region $ | \eta | <$ 1.37 and $1.52 < | \eta | <$ 2.4, is measured to be
\[
A_\ell \ = \ \frac{(\sfidWp-\sfidWm)}{(\sfidWp + \sfidWm)} \ = \ 0.20 \pm 0.02 \rm{(stat)} \pm 0.01 \rm{(syst)}.
\]
This measurement demonstrates clearly the expected charge
asymmetry for $W$ boson production in proton-proton collisions. Theoretical predictions are in agreement with this measurement, which, at present, is still statistically limited.

Despite the rather low integrated luminosity used in the analyses presented here, the accuracy of the
cross-section measurements is, however, already dominated by systematic uncertainties, most prominently by the luminosity uncertainty of $\pm$11\% and to a lesser
degree by the experimental uncertainties on lepton identification. The latter uncertainties are expected to improve
significantly with more data. In particular,  high-statistics $Z$-boson samples can be used to perform measurements of
efficiencies and to reduce the corresponding uncertainties. The luminosity uncertainty is dominated by
the~$\pm$10\%~uncertainty on the beam currents in the machine and is also expected to improve with more
precise and dedicated measurements.

\section*{Acknowledgements}

We wish to thank CERN for the efficient commissioning and operation of the LHC during this initial high-energy data-taking period as well as the support staff from our institutions without whom ATLAS could not be operated efficiently.

We acknowledge the support of ANPCyT, Argentina; YerPhI, Armenia; ARC, Australia; BMWF, Austria; ANAS, Azerbaijan; SSTC, Belarus; CNPq and FAPESP, Brazil; NSERC, NRC and CFI, Canada; CERN; CONICYT, Chile; CAS, MOST and NSFC, China; COLCIENCIAS, Colombia; MEYS (MSMT), MPO and CCRC, Czech Republic; DNRF, DNSRC and Lundbeck Foundation, Denmark; ARTEMIS, European Union; IN2P3-CNRS, CEA-DSM/IRFU, France; GNAS, Georgia; BMBF, DFG, HGF, MPG and AvH Foundation, Germany; GSRT, Greece; ISF, MINERVA, GIF, DIP and Benoziyo Center, Israel; INFN, Italy; MEXT and JSPS, Japan; CNRST, Morocco; FOM and NWO, Netherlands; RCN, Norway;  MNiSW, Poland; GRICES and FCT, Portugal;  MERYS (MECTS), Romania;  MES of Russia and ROSATOM, Russian Federation; JINR; MSTD, Serbia; MSSR, Slovakia; ARRS and MVZT, Slovenia; DST/NRF, South Africa; MICINN, Spain; SRC and Wallenberg Foundation, Sweden; SER,  SNSF and Cantons of Bern and Geneva, Switzerland;  NSC, Taiwan; TAEK, Turkey; STFC, the Royal Society and Leverhulme Trust, United Kingdom; DOE and NSF, United States of America.

\bibliographystyle{atlasnote}
\bibliography{WZcrossPub_3}

\clearpage

% ATLAS Collaboration author list for 10-SEP-2010
% Data extracted on 28-Sep-2010 for paperid 38
%\documentclass[11pt]{article}
%\usepackage{a4wide}\begin{document}
\begin{flushleft}
\label{app:collab} 
{\Large The ATLAS Collaboration}

\bigskip

G.~Aad$^{\rm 48}$,
B.~Abbott$^{\rm 111}$,
J.~Abdallah$^{\rm 11}$,
A.A.~Abdelalim$^{\rm 49}$,
A.~Abdesselam$^{\rm 118}$,
O.~Abdinov$^{\rm 10}$,
B.~Abi$^{\rm 112}$,
M.~Abolins$^{\rm 88}$,
H.~Abramowicz$^{\rm 153}$,
H.~Abreu$^{\rm 115}$
% ...
\bigskip

% ...
$^{10}$ Institute of Physics, Azerbaijan Academy of Sciences, H. Javid Avenue 33, AZ 143 Baku, Azerbaijan\\
$^{11}$ Institut de F\'isica d'Altes Energies, IFAE, Edifici Cn, Universitat Aut\`onoma  de Barcelona,  ES - 08193 Bellaterra (Barcelona), Spain\\
% ...
$^{48}$ Albert-Ludwigs-Universit\"{a}t, Fakult\"{a}t f\"{u}r Mathematik und Physik, Hermann-Herder Str. 3, D - 79104 Freiburg i.Br., Germany\\
$^{49}$ Universit\'e de Gen\`eve, Section de Physique, 24 rue Ernest Ansermet, CH - 1211 Geneve 4, Switzerland\\
% ...
$^{88}$ Michigan State University, Department of Physics and Astronomy, High Energy Physics Group, East Lansing, MI 48824-2320, United States of America\\
% ...
$^{111}$ University of Oklahoma, Homer L. Dodge Department of Physics and Astronomy, 440 West Brooks, Room 100, Norman, OK 73019-0225, United States of America\\
$^{112}$ Oklahoma State University, Department of Physics, 145 Physical Sciences Building, Stillwater, OK 74078-3072, United States of America\\
% ...
$^{115}$ LAL, Univ. Paris-Sud, IN2P3/CNRS, Orsay, France\\
% ...
$^{118}$ Oxford University, Department of Physics, Denys Wilkinson Building, Keble Road, Oxford OX1 3RH, United Kingdom\\
% ...
$^{153}$ Tel Aviv University, Raymond and Beverly Sackler School of Physics and Astronomy, Ramat Aviv, IL - Tel Aviv 69978, Israel\\
% ...
\end{flushleft}

%\end{document} 

\end{document}